\documentclass[final,1p,times]{elsarticle}

 \usepackage{graphics}

\usepackage{amsmath}
\usepackage{amssymb}
\usepackage{subfigure}
\usepackage{float}

\usepackage{color}

\usepackage{rotating}
\usepackage{multirow}
\usepackage{comment}

\journal{J. Comput. Phys.}
\graphicspath{{./pics/}}
\newcommand \be{\begin{equation}}
\newcommand \ee{\end{equation}}
\newcommand \ba{\begin{eqnarray}}
\newcommand \ea{\end{eqnarray}}

\begin{document}

\begin{frontmatter}



\title{
Spatially hybrid computations for streamer discharges
: II. Fully 3D simulations
}


\cortext[cor1]{Corresponding author, Phone: 0031-20-5924208, Fax: 0031-20-5924200}

\author[CWI,TUE]{Chao Li\corref{cor1}}
\ead{Li@cwi.nl}
\author[CWI,TUE]{Ute Ebert}
\ead{Ebert@cwi.nl}
\author[CWI,RUN]{Willem Hundsdorfer}
\ead{Willem@cwi.nl}

\address[CWI]{Centre for Mathematics and Informatics (CWI), P.O.~Box~94079, 1090~GB Amsterdam, The
Netherlands}
\address[TUE]{Department of Applied Physics, Eindhoven University of Technology, P.O.~Box~513,
5600~MB, Eindhoven, The Netherlands}
\address[RUN]{Department of Science, Radboud University Nijmegen, Heyendaalseweg 135, 6525 AJ Nijmegen, The Netherlands}

\begin{abstract}
We recently have presented first physical predictions of a spatially hybrid model that follows the evolution of a negative streamer discharge in full three spatial dimensions; our spatially hybrid model couples a particle model in the high field region ahead of the streamer with a fluid model in the streamer interior where electron densities are high and fields are low. Therefore the model is computationally efficient, while it also follows the dynamics of single electrons including their possible run-away. Here we describe the technical details of our computations, and present the next step in a systematic development of the simulation code. First, new sets of transport coefficients and reaction rates are obtained from particle swarm simulations in air, nitrogen, oxygen and argon. These coefficients are implemented in an extended fluid model to make the fluid approximation as consistent as possible with the particle model, and to avoid discontinuities at the interface between fluid and particle regions. Then two splitting methods are introduced and compared for the location and motion of the fluid-particle-interface in three spatial dimensions. Finally, we present first results of the 3D spatially hybrid model for a negative streamer in air.
\end{abstract}


\end{frontmatter}

\section{Introduction}

\subsection{Streamer discharges and single electron dynamics}

Streamers are ionized fingers that penetrate into a previously non-ionized region under the action of an electric field; they pave the way of sparks and lightning, but they also appear without these subsequent stages in corona reactors or in the form of huge sprites high above thunderclouds; for recent reviews on streamer physics we refer, e.g., to~\cite{Ebert2006/PSST,Ebert2010/JGR}. Streamers are characterized by a thin charge layer around their head, which screens the electric field inside the streamer body and enhances it at the tip. This local enhancement of the electric field allows them to generate additional plasma ahead of the already ionized channel through impact of accelerated electrons onto neutral atoms and molecules.

As streamers in air at standard temperature and pressure contain at least of the order of $10^7$ electrons already at the moment when they emerge from avalanches through the build-up of a space charge layer~\cite{Mon2006:3,Li2008:2}, it is extremely challenging to impossible to follow the growing number of electrons individually inside the streamer discharge during the further evolution. Therefore most streamer simulations are performed in a density approximation for the electron and ion dynamics in so-called fluid models. As reviewed below and in another paper in this issue~\cite{Luque2010/JCPprep}, the numerical solution of fluid streamer models is very challenging as well due to the widely differing length scales of the problem; these simulations give realistic results for the macroscopic dynamics.

But there are certain physical phenomena that are difficult to impossible to simulate with a fluid model. These are related to either small electron numbers or to high electron energies.

\subsubsection{Small electron densities and density fluctuations}

Regions with very low electron densities can dominate the macroscopic discharge evolution in two cases, namely during the initial avalanche regime, and when streamer propagation is unstable and susceptible to branching. In these cases, the presence or absence of single electrons can be visible in the macroscopic evolution. A single electron can start a discharge; and the stochastic distribution of discharge inception times at low voltages can probably be traced back to the stochastic distribution of single electrons during the inception phase. Furthermore,  recent streamer experiments in very pure gases gave the first experimental evidence of avalanches created by single electrons~\cite{Nijdam2010/JPD,Wormeester2010/JPD}; these avalanches do not necessarily lead to branching, but rather can give the discharge a feather-like structure with many parallel "hairs" on the main channel. (We remark in passing that avalanches are neither necessary nor sufficient for streamer branching, as discussed in more detail in~\cite{Ebert2006/PSST,Ebert2011/Nonlinearity}, but they can accelerate the branching of an already unstable streamer head~\cite{Luque2010/JCPprep}.

\subsubsection{High electron energies and electron run-away}

The electron energy distribution in the ionization front can develop a long tail at high energies that is not appropriately modeled by a fluid model, even if the fluid model is extended by additional terms as in~\cite{Li2010:1}. In particular, individual electrons can run away~\cite{Mos2006,Li2009,Cha2010}. And when the electric field exceeds a threshold, the majority of electrons will run away and the density or fluid model will break down completely~\cite{Phe1987,Vrh1992,Vrh2001}.

Much recent interest in extreme electron energies comes from the observation of Terrestrial Gamma-Ray Flashes. These flashes were first detected by the  Compton Gamma Ray Observatory satellite in 1994, and soon a correlation with lightning discharges was found~\cite{Fis1994}. This energetic radiation is now generally believed to be the bremsstrahlung of electrons with extremely high energies. Meanwhile, even a considerable content of positrons is found in these flashes~\cite{Briggs2011/JRL}. But how the electrons gain these energies in the discharge, is under debate, candidates are either relativistic run-away electron avalanches or run-away electrons emitted from streamers, leaders or other lightning discharge processes. Hard radiation was also observed from rocket triggered lightning near the ground~\cite{Dwy2003:2,Sal2009}, and from long sparks generated in laboratory with MegaVolt pulses~\cite{Ngu2008,Dwy2008,Rep2008} as well from streamer coronas generated by a pulsed voltage of 85~kV~\cite{Ngu2010}. In the laboratory experiments and in the approaching lightning leader, it is clear that local field enhancement at electrodes and/or streamer tips is responsible for high electron energies. The same mechanism could be at work in Terrestrial Gamma-Ray Flashes.

\subsubsection{Why to develop a spatially hybrid model}

Discharge inception, streamer feathers and streamer branching on the one hand and the run-away and continued acceleration of electrons in streamer discharges on the other hand give plenty of motivation to study the dynamics of single electrons in a streamer. But --- as remarked above and further elaborated in the next subsection --- neither particle nor fluid model are presently able to deal with these challenges.

In this paper therefore the next step is taken in the development of a spatially hybrid simulation tool. The hybrid simulation treats the majority of electrons inside the streamer in a density approximation, which is appropriate and efficient as the densities are high and the local field is low; the fewer electrons in regions with lower densities and high field are treated in a full particle model. New steps in the present paper concern the consistent density approximation of the particle model, and the construction and motion of the interface between particle and fluid region in full three dimensions. This model will allow to calculate how the dynamics of single electrons in the high field region influences the streamer motion, and how this high field region in turn is generated by the growing streamer body.

\subsection{The state of particle and fluid models of streamers}

\subsubsection{Particle models}

Most particle simulations of plasmas in general, and of streamers in particular, are carried out with a particle-in-cell/Monte Carlo collision (PIC/MCC) model. The PIC/MCC model follows the free flight of single electrons and treats all the (important) collisions stochastically~\cite{Kun1988:2,Dow2003,Mos2006,Cha2008}.

An important constraint of particle models is the number of electrons they can treat;  the electron number can easily exceed limitations of computer memory and computational power. Particle simulations with real particles can only be performed for the early phase of streamers, such as the avalanche phase or the avalanche-to-streamer transition. To follow the streamers in later phases, super-particles are typically used, where one computational particle stands for the mass and charge of many real particles. However super-particles have their own drawbacks, not only through the lower resolution, but more importantly, because they cause numerical heating and stochastic errors, as shown in~\cite{Vah1993,Li2008:2}, which can have a detrimental influence on the simulation results.

Due to the heavy computational cost for a large amount of particles as well as for solving the 3D Poisson equation, a full 3D particle model of a streamer is difficult or in many situations impossible to realize. A compromising approach is to keep the positions and velocities of the particles defined in 3D space as in the standard PIC/MCC procedure, while calculating the electric field in 1D~\cite{Kun1986:2,Kun1988:2} or 2D~\cite{Cha2008,Dow2003}, and then to describe the electron motion in the electric field by interpolation. Runaway electrons generated by streamers have been addressed in the work of Moss $et$ $al.$~\cite{Mos2006} where the 3D electric field is simplified into a two-step function in 1D: the high field value stands for the field at the streamer head and a low field value stands for the field far ahead of the streamer head. Chanrion $et$ $al.$~\cite{Cha2010} recently published runaway simulation results in a 2D geometry with radial symmetry. Superparticles are used in both simulations, but an energy-dependent re-sampling technique for the superparticles is developed in~\cite{Cha2010}, and a low weight can be attributed to energetic electrons which allows to study electron runaway in negative streamers with higher precision.

In our particle model, that constitutes a part of our hybrid model, both the electric field and the electrons are defined and calculated in 3D.

\subsubsection{Fluid models}

The fluid model approximates electrons by continuous densities and is therefore computationally much more efficient. The fluid model is in most cases derived by averaging over the Boltzmann equation~\cite{Hir1964,Shk1966,Cha1974}. The transport coefficients such as mobilities and diffusion rates, and the reaction rates are approximated as functions of the local electric field or the local mean electron energy. (More details on the local field approximation and on the energy equation can be found in~\cite{Hag2005,Li2010:1}.)

3D fluid models have become available only recently~\cite{Pan2005:2,Luque2008:2}; progress was from 1D~\cite{War1965} and 1.5D~\cite{Dav1964} density simulations to various 2D~\cite{Dha1987,Guo1993,Bab1996} fluid descriptions developed in the last 20 years. While in 2D, the radial nonuniformity of particle densities and the local field enhancement are included in the fluid description, the streamer evolution after branching or the interaction of streamers with another~\cite{Luque2008/PhRvL} or with lateral walls requires a model in full 3D.

Most streamer simulations are carried out with fluid models~\cite{Dha1987,Guo1993,Vit1994,Kul1995,Bab1996,Nai1997,Arrayas2002/PhRvL,Roc2002,Pan2003,Liu2004,Mon2006:3,Luque2008:3,Cel2009,Wormeester2010/JPD}.
Solving the fluid equations numerically is not an easy task due to the multiscale nature of streamers. New simulation techniques have been developed~\cite{Mon2006:3,Luque2008:3,Pan2008,Pasko2007}, a further review can be found~\cite{Luque2010/JCPprep}. A high resolution for the inner structure of the streamer head is needed, where the electron and ion densities decay sharply and where the electric field changes rapidly in space and time. These numerical challenges have been met by adaptive grid refinement~\cite{Mon2006:3,Pan2008,Luque2008/PhRvL}. Improvement of the numerical techniques is only one aspect of the recent development in the fluid models; more physics is included by using more complicated and realistic plasma-chemical models~\cite{Kos1992,Pan2005:1,Pet2007,Luque2008:3}, better techniques of modeling electrode geometries~\cite{Bab1996,Luque2008:3,Cel2009} and efficient descriptions of non-local photo-ionization sources~\cite{Pan2001,Bog2007,Liu2004,Liu2006,Luque2007,Bourdon2007/PSST}.

\subsection{Developing the hybrid model}

To combine the computational efficiency of the fluid model with the detailed physical description of the particle model, we here take the next steps in the development of a model that is hybrid in space, coupling a particle description of the electrons in the region of high field and low electron density with a fluid description in the region with low field and high densities. We refer to~\cite{Li2007,Li2009,Li2010:1} for the detailed concept of the spatially hybrid model.

In previous work, we have identified deviations between particle and naive fluid descriptions of planar streamer ionization fronts~\cite{Li2007}, and we have described how to construct the numerical interface between a particle and a fluid model in 1D, and how to perform the hybrid simulation of a 1D streamer front~\cite{Li2008:1,Li2010:1}. In~\cite{Li2009} we have presented first results of hybrid calculations in full 3D, emphasizing on the acceleration of electrons to energies above 3.5 keV in a streamer head.

In this paper, we focus on two aspects in the model development, namely on the calculation of transport and reaction coefficients for the fluid model from the particle model, and on the construction of the interface between particle and fluid model in a 3D simulation.

First, a set of transport coefficients and reaction rates is derived from the particle model for the fluid model; this is done for air in section~\ref{sec:swarm_tran_rea}, and for pure nitrogen, oxygen and argon in the appendix. In our previous papers~\cite{Li2007,Li2010:1}, we already had calculated these coefficients for pure nitrogen, but there we fixed the elastic total cross section rather than the elastic momentum transfer cross section. Therefore the calculated coefficients were inconsistent with each other when different distribution for the scattering angles were used in the electron-neutral collisions.

Second, two splitting methods are developed in section~\ref{sec:1st_interface} and~\ref{sec:2nd_interface} to determine the interface between the particle and the fluid region. The first one, the column based splitting, is a 3D extension of the splitting method developed for planar fronts~\cite{Li2010:1}. The first 3D results of this method were reported in~\cite{Li2009}, here we give the numerical details. The second splitting method, the full 3D splitting, is an improvement of the first method. It generates model interfaces wherever needed, e.g., at both ends of a double headed streamer or even when avalanches or streamers break up in several parts.

We also present new results for the 3D fluid model as well as for the 3D hybrid model in air.

\subsection{Organization of the paper}

Particle and fluid model are the basic components of the 3D hybrid model. We first give the details of numerical methods and their implementation in Section~\ref{sec:3d_particle} for the particle model and in Section~\ref{sec:3dfluid} for the fluid model.
The coupling of the two models is discussed in Section~\ref{sec:3D_hyb_model}
in which the general coupling procedure is given in Section~\ref{sec:gen_coupling_pro}, the methods to split the simulation domain into particle regions and fluid regions are then discussed in Section~\ref{sec:1st_interface} and Section~\ref{sec:2nd_interface}, and the definition of the buffer region is discussed in Section~\ref{sec:buffer_region}.
In Section~\ref{sec:sim_result}, we present the 3D hybrid simulation result for a negative streamer in air.
We finish with some concluding remarks and suggestions for further research in Section~\ref{sec:conclusion}.
The transport coefficients and reaction rates for nitrogen, oxygen, and argon are given in Appendix~\ref{app:parameters}. 



\section{Particle and fluid model} \label{ch:parAndFluid}

We here lay the basis for constructing the hybrid model in 3D in section~\ref{sec:3D_hyb_model}. In subsection A, we recall the essentials of the particle model with Monte Carlo procedure and input data for the differential cross sections; and we explain how we calculate transport and reaction coefficients for the fluid model in the most consistent manner from the particle model. We recall that the fluid model needs an extension in high fields to fit the particle model~\cite{Li2010:1}. In subsection B, we discuss the fluid model and its numerical implementation and how to solve the Poisson equation. Finally, we present first simulation results of the fluid model in 3D on a Cartesian grid.


\subsection{Particle model}\label{sec:3d_particle}

\subsubsection{The Monte Carlo procedure}

\begin{figure}
 \begin{center}
     \includegraphics[width=0.35\textwidth]{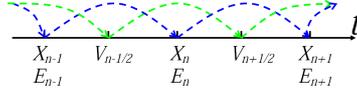}
\caption{The schematic illustration of the Leapfrog scheme.} \label{fig:leapfrog}
\end{center}
\end{figure}

The particle model follows the standard procedure of Monte Carlo (MC) particle simulations in plasma physics; it is summarized here. The particle model tracks the motion of all electrons while treating the neutral atoms and molecules as a random background. Since the mobility of ions is two orders of magnitude smaller than that of electrons, ions are treated as immobile. As the ionization density in streamers is low at standard temperature and at standard pressure or below \cite{Ebert2010/JGR}, only collisions with neutrals need to be taken into account. During a collision with a neutral, the electrons change velocity and energy (in the case of an elastic collision only velocity); the collisions are modeled as instantaneous. Between collisions, electrons follow the equation of motion \( \ddot{\bf x}=q{\bf E}/m\); numerically this motion is solved with the leapfrog method
\begin{eqnarray}
{\bf x}_{n+1} & = & {\bf x}_n+\Delta t {\bf v}_{n+\frac{1}{2}} ,  \\
{\bf v}_{n+\frac{1}{2}} & = & {\bf v}_{n-\frac{1}{2}} + \Delta t \frac{q}{m}{ \bf E}({ \bf x}_n,t_n) ,
\end{eqnarray}
 as illustrated in Fig.~\ref{fig:leapfrog}. Here $q$ and $m$ are elementary charge and electron mass, ${ \bf x}_n = (x,y,z)_n$ is the electron position at time $t_n$, and ${\bf v}_{n+\frac{1}{2}}=(v_x,v_y,v_z)_{n+\frac{1}{2}}$ is the electron velocity at time $t_{n+\frac{1}{2}}$.

The collisions of electrons with the neutral background are characterized by probability distributions for the collision time, for the type of collision (elastic, inelastic and ionizing processes) and for the scattering angle, and in the case of an ionizing collision also for the distribution of energy between the two outcoming electrons. The actual collision events are sampled from the probability distributions through random numbers in a Monte Carlo procedure. The probability distributions are determined by the differential cross sections for the respective gas composition and by the gas density. The cross sections used in the present paper are summarized in subsection~\ref{sec:Data_DCS}.

\begin{figure}
 \begin{center}
     \includegraphics[width=0.6\textwidth]{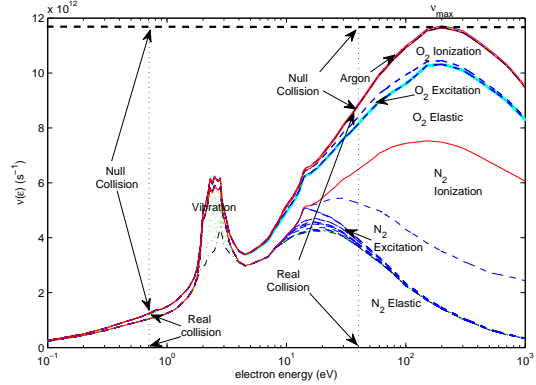}
\caption{Collision frequencies $\nu(\epsilon)$ in air at standard temperature and pressure as a function of electron energy $\epsilon$. The frequencies of different collision types are plotted on top of each other such that they add up to the total collision frequency. From below, the collision frequencies are successively elastic, rotational and vibrational, electron exciting and ionizing collisions with nitrogen, elastic, attachment, rotational and vibrational, exciting and ionizing collisions with oxygen, and elastic, exciting and ionizing collisions with argon. When the collision frequency is smaller than maximal $\nu(\epsilon)<\nu_{max}$, artificial "null collisions" without physical effect are added to make the Monte Carlo procedure more efficient.}
\label{fig:fre_total}
\end{center}
\end{figure}

The cross sections and therefore the collision frequencies in general depend on the electron energy. But if the next collision time is calculated for each electron individually depending on its present energy, and if the change of energy during the flight is not taken into account, the calculation is both numerically expensive and inaccurate. Therefore the ``null collision''~\cite{Boe1982} is used in most MC codes, which is actually a pseudo collision because nothing changes in this collision process. With the null collision technique, the next collision time is sampled by drawing a random number right after the previous collision; the probability distribution is based on the maximum of the collision frequency $\nu_{max}$ over all reasonably occurring electron energies. Then when the collision time is reached, the actual energy $\epsilon_c$ of the incident electron is calculated. Another random number is drawn to determine the type of collision process (elastic, inelastic, ionizing or null) for an electron with $\epsilon_c$. The null collision accounts for the probability that no collision occurs at the previously determined collision time, because the actual collision frequency for electrons with energy $\epsilon_c$ is smaller than $\nu_{max}$. The loss of electron energy is determined by the type of collision process, and the scattering angle is determined by the incident electron energy and sampled with another random number. If the collision is ionizing, another random number is drawn to determine the energy distribution between the two out-coming electrons. The scattering angles of the out-coming electrons are determined by their energies as described in the next subsection. The electrons then follow their Newtonian (or eventually relativistic) trajectory up the next collision.

Fig.~\ref{fig:fre_total} shows the total collision frequencies as a function of electron energy in air (modeled as 78.12\% nitrogen, 20.946\% oxygen, and 0.934\% argon) at standard temperature and pressure; the frequencies of elastic, rotational and vibrational, different exciting and finally ionizing collisions are plotted on top of each other such that they add up to a total collision frequency.  This total collision frequency reaches a first peak around $\epsilon=$ 2 eV due to the vibrational collisions. For growing $\epsilon$ $>$ 2 eV, it then first drops and then increases again up to a maximum $\nu_{max}$ at $\epsilon=$ 150 to 200 eV, after which it decreases. When the collision frequency decreases, the mean free path length of the electrons increases. If the electric field is so high, that the mean energy gain between collisions exceeds the mean energy loss during collisions, the electrons can continuously gain energy and run away. Electron run-away is one of the effects we want to track with our model.

An important difference with most other PIC/MCC codes is that we use real particles instead super-particles. Since our particle model is designed for the region with high field enhancement where the electron density is relatively low, we intend to use our particle model with real particles. Therefore, as long as one particle is identified with one real electron, the common problems of a PIC/MCC procedure such as numerical heating, stochastic errors and low resolution at essential regions cannot occur.


The electric field is calculated using a fast Poisson solver, see more discussion in Section.~\ref{sec:test_fishpack}.

\subsubsection{Data for differential cross sections}\label{sec:Data_DCS}

The cross section data can be obtained through either one or more of the following means: (1) measurement from single-scattering beam experiments~\cite{Bru2002}, (2) ab initio quantum theoretical calculations~\cite{Bettega1993/PRA,Winstead2000/AAMOP} or (3) inversion of swarm experimental data~\cite{Phe1985,Cro1994}. Different cross section databases have been used by different authors, depending also on the stage of streamer development, for example, Dowds $et$ $al.$ uses the National Institute of Standards and Technology (NIST) electron impact cross section~\cite{NIST_impa} in a particle avalanche model~\cite{Dow2003}, Babich $et$ $al.$ uses the Evaluated Electron Data Library (EEDL)~\cite{EPDL97} for the simulation of high energy electron avalanches~\cite{Bab2001}, and Li $et$ $al.$ and Chanrion $et$ $al.$ use the Siglo database in the streamer simulations~\cite{Li2008:2,Cha2008}. None of the cross section databases mentioned above covers the energy range from low energy (less than 1 eV) to very high energy (above 1 MeV). 

In our model, the cross sections for electrons with energy up to 1~keV colliding with nitrogen, oxygen or argon particles are taken from the {\sc siglo} database~\cite{Siglo}, which includes all important collision processes in the streamer development. Above 1~keV, the Born approximation~\cite{Liu1987} is used for elastic collisions, a fit formula in~\cite{Mur1988:1} is implemented for the electronically exciting collisions and the Born-Bethe approximation~\cite{Ino1971,Gar1988} is used for ionizing collisions; these approximations continuously extend the {\sc siglo} database. The distribution of scattering angles is discussed further below.

\begin{figure}
 \begin{center}
     \includegraphics[width=0.5\textwidth]{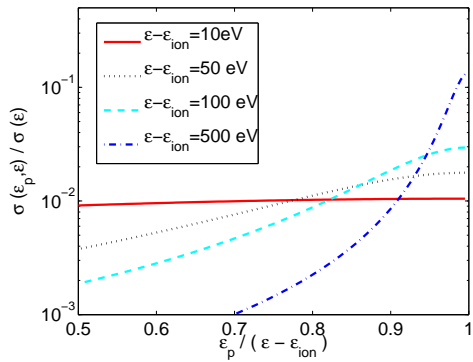}
\caption{Normalized probability that the primary out-coming electron carries a fraction r$=\epsilon_p/(\epsilon-\epsilon_{ion})$ of the total available energy. The incoming electrons have energies $\epsilon=$ 10 (solid), 50 (dotted), 100 (dashed), 500 (dash-dotted) eV; $\epsilon_{ion} = 15.6$ eV is the ionization energy of nitrogen. The probabilities are normalized over 100 $\Delta$r from r=0 to 1.}
\label{fig:Opal_ejected}
\end{center}
\end{figure}

In an ionizing collision, the incident electron loses energy to liberate an electron from the neutral molecule or atom. There are several measurements of the angular and energy distribution, the so called doubly differential cross sections for ionization. The energy distribution of the secondary electrons ejected upon ionizing collisions also has a clear influence on the simulated particle swarms or streamer avalanches~\cite{Kun1986:3}, since clearly electron run-away is more likely if one electrons carries away most of the energy after the collision while the other one is slow. The measurements of the secondary electron spectra by Opal $et$ $al.$ ~\cite{Opa1971,Gor1987} have become a standard, and their empirical fitting equation has been widely used in different simulation groups. Opal's empirical fitting is used in our calculation as well.

In Fig.~\ref{fig:Opal_ejected}, we show the probability distribution $\sigma(\epsilon_p,\epsilon)$/$\sigma(\epsilon)$ that the primary out-coming electron carries the fraction r$=\epsilon_p/(\epsilon-\epsilon_{ion})$ of the available energy, for given incident electron energies $\epsilon=$ 10 (solid), 50 (dotted), 100 (dashed), 500 (dash-dotted) eV; here $\epsilon_p$ is the energy of the primary electron (that takes more energy than the secondary electron), and $\epsilon_{ion}$ is the ionization threshold energy of nitrogen. $\sigma(\epsilon)$ is the total ionization cross section for electrons with incident energy $\epsilon$, and $\sigma(\epsilon_p,\epsilon)$ is the cross section for incident energy $\epsilon$ and out-coming primary electron energy $\epsilon_p$. The probabilities are normalized over 100 $\Delta$r for $r\in(0,~1)$. The plot shows that an incident electron with high energy is likely to keep most of its energy.

The scattering angles of primary and secondary electron are determined by their energies~\cite{Boe1982} with
\be \cos \chi_i= \sqrt{ \epsilon_i/(\epsilon-\epsilon_{ion})  }, \ee
where $i=1,2$ indicates the primary or secondary electron. If the primary electron takes most of the energy, i.e., $\cos \chi_i \approx 1$, it leaves the collision mostly  in a forward direction; the secondary electron with low energy $\cos \chi_i \approx 0$ then flies in a direction perpendicular to the primary one.


For the distribution of scattering angles in elastic collisions, analytic approximations are used as well in our model due to the relative lack of experimental data.
Okhrimovskyy $et$ $al$. has derived a formula from the integrated and momentum transfer elastic cross section data from Phelps and Pitchford~\cite{Phe1985} for nitrogen; it has been implemented in our model for the electron-N$_2$ elastic collisions
\be \label{equ:Okhr} I(\epsilon,\chi) = \frac{1}{4 \pi}\frac{1-\xi^2(\epsilon)}{[1-\xi(\epsilon)\cos \chi]^2} , \ee
where $I(\epsilon,\chi)$ is the normalized differential cross section, $\epsilon$ stands for the electron energy in eV and $\chi$ is the scattering angle with respect to the direction of the incident electron.
An empirical fit of the parameter $\xi$ for nitrogen is
\be \xi(\hat{\epsilon})=\frac{0.065~\hat{\epsilon} +0.26~\sqrt{\hat{\epsilon}}}{1+0.05~\hat{\epsilon} +0.2~ \sqrt{\hat{\epsilon}}}-\frac{12~\sqrt{\hat{\epsilon}}}{1+40~\sqrt{\hat{\epsilon}}}, \label{equ:Okhr_xi} \ee
where $\hat{\epsilon}=\epsilon/{\rm eV}$ is the electron energy in dimensionless units. The formula derived by Surendra $et$ $al.$~\cite{Sur1990b} has been used in a particle simulation for electron-O$_2$ and electron-Ar collisions in~\cite{Sur1990b,Vah1995}, and is also used here in our model for electron-O$_2$ and electron-argon elastic collisions,
\be I(\epsilon,\chi) = \frac{\epsilon}{4 \pi ~[1+\epsilon~ \sin^2(\chi/2) ] \ln(1+\epsilon)}. \label{equ:Surendra}
 \ee

The experiment has shown that the electron scattering angles are different for different collision processes~\cite{Bru2002}. A complete particle model could apply different differential cross sections for different excitational collision process. However, the differential cross sections are not available for all essential collision processes in the literature, and the available data are mostly measured for electrons at a few discrete energies and not in the form ready to be implemented in the particle model.
Therefore, the electron scattering from the rotational, vibrational and electronic excitations is assumed to have the same scattering probabilities as from the elastic collisions.
In an excitational collision, the electron scattering angle is first sampled based on the incident electron energy, then the energy loss in the excitation collision is subtracted from this energy.

A methodological difference with our previous swarm calculation for nitrogen~\cite{Li2007} should be noted.  In~\cite{Li2007} we used the elastic momentum transfer cross section
\be \sigma_m(\epsilon)= 2 \pi \int_0^{\pi} (1-\cos \chi)~I(\epsilon,\chi) \sin \chi d\chi \label{Eq:sigma_m} ,
\ee
from the {\sc siglo} database, but we fixed it as the total elastic cross section. This lead to inconsistent transport and reaction coefficients between models with isotropic and anisotropic scattering. Now the total elastic cross section $\sigma_t$ 
\be \sigma_t(\epsilon)= 2 \pi \int_0^{\pi} ~I(\epsilon,\chi) \sin \chi d\chi  \label{Eq:sigma_t} 
\ee
is adjusted depending on the applied scattering model, such that a constant momentum and energy transfer rate is obtained independently of the angular scattering model. The change has negligible influence on the electron transport and reactions for low energy electrons, but influences become significant for electrons with higher energies, as shown in~\cite{Kun1986:1,Sur1990b} and also found in our study.

When anisotropic scattering is applied in the particle model, the ratio of elastic momentum transfer cross section $\sigma_m$ and total elastic cross section $\sigma_t$ can be obtained from Eq.~(\ref{Eq:sigma_m},\ref{Eq:sigma_t}). The total elastic cross section can then be calculated from the elastic momentum transfer cross section with these ratios. The ratio for nitrogen is written as~\cite{Okh2002}
\be \frac{\sigma_m(\epsilon)}{\sigma_t(\epsilon)}= \frac{1-\xi(\epsilon)}{2\xi(\epsilon)^2} \left(  (1+\xi(\epsilon)) \ln \frac{1+\xi(\epsilon)}{1-\xi(\epsilon)} -2\xi(\epsilon) \right)  \ee
where $\xi$ is the same as Eq.~\ref{equ:Okhr_xi}, and for oxygen and argon the ratio is written as~\cite{Sur1990b,Vah1995}
\be \frac{\sigma_m(\epsilon)}{\sigma_t(\epsilon)}= \frac{2}{\ln(1+\epsilon)} \left( 1-\frac{\ln(1+\epsilon)}{\epsilon} \right), \ee
where the function $I(\epsilon, \chi)$ are from Eq.~\eqref{equ:Okhr} and ~\eqref{equ:Surendra} respectively.

\subsubsection{Calculation of transport and reaction coefficients in air} \label{sec:swarm_tran_rea}

The electron transport coefficients and reaction rates are calculated in particle swarm experiments and used in the fluid model to keep the two models consistent. Two sets of transport coefficients and reaction rates have been calculated by particle swarm experiments: flux coefficients and bulk coefficients~\cite{Rob2005}. (The use of flux or bulk coefficients is discussed in the next section.)
In a constant field $E$, the bulk coefficients are calculated from the swarm using the following equations:
\begin{eqnarray}\label{equ:tof_bulk}
\mu(E)|E|&=&
\frac{\langle z(t_2)\rangle -\langle z(t_1)\rangle}{t_2-t_1},\\
\alpha(E) & = & \frac1{\mu(E)|E|}\;\frac{\ln N_e(t_2) - \ln N_e(t_1)}{t_2-t_1},\\
{\bf D}(E) & = & \langle {\bf r} {\bf v} \rangle -\langle {\bf r}  \rangle \langle {\bf v} \rangle, \nonumber\\
\end{eqnarray}
and for the flux coefficients:
\begin{eqnarray}\label{equ:tof_bulk}
\mu^*(E)|E|&=& \langle v_z \rangle \\
\alpha^*(E) & = & \frac1{\mu^*(E)|E|}\;\frac{\ln N_e(t_2) - \ln N_e(t_1)}{t_2-t_1},\\
{\bf D}^*(E) & = & \frac{( \langle {\bf r}(t_2)^2 \rangle-\langle {\bf r}(t_2)\rangle ^2) -(\langle {\bf r}(t_1)^2 \rangle -\langle {\bf r}(t_1)  \rangle ^2)  }{2(t_2-t_1)}, \\
k_1(E) & = & \frac{\mu(E)-\mu^*(E)}{\alpha^*(E)\;\mu^*(E)} =  \frac{\mu(E)-\mu^*(E)}{\alpha(E)\;\mu(E)}  ,
\end{eqnarray}
where $E$ is applied along $z$~axis, $N_e(t)$ is the total number of electrons at time $t$, ${\bf r} = (x, y, z) $ and ${\bf v} = (v_x, v_y, v_z) $ are the position and velocity of electrons and $\langle \cdots \rangle$ denotes the average over all particles. $k_1$ is the nonlocal parameter in the extended fluid model (see Section.~\ref{sec:3dfluid} or~\cite{Li2010:1} for the definition of $k_1$). The particle swarm experiments have been done for a range of electric field from $E=5$ to $250$ kV/cm for air at standard temperature and pressure. By using similarity laws, they can be easily converted to other temperature and pressures. 

 The flux coefficients represent the real motions of electrons, for example, the flux mobility times electric field $\mu E$ represents the mean drift velocity of the existing electrons neglecting creation or loss of electrons.
The bulk coefficients describe the averaged change of the electron bulk as a whole. For example, the bulk drift velocity is the displacement of the mean position of the electron swarm. The mean displacement of the swarm is not only the result of the electron drift, but also of the spatially varying mean electron energies, ionization and attachment rates throughout the swarm~\cite{Li2007,Li2010:1}.
Bulk transport coefficients are experimentally measurable transport quantities, while the measurement of the flux transport coefficients is rather difficult for swarms at high electric field at normal pressure.
For more discussion of flux and bulk coefficients, we refer to Section~\ref{sec:imp_3dfluid} and to~\cite{Rob2005,White2009/JPD,Petrovic2009/JPD,Li2010:1}.

\begin{figure}
\centering
   \subfigure[\label{fig:coe_mu} ~Mobility $\mu(E)$ ]{
         \includegraphics[width=.40\textwidth]{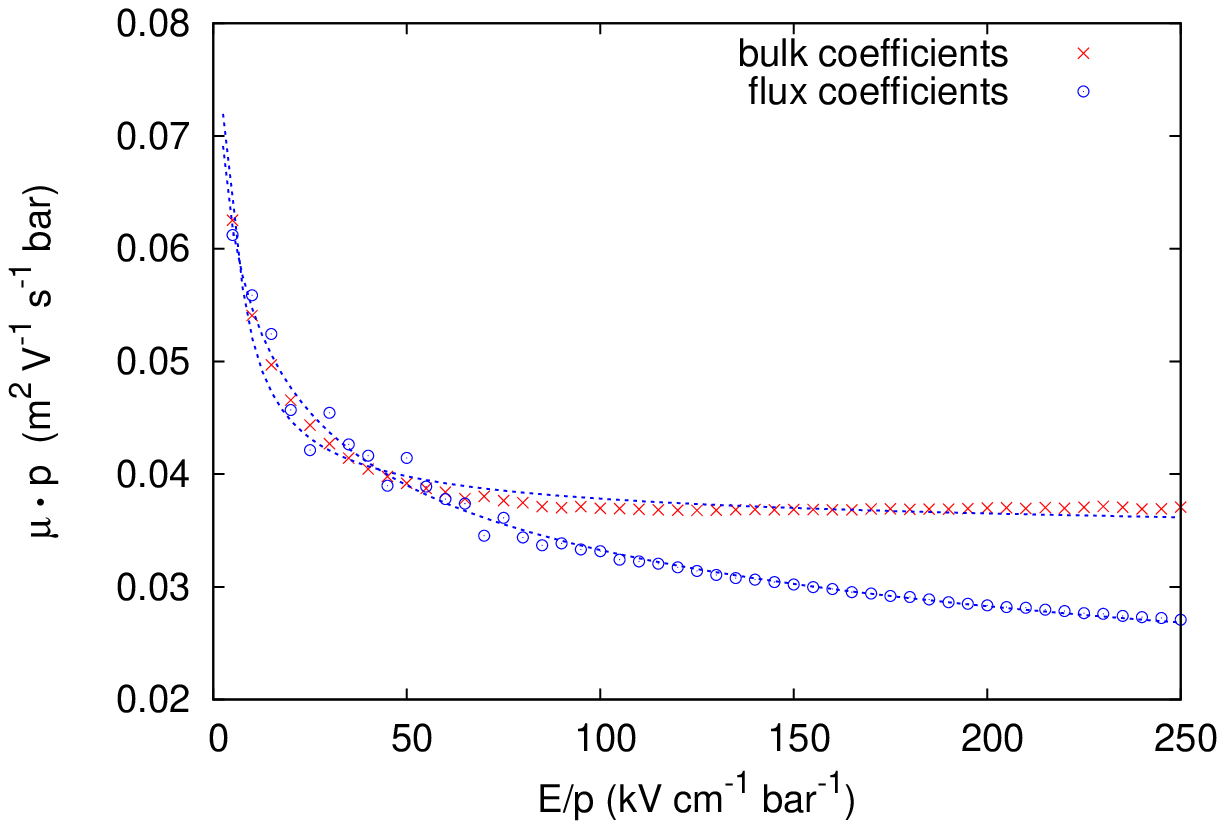}
}
   \subfigure[\label{fig:coe_al} ~Ionization and attachment rates $\alpha(E)$ ]{
         \includegraphics[width=.40\textwidth]{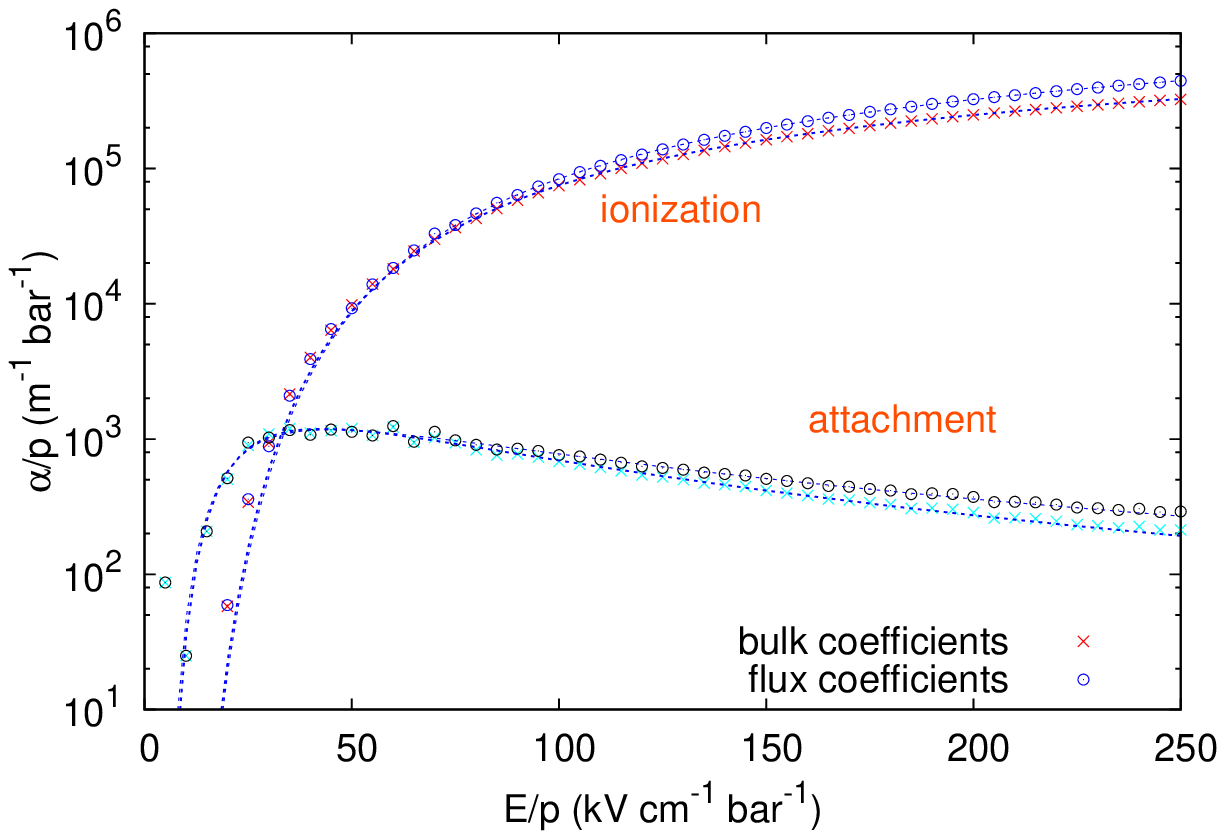}
}
   \subfigure[\label{fig:coe_dr} ~Transversal diffusion $D_T(E)$]{
         \includegraphics[width=.40\textwidth]{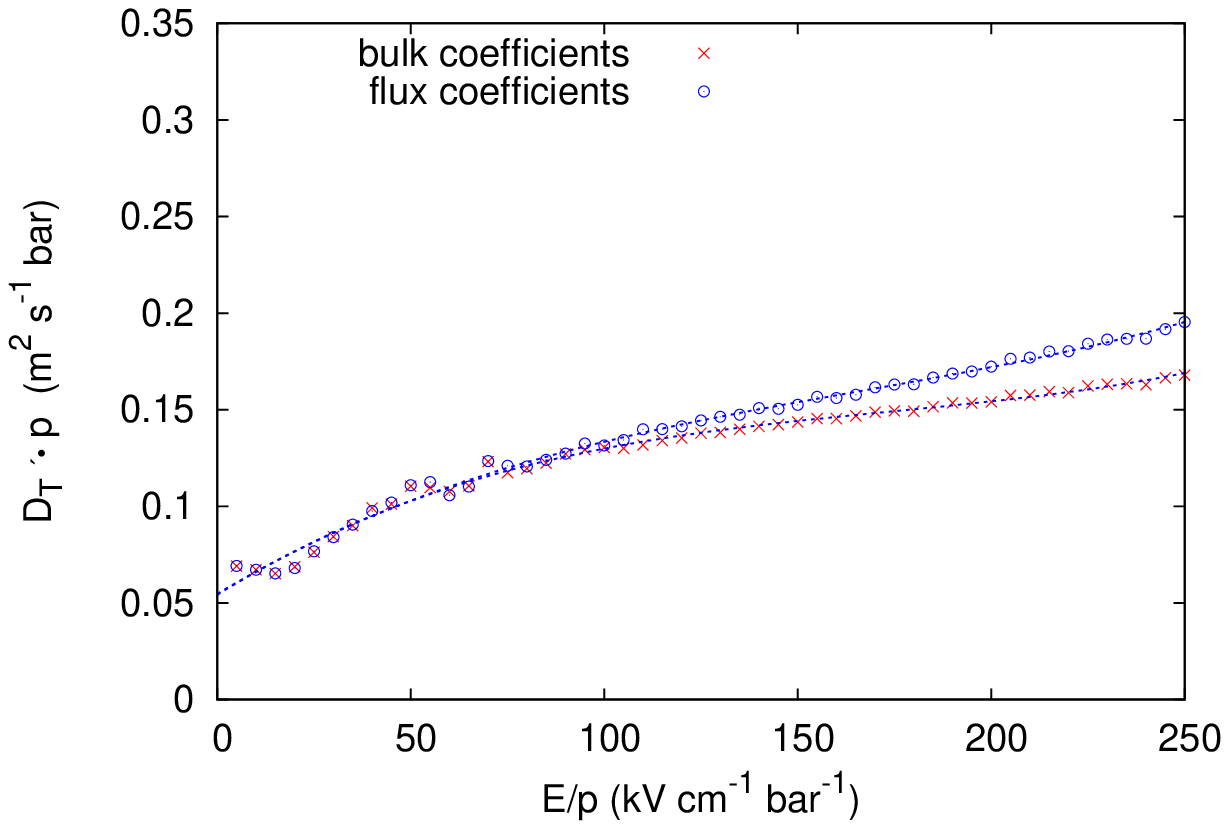}
}
    \subfigure[\label{fig:coe_dl} ~Longitudinal diffusion $D_L(E)$ ]{
         \includegraphics[width=.40\textwidth]{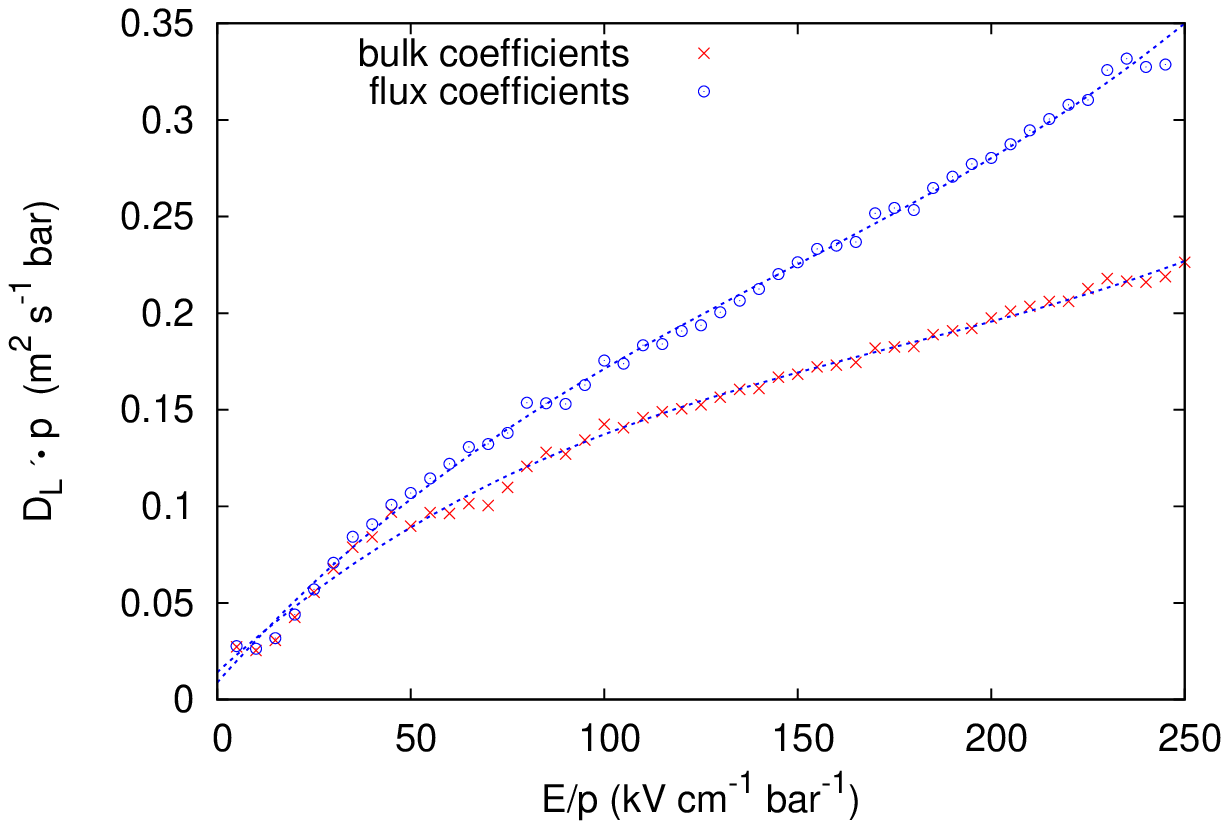}
}
   \subfigure[\label{fig:coe_en} ~Average energy $\varepsilon(E)$]{
         \includegraphics[width=.40\textwidth]{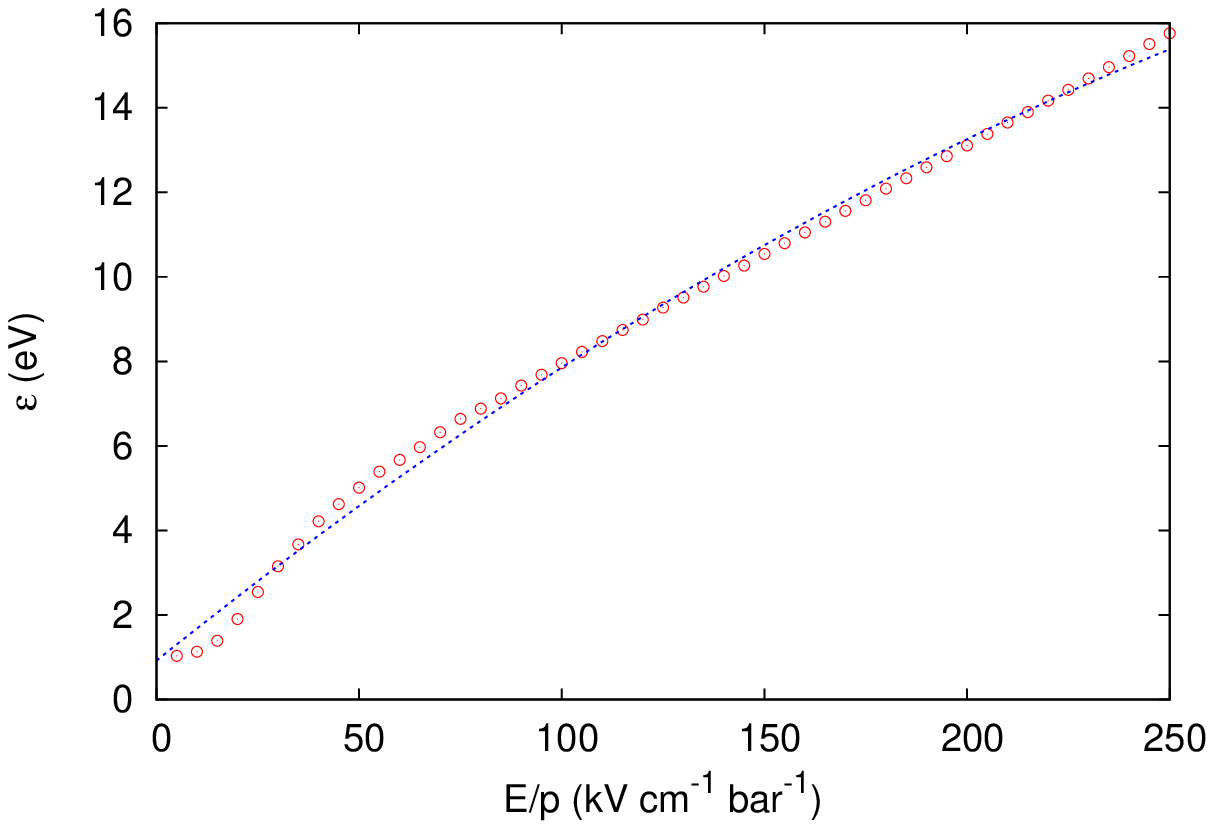}
}
   \subfigure[\label{fig:coe_k1} ~Coefficient k$_1$(E)]{
         \includegraphics[width=.40\textwidth]{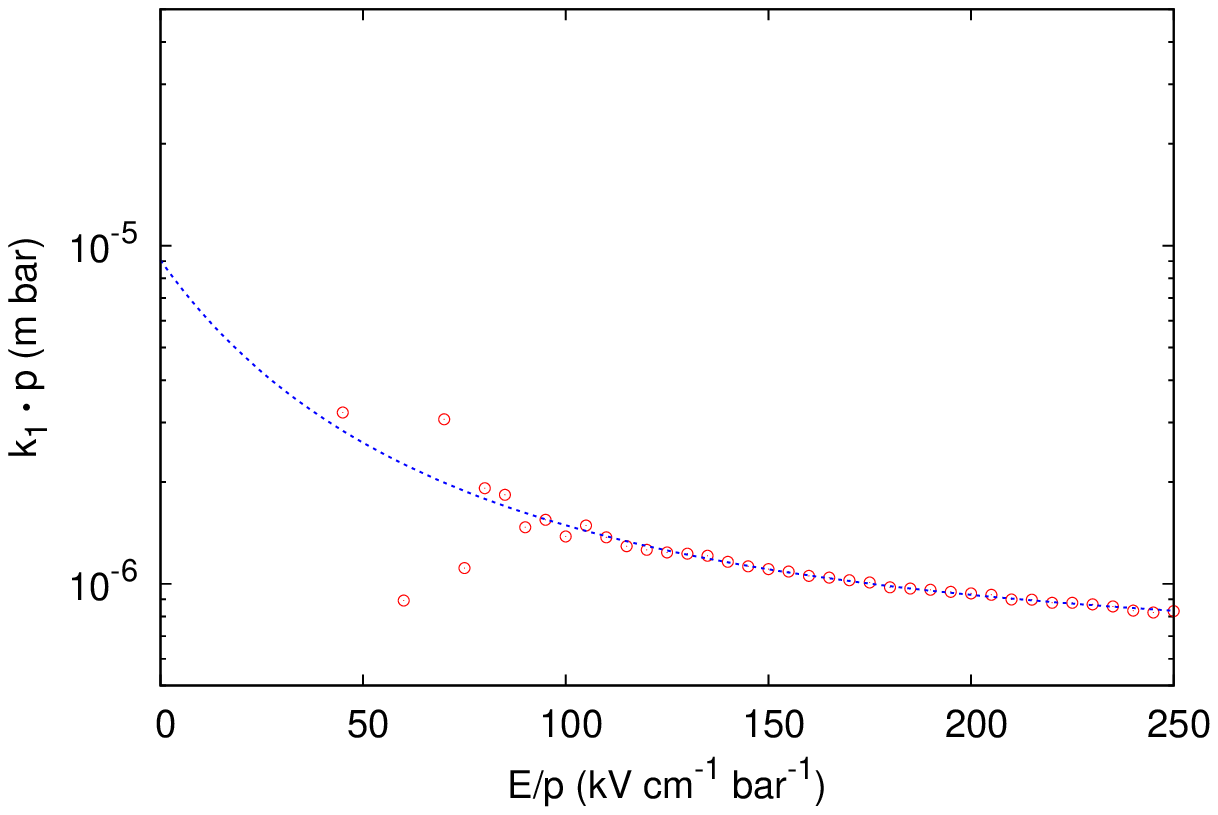}
}
\caption{Shown are the transport coefficients, ionization and attachment rate, and the mean energy of electrons in air as a function of the reduced electric field at room temperature. The presented coefficients are the bulk coefficients  (marked with ``x'') and the flux coefficients (marked with ``o'') calculated from the particle swarm experiments. Their empirical fits are indicated with dashed lines.}
\label{fig:rate_air}
\end{figure}

In Fig.~\ref{fig:rate_air}, the transport parameters: electron mobility $\mu$, electron diffusion tensor ${\bf D}$, and the ionization rate $\alpha_i$ together with the attachment rate $\alpha_{att}$ of electron ensembles are presented for both bulk coefficients (marked with ``x'') and flux coefficients (marked with ``o''). In Fig.~\ref{fig:Com_with_Bolsig_Air}.e and f, the mean energies $\bar{\epsilon}$ and the nonlocal parameter $k_1$ are also presented. For simplicity of notation, we fix standard temperature T$_0$ and vary the pressure $p$. Quantities really scale with gas density $n_0=p/KT$.

For the particle swarm generated coefficients presented in Fig.~\ref{fig:rate_air}, we generated empirical fits. They are indicated by dashed lines in Fig.~\ref{fig:rate_air}.
The fitting functions for the bulk coefficients are
\ba
\mu(\bar{E})  ~/  {\rm m^2  V^{-1}s^{-1}} & = &  \exp \left[   {-3.18-2.65\times10^{-2}\cdot \ln{\bar{E}}+3.44 /{\bar{E}} - (4.99/{\bar{E}})^2} \right] \nonumber \\
\alpha_i(\bar{E})~ / {\rm m^{-1}}  & = & \exp\left[1.14\times10+3.73\times10^{-1}\cdot \ln{\bar{E}}-1.88\times10^{2}/{\bar{E}}  \right] \nonumber \\
\alpha_{att}(\bar{E})~ / {\rm m^{-1}} & = & \exp\left[1.63\times10-1.95\cdot \ln{\bar{E}}-8.32\times10/{\bar{E}}  \right] \nonumber \\
D_T(\bar{E}) ~/{\rm m^2s^{-1}} & = &   {5.43\times10^{-2}+  1.24\times10^{-3}\cdot {\bar{E}}-6.01\times10^{-6} \cdot \bar{E}^2 + 1.15\times10^{-8} \cdot \bar{E}^3} \nonumber \\
D_L(\bar{E})  ~/ {\rm m^2s^{-1}} & = &   {1.42\times10^{-2}+  1.84\times10^{-3}\cdot {\bar{E}}-7.46\times10^{-6} \cdot \bar{E}^2 + 1.41\times10^{-8} \cdot \bar{E}^3}
\ea
and for the flux coefficients:
\ba
\mu^*(\bar{E}) ~/ {\rm m^2V^{-1}s^{-1}}  & = & \exp\left[{-2.30-2.39\times10^{-1}\cdot \ln{\bar{E}}-6.57\times10^{-1}/{\bar{E}} + (6.71\times10^{-1}/{\bar{E}})^2} \right] \nonumber \\
\alpha^*_i(\bar{E})~ /{\rm m^{-1}} & = & \exp\left[1.04\times10+6.01\times10^{-1}\cdot \ln{\bar{E}}-1.86\times10^{2}/{\bar{E}}  \right] \nonumber \\
\alpha^*_{att}(\bar{E}) ~/ {\rm m^{-1}} & = & \exp\left[1.49\times10-1.63\cdot \ln{\bar{E}}-7.30\times10/{\bar{E}}  \right] \nonumber \\
D_T^*(\bar{E}) ~/ {\rm m^2s^{-1}} & = &   {5.47\times10^{-2}+  1.19\times10^{-3}\cdot {\bar{E}}-5.06\times10^{-6} \cdot \bar{E}^2 + 1.02\times10^{-8} \cdot \bar{E}^3} \nonumber \\
D_L^*(\bar{E}) ~/ {\rm m^2s^{-1}}  & = &  {8.87\times10^{-3}+  2.26\times10^{-3}\cdot {\bar{E}}-8.30\times10^{-6} \cdot \bar{E}^2 + 1.88\times10^{-8} \cdot \bar{E}^3}   \nonumber \\
k_1(\bar{E})  ~/ {\rm m} & = & 6.19\times10^{-7}+1.89\times10^{-2}/(\bar{E}+4.73\times10)^2.
\ea
where $\bar{E}=E/ ({\rm kV ~cm^{-1}~ bar^{-1}})$ is the electric field in dimensionless units, and $\alpha_{i,att}$ are the ionization and attachment rate.
These fitted coefficients will be used in the fluid model to reach optimal agreement between particle and fluid model in the hybrid computation. In Appendix~\ref{app:parameters}, we also compared our calculation with the {\sc Bolsig+}~\cite{Siglo,Hag2005} generated transport coefficients and reaction rates for air, nitrogen, oxygen, and argon.

We refer to~\cite{Li2008:2} for the simulation results of the particle model.                                                                             
\subsection{Fluid model}\label{sec:3dfluid}

\subsubsection{The extended fluid model with flux coefficients}\label{sec:imp_3dfluid}

A fluid model for streamers consists of continuity equations for electron and ion densities $n_{e,p}$ coupled to the electric field ${\bf E}$
\ba
\frac{\partial n_e}{\partial t} + {\nabla}\cdot {\bf j}_e &=& {\cal S}, \label{ch6_eq:fluid1}\\
\frac{\partial n_p}{\partial t} \hspace{1.35cm} &=& {\cal S}, \label{ch6_eq:fluid2} \\
{\bf j}_e &=& - \mu({E}) {\bf E} n_e - {\bf D}({\bf E})\cdot {\nabla} n_e =: -{\bf j}^a-{\bf j}^d,  \label{ch6_eq:flux}\\
\nabla \cdot{\bf E} &=& \frac{{\rm q}\;(n_p-n_e)}{\epsilon_0}.  \label{ch6_eq:Poisson}
\ea
${\bf j}_e$ is the electron flux; it consists of an advection part $-{\bf j}^a=- \mu({E}) {\bf E} n_e$ and a diffusion part $-{\bf j}^d=- {\bf D}({\bf E})\cdot {\nabla} n_e$. The source term ${\cal S}$ accounts for the ionization or attachment  reactions that change the density of electrons or ions. The ions are approximated as immobile on the time scale of electron motion, therefore the total space charge created by different types of positive and negative ions can be summarized into one ion density $n_p$. {\bf q} is the elementary charge, and $\epsilon_0$ is the permittivity of the gas. The electric field is typically calculated in electrostatic approximation ${\bf E}=-\nabla \phi$.

The source term for impact ionization and attachment is traditionally written in local form as
\be {\cal S}  =  |n_e\;\mu(E)\;{\bf E}|\;\alpha(E), \ee
where the Townsend coefficient $\alpha(E)$ ($E=|{\bf E}|$) depends on gas density, collision cross sections and the local electric field. It is also tempting to simply insert transport and reaction coefficients from electron swarm experiments, i.e., the so-called bulk coefficients from section~\ref{sec:swarm_tran_rea}, into the fluid equations. However, in~\cite{Li2010:1} we showed that in high electric fields (above 75 kV/cm in nitrogen at standard temperature and pressure) this classical fluid model does not approximate the MC particle model well. But when particle and fluid model are inconsistent, discontinuities are building up in time at the interface between fluid and particle model when simulating a planar streamer ionization front~\cite{Li2010:1}. Therefore, the flux coefficients of section~\ref{sec:swarm_tran_rea} have to be used, and the fluid model has to be extended by writing the source term as
\be \label{equ:source_extended} {\cal S}  =  |n_e\;\mu(E)\;{\bf E}|\;\alpha(E) \left(1+k_1(E) \frac{{\bf E}}{E} ~ \nabla \ln n_e \right). \ee
(Instead of extending the source term, one can also let the source term depend on the mean electron energy and introduce another equation for the electron energy, as discussed in the appendix of~\cite{Li2010:1}).
Although only the source term differs between the classic and the extended fluid model and the models describe the same experiments, all transport and reaction coefficients differ between the two models.

There are two reasons to use the extended fluid model with flux coefficients in our hybrid model, one is based on the macroscopic simulation results, the other on the microscopic understanding of the electron fluxes.

First, the extended fluid model approximates the particle model better than the classic fluid model. The particle density profiles generated by these three models were compared in~\cite{Li2010:1} for electron swarms as well as for planar ionization fronts. The electron and ion density profiles are almost identical in the swarm simulations for all three models, but for the classical fluid model this is due to a cancellation of terms, as the electrons move too fast, but the ionization rate at the tip of the swarm is too low. For the planar front, the electron and ion densities behind the ionization front agree well between particle model and extended fluid model, but they are too low in the classical fluid model.

Second, the extended fluid model uses the flux coefficients, that characterize the average electron motion. A consistent definition of electron fluxes is important when coupling particle and fluid model in space. The definitions of 1) the number of electrons crossing the model boundaries in the particle model and of 2) the electron density flux rate at the model boundaries in the fluid model have to be consistent. Since the flux coefficients represent the real motion and reaction rate of electrons, while the bulk coefficients also depend on the macroscopic electron-density-profile, a consistent coupling of particle and fluid model can only be achieved with flux coefficients.

From now on, the term ``fluid model'' will refer to the extended fluid model with flux coefficients.



\subsubsection{Numerical discretization of densities and fluxes}

In our implementation of the fluid model, the electron and ion densities are discretized in space with a finite volume method based on the mass balances for all cells, and the electron density is updated in time using a third order upwind-biased advection scheme combined with a two-stage Runge-Kutta method.
The numerical algorithm and the spatial discretizations of the continuity equations in a 2D radially symmetric system have been discussed in~\cite{Mon2006:3,Mon2005p}.

We now focus on the numerical discretization of the electron fluxes in the fluid model in 3D.

The flux of electrons is given in Eq.~(\ref{ch6_eq:flux}), where the ${\bf j}^a =\mu(E){\bf{E}} n_e$ and ${\bf j}^d={\bf{D}}({\bf E})\cdot\nabla n_e$ denote the advective and diffusive electron fluxes through the cell boundaries, and ${\bf{D}}$ is a tensor in the form of
\begin{equation}
\label{equ:D_tensor_normal}
{\bf D}({\bf E}) = D_L(E) \frac{{\bf{E E}}^T}{E^2} +D_T(E) \left({\bf I}-\frac{{\bf{E E}}^T}{E^2}\right),
\end{equation}
 where {\bf I} is the identity matrix.

Electron diffusion in electrostatic fields in gas discharges is frequently approximated as isotropic in fluid simulations~\cite{Dha1987,Kul1995,Bab1996,Ebe1996,Arr2002,Mon2005}, while it, of course, is anisotropic~\cite{Wag1967,Low1969}. For example, in the flux diffusion coefficients presented in Fig.~\ref{fig:rate_air}, the longitudinal diffusion rate $D_L$ and transversal diffusion rate $D_T$ relative to the direction of the electric field differ due to the anisotropic collision processes in the particle model: $D_L<D_T$ at low fields $E < 50$ kV/cm, and $D_L > D_T$ when the field strength is above 50 kV/cm. The discretization of the flux terms requires extra care when the diffusion tensor is anisotropic~\cite{Ric1983}.

Note that the electron and ion densities calculated at cell centers on a uniform grid, can also be viewed as averages over the cell. The electric potential $\phi$ and the field strength $E$ are taken also in the cell centers, where $E$ determines the electron and ion growth in the cell. The electric field components ${\bf E}=(E_x, E_y, E_z)$ are taken on the cell vertices, where they determine the mass fluxes.
The flux coefficients $\mu$ and $\bf{D}$, and ionization rate $\alpha_{i}$ and attachment rate $\alpha_{att}$ with anisotropic scattering are from Section.~\ref{sec:swarm_tran_rea}.

\begin{figure}
 \begin{center}
     \includegraphics[width=0.4\textwidth]{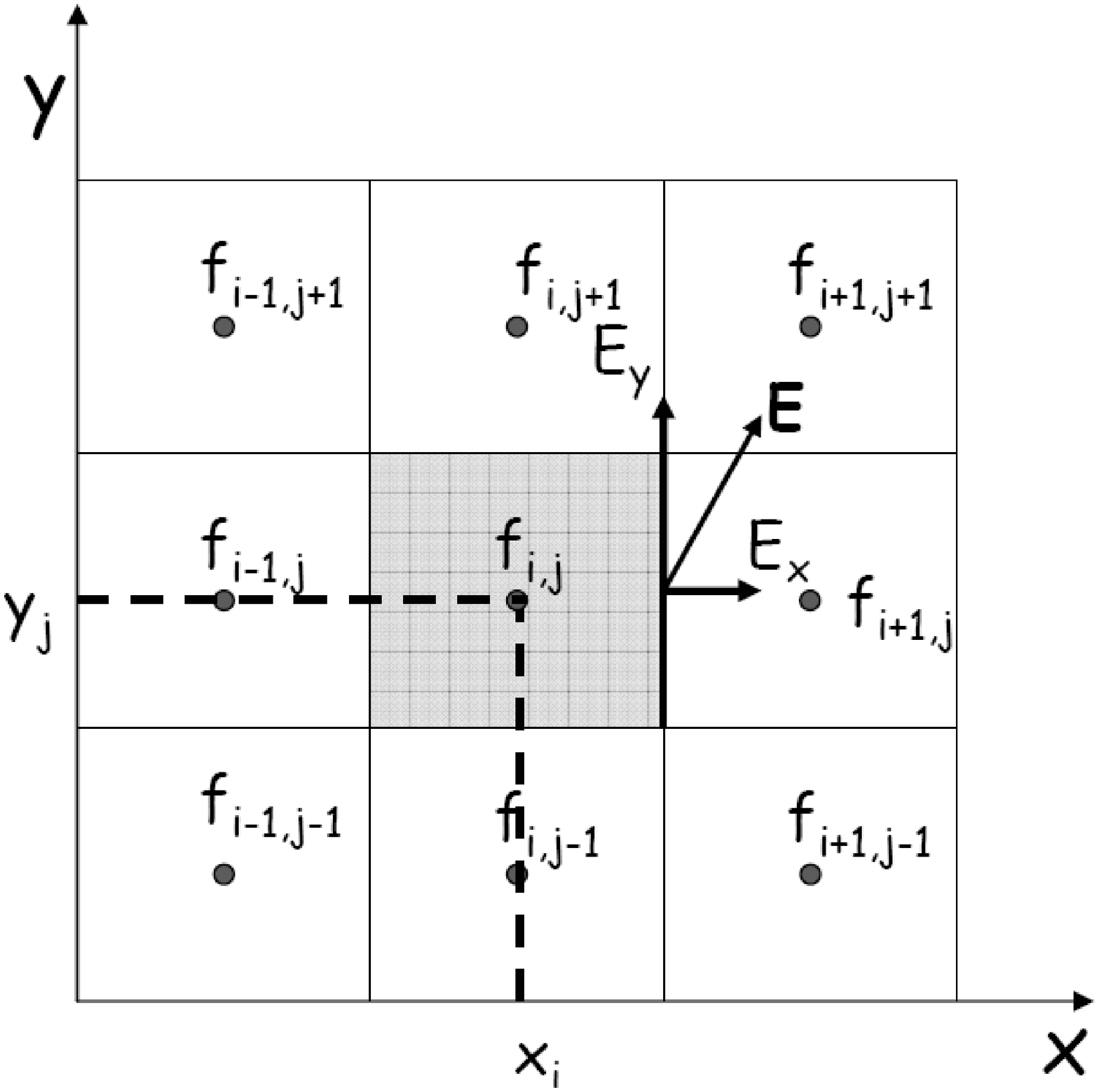}
  \caption{Evaluation of the electron flux on the boundary between two cells. \label{fig:tensor_diffusion}
}
\end{center}
\end{figure}


We first consider the 2D example to explain how the electron flux is numerically discretized. As shown in Fig.~\ref{fig:tensor_diffusion}, we have a uniform grid with the quantities $f$ evaluated in the centers of the grid cells ($f$ can be $n_e$, $\phi$ or $E$). Consider the flux on the boundary between cell $(i,j)$ and $(i+1,j)$. The advective flux $j_{(i+\frac{1}{2},j),x}^a$ can be written as
\be j_{x,(i+\frac{1}{2},j)}^a= {\bf n}^T \mu(E_{i+\frac{1}{2},j}) {\bf E}_{i+\frac{1}{2},j} n_{e,{i+\frac{1}{2},j}} \label{eq:flux_in_Fa}\ee
where ${\bf n}^T  =  (1, 0)$, ${\bf {E}}_{i+\frac{1}{2},j}^T  =  (E_{x,(i+\frac{1}{2},j)}, E_{y,(i+\frac{1}{2},j)})$ given by Eq.~(\ref{eqn:Ex_Ey}), and $n_{e,{i+\frac{1}{2},j}}$ is approximated by a third-order upwind-biased scheme. This gives mass conservation and monotone solutions without introducing too much numerical diffusion~\cite{Mon2005p}. The Koren limiter function is used here. Denote $E^L=\max(-E,0)$ and $E^R=\min(-E,0)$ to distinguish the upwind direction for the field components, the advective flux $j^a$ at cell face $(x_{i+\frac{1}{2}},y_j,z_k)$ then can be rewritten as:
\begin{eqnarray}
\label{eqn:Fa}
j_{x,(i+\frac{1}{2},j)}^a & = & \mu(E) E_x^{L}[n_{e,(i,j)}+\psi(p_{i,j})(n_{e,(i+1,j)}-n_{e,(i,j)})] \nonumber \\
   & & + \mu(E) E_x^{R}[n_{e,(i+1,j)}+\psi(1/p_{i+1,j})(n_{e,(i,j)}-n_{e,(i+1,j)})]
\end{eqnarray}
where $ p_{i,j}=\frac{n_{e,(i,j)}-n_{e,(i-1,j)}}{n_{e,(i+1,j)}-n_{e,(i,j)}}$, $\psi$ is the limiter function $\psi(\theta)=\max\left(0, \max\left(1,\frac{1}{3}+\frac{\theta}{6},\theta\right)\right)$, and here $E$ and $E_x$ are taken at cell face ($x_{i+\frac{1}{2}},~y_j$).

To obtain the electron diffusive flux at the cell boundary $(x_{i+\frac{1}{2}},y_j)$, we need the electric field and the electron density gradients at the boundary. The electric field is taken as
\begin{eqnarray}
\label{eqn:Ex_Ey}
E_{x,(i+\frac{1}{2},j)} & = & \frac{1}{\triangle x} (\phi_{(i,j)}-\phi_{(i+1,j)})  \nonumber \\
E_{y,(i+\frac{1}{2},j)} & = & \frac{1}{2}\left[\frac{1}{2\triangle y}(\phi_{(i,j-1)}-\phi_{(i,j+1)}) +
\frac{1}{2\triangle y}(\phi_{(i+1,j-1)}-\phi_{(i+1,j+1)})\right].
\end{eqnarray}
The diffusive flux $j^d$ is calculated as:
\be
\label{equ:Fd}
j_{x,(i+\frac{1}{2},j)}^d = {\bf n}^T \left[ D_L(E) \frac{{\bf{E E}}^T}{E^2} +D_T(E) \left({\bf I}-\frac{{\bf{E E}}^T}{E^2}\right) \right]_{i+\frac{1}{2},j} {\nabla} n_{e,i+\frac{1}{2},j}
\ee
where the field strength is taken at the cell boundaries and the density gradient $\nabla n_e$ 
at cell face $(i+\frac{1}{2},j)$ is defined in the same way as the electric field,
\begin{eqnarray}
\label{eqn:nex_ney}
\frac{\partial n_e}{\partial x}_{(i+\frac{1}{2},j)} & = & \frac{1}{\triangle x} (n_{e,(i,j)}-n_{e,(i+1,j)}) \nonumber  \\
\frac{\partial n_e}{\partial y}_{(i+\frac{1}{2},j)} & = & \frac{1}{2}\left[\frac{1}{2\triangle y}(n_{e,(i,j-1)}-n_{e,(i,j+1)}) +  \frac{1}{2\triangle y}(n_{e,(i+1,j-1)}-n_{e,(i+1,j+1)})\right].
\end{eqnarray}
The flux in the $x$-direction in 2D therefore can be written as
\be\label{equ:flux0_flux_2_2d}
j_{e,x} = -j^a_x- j^d_x ={\bf n}^T \cdot (-\mu(E){\bf E} n_e-{\bf D}({\bf E}) \cdot{\nabla} n_e ).
\ee
In the $y$-direction, Eq.~(\ref{equ:flux0_flux_2_2d}) applies as well with $x$ exchanged by $y$ and with ${\bf n}^T=$(0 1).

The electron flux on a cell face can be written in the same way for 3D as Eq.~(\ref{equ:flux0_flux_2_2d}) for 2D,
where {\bf D} is a tensor defined in Eq.~(\ref{equ:D_tensor_normal}), ${\bf n}$ is a vector normal to the cell face, ${\bf E}^T= (E_x, E_y, E_z)$, and $({\nabla} n_e)^T=(   \partial n_e/\partial x ,  \partial n_e/\partial y ,  \partial n_e/\partial z)$ are taken at the cell face.
For example, the diffusive flux is calculated with second-order central differences as
\be
j_{x,(i+\frac{1}{2},j,k)}^d  =  \frac1{E^2} \Bigg(D_L(E) - D_T(E) \Bigg) E_x \left( E_x {\partial n_e \over \partial x} +E_y {\partial n_e \over \partial y}+ E_z {\partial n_e \over \partial z} \right) + D_T(E) {\partial n_e \over \partial x}
\ee
where $E$, {\bf E}, and ${\nabla} n_e$ are taken at the cell face ($x_{i+\frac{1}{2}},~y_j,~z_k$). Here {\bf E} at the cell face is defined in the same way as in Eq.~(\ref{eqn:Ex_Ey}) and ${\nabla} n_e$ is defined as in Eq.~(\ref{eqn:nex_ney}).

\subsubsection{The Poisson equation}\label{sec:test_fishpack}

In the particle as well as in the fluid model, the electric field has to be calculated at each time step, therefore this must be done in a very efficient way. A fast 3D Poisson solver {\sc fishpack} has been chosen for this. It uses second-order finite differences. The {\sc fishpack} is a direct solver, based on cyclic reduction with Fast Fourier Transform (FFT) in the third dimension. Although it is not versatile (it can only solve Poisson and Helmholtz problems on rectangular grids), it is very fast.

\begin{figure}
\centering
 \includegraphics[width=0.4\textwidth]{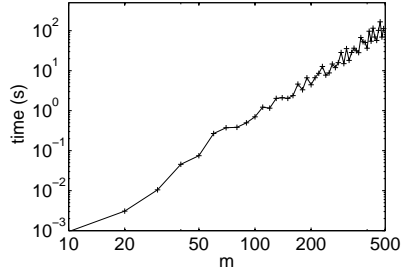}\\
 \caption{The computation time of the {\sc Fishpack} 3D Poisson solver on $m\times m\times m$ grids, as a function of $m$. The test is carried out on a desktop computer with {\rm AMD} Athlon 1.6GHz CPU and 1G memory.}\label{fig:workload_fishpack}
\end{figure}

The computing times are evaluated for a test problem with a Gaussian source term on various $m\times m\times m$ grids. In Fig~\ref{fig:workload_fishpack}, we plot the computing time as a function of $m$, when double precision is used. Since the solution is direct, the actual shape of the solution does not influence these computing times. Another advantage of {\sc fishpack} is that it needs, apart from arrays to store the solution, almost no additional memory. For details and additional tests with this solver we refer to~\cite{Swa1975,Sch1976,Bot1997}.

\subsubsection{Results of a 3D fluid simulation}\label{sec:fluid_results}

\begin{figure}
\begin{center}
$\begin{array}[c]{p{1.0cm}c}
0.48~\text{ns} &
\includegraphics[width=1.0\textwidth] {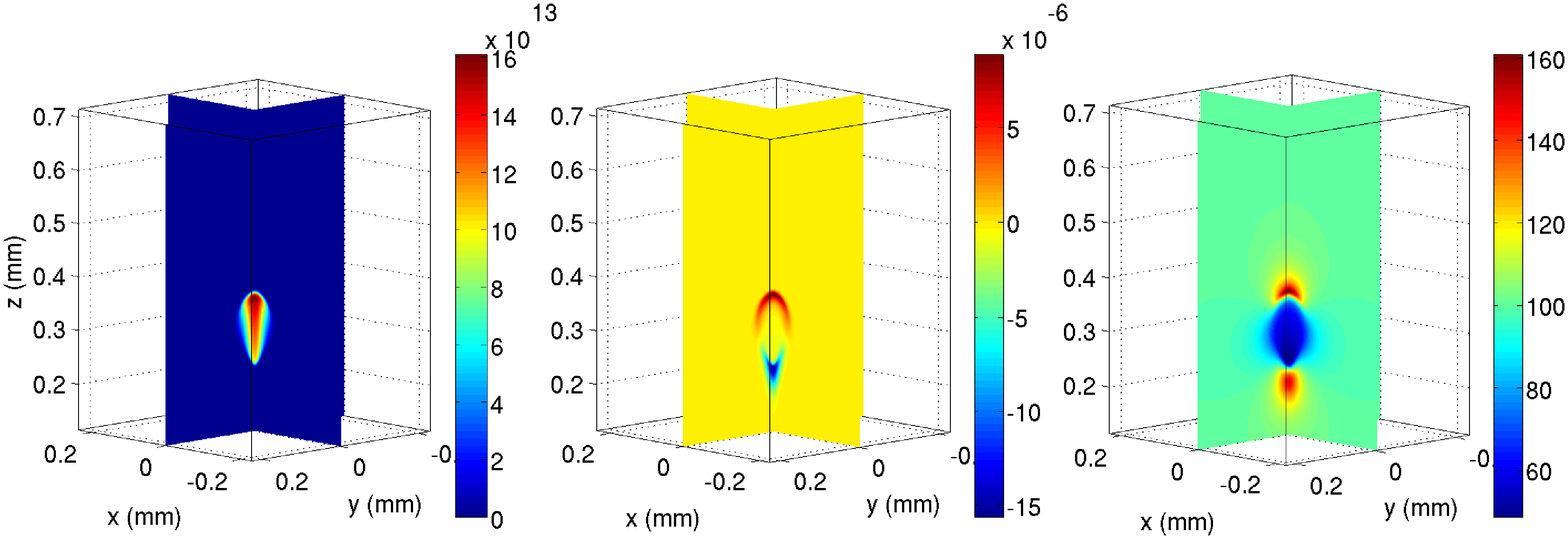}
\\
0.56~\text{ns} & \includegraphics[width=1.0\textwidth] {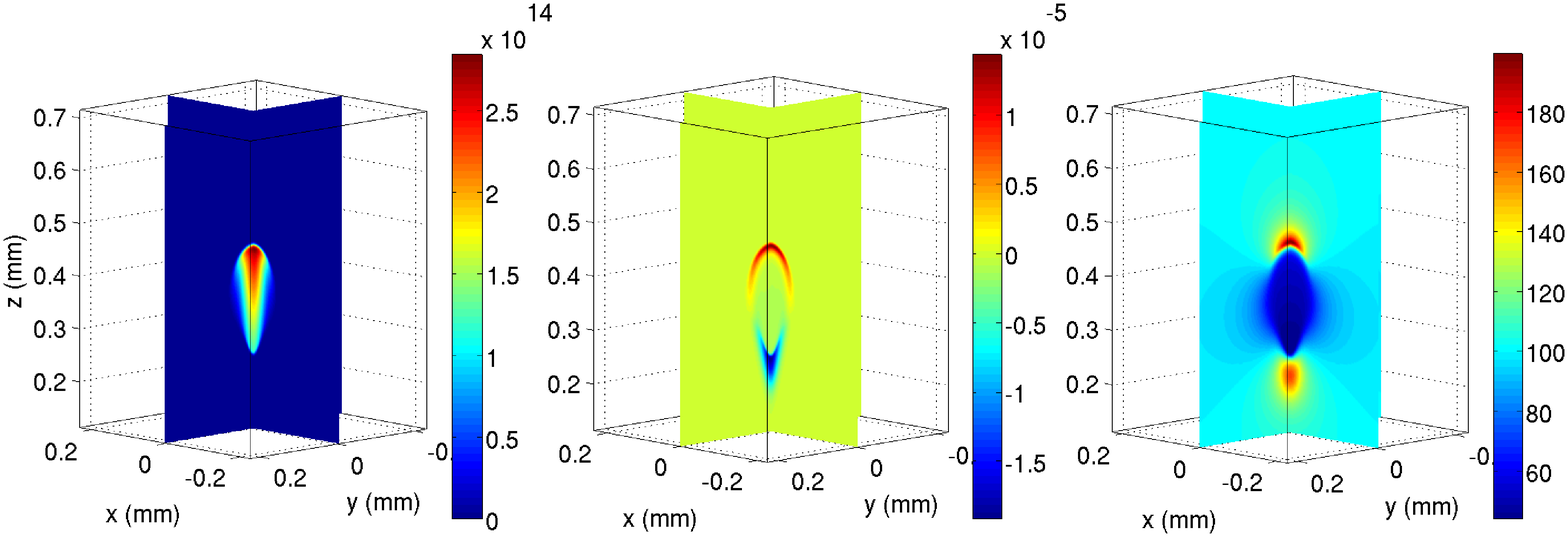}
\\
0.72~\text{ns} &\includegraphics[width=1.0\textwidth] {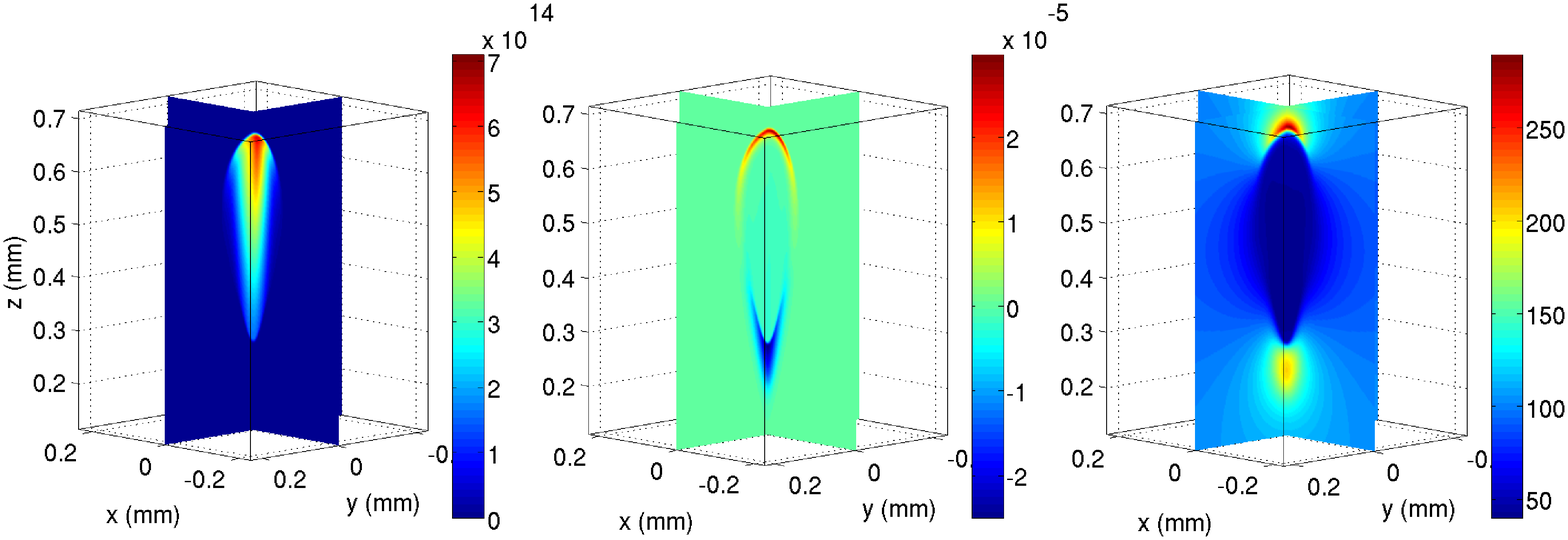}
\\
& \text{electron~density}~(1/\text{cm}^3) \hspace{1.2cm} \text{charge~density}~ (\text{C}/\text{cm}^3) \hspace{1.7cm} \text{E}_z~ \text{(kV/cm)\hspace{1.7cm}}
\end{array}$
\end{center}
\caption{3D simulation results of the extended fluid model for a negative streamer developing in air in a background field of -100 kV/cm.  First row: streamer at t = 0.48 ns, second row: streamer at t = 0.56 ns, third row: streamer at t = 0.72 ns. The columns show from left to right: electron density, charge density, and electric field $E_z$ in the $z$ direction. Particle densities and fields are represented on two orthogonal planes that intersect with the 3-dimensional structure. \label{fig:fl_3d} }
\end{figure}

Here we present 3D fluid simulation results for a negative streamer in air, propagating in a background field of -$100$ kV/cm. Initially 500 electrons and ions are each distributed within the same Gaussian distribution around a spot near the cathode. The simulation was carried out on a grid of $256 \times 256 \times 512 $ points with $\Delta x= \Delta y=\Delta z=2.3~\mu$m, on a system with $x\in$ [-0.29 0.29] mm, $y\in$ [-0.29 0.29] mm and $z\in$ [0 1.17] mm. The time step is $\Delta t= 3 \times 10^{-13}$ s.


In Fig.~\ref{fig:fl_3d}, the electron density (left column), charge density (middle column), and the electric field in $z$-direction (right column) are shown at 0.48 ns (first row), 0.56 ns (second row), and 0.72 ns (third row). The simulation starts with an avalanche of electrons present initially. After t= 0.48 ns, a charge layer is clearly formed and the electric field is altered by the space charge. From the 0.48 ns to 0.72 ns, the maximal electron density increases from $1.6\times10^{13}$ to $7\times10^{14}$/cm$^3$, while the maximal field strength increases from 160 kV/cm to about 290 kV/cm.

\newpage

\section{3D hybrid model}  \label{sec:3D_hyb_model}

\subsection{General coupling procedure} \label{sec:gen_coupling_pro}


The simulations begin with a few electron and ion pairs followed only by the particle model. As new electrons are generated, the number of particles eventually reaches a given threshold, after which the simulation switches from the particle simulation to the hybrid approach. If the computation time is not a concern, the threshold number can be set as the maximum allowed by a real particle simulation. The threshold in our simulation is normally set to $20$ million electrons.

After the number of electrons reaches the threshold, suitable regions to apply the particle or the fluid model are found with given criteria to initiate the hybrid computing.
To ensure a reasonable interaction between the models, buffer regions are built by extending the particle regions into the fluid region. In the buffer region, the particle movements and the particle densities are followed both by the particle model and by the fluid model.

To update the particle densities and the electric field from one time step to the next, we first follow the movement of electrons
in the particle region and the buffer region, and the number of electrons crossing the model interface
is recorded during this time step. The electron fluxes on the interface are used as the boundary condition to update the densities in the fluid region.
We then map the particles in the particle region to the densities on the fluid grid.
The electron and ion densities are now known in both particle and fluid region, and the new electric field is then calculated.
With the updated electric field and the particle densities, the new model interface is determined. This procedure is repeated in each time step.

The coupling procedure in 3D has some similarities to the 1D coupling.
As for planar fronts, i) we use the zero-order mapping (see~\cite{Li2010:1}) around the model interface in the particle model to avoid particle leaking, and ii) we use forward Euler time stepping instead of two-stage Runge-Kutta method in the fluid model to save computer memory and time.
However, additional problems appear in a real 3D problem. For example,
in a planar front, only one point needs to be specified to determine the position of the model interface; in 3D, the model interface is a 2D surface and
extra care is needed; a strongly fluctuating model interface will create large buffer regions and dramatically increase the computation cost. The way to count the electron flux over the model interface is also different in 3D. In a planar front,
the model interface is a straight line and all crossing electrons will contribute to the density flux; in 3D,
the shape of the model interface is more complicated, and crossing electrons need to be defined carefully.

In the following sections, we consider a particle simulation that starts with the same initial condition as the fluid simulation in Section.~\ref{sec:fluid_results}, and the simulation is carried out on the same grid. The particle simulation runs until t $\approx$ 0.46 ns, when the number of electrons reaches $2\times10^7$. The particle simulation then switches to the hybrid simulation.
The problems and our solutions during this transition are described in detail in the following sections.

\subsection{First interface construction: column based splitting} \label{sec:1st_interface}


Aiming at following the high energy electrons in a fully developed streamer, fluid and particle model will be applied adaptively in suitable regions in the 3D hybrid model.
We would like to couple the particle and fluid model in such a way that: $i$) the hybrid model represents the correct physics, $ii$) high energy particles will be included in the
particle region, and $iii$) the model is computationally as efficient as possible.

The coupling of fluid and particle model was first realized in a planar front as discussed in~\cite{Li2008:1,Li2010:1}. Two splitting criteria were studied for the planar front. The model interface can either be set at the position where the electron density reaches $n_e=\eta~n_{e,max}$ when approaching from the nonionized region, or where the electric field is $E=\xi~ E_{max}$ ($E^+$ in~\cite{Li2010:1}); here $n_{e,max}$ and $E_{max}$ stand for the maximal electron density and electric field, and $\xi$ and $\eta$ are real numbers between $(0, 1)$, and chosen in such a manner that the calculation is efficient  and the error is small.  



A 3D streamer can be decomposed into columns with small transversal area parallel to the streamer axis.
For each column, the previously derived results for the planar front can be used, such as the best position of the model
interface for a given field ahead of the ionization front. When this is implemented numerically, we take a column of mesh cells in the direction of the background electric field, and try to locate the position of the model interface. Note that the cell-columns near the streamer axis cross through a rather planar ionization front, while the cell-columns at the streamer edges are further from this situation.

\subsubsection{Density criteria}

\begin{figure}
    \centering
    \subfigure[$n_e=n_{e,max}$]{
         \label{fig:where_mi_first_1}
         \includegraphics[width=.30\textwidth]{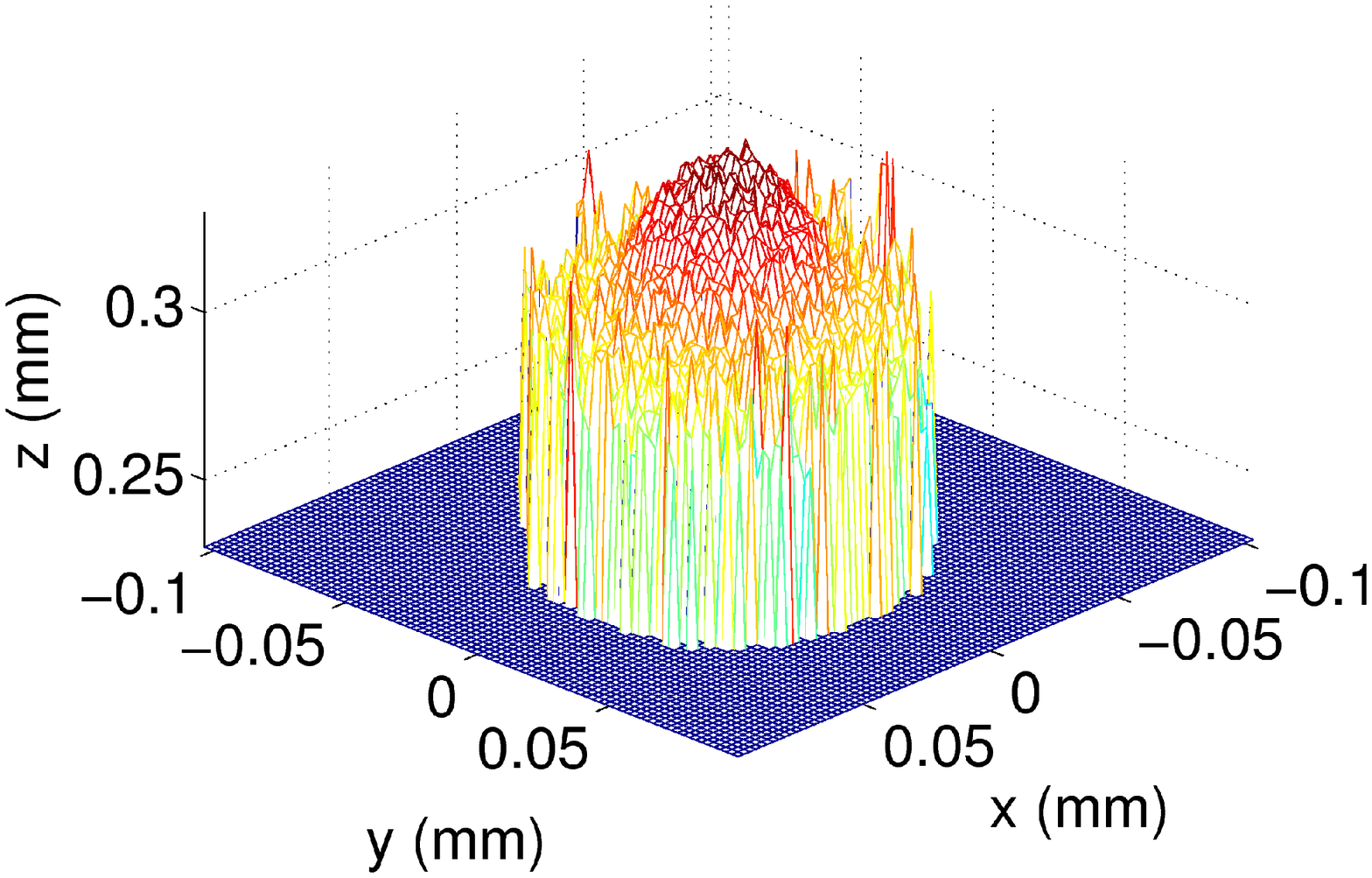}
        }
    \subfigure[$n_e=0.7~n_{e,max}$]{
         \label{fig:where_mi_first_2}
         \includegraphics[width=.30\textwidth]{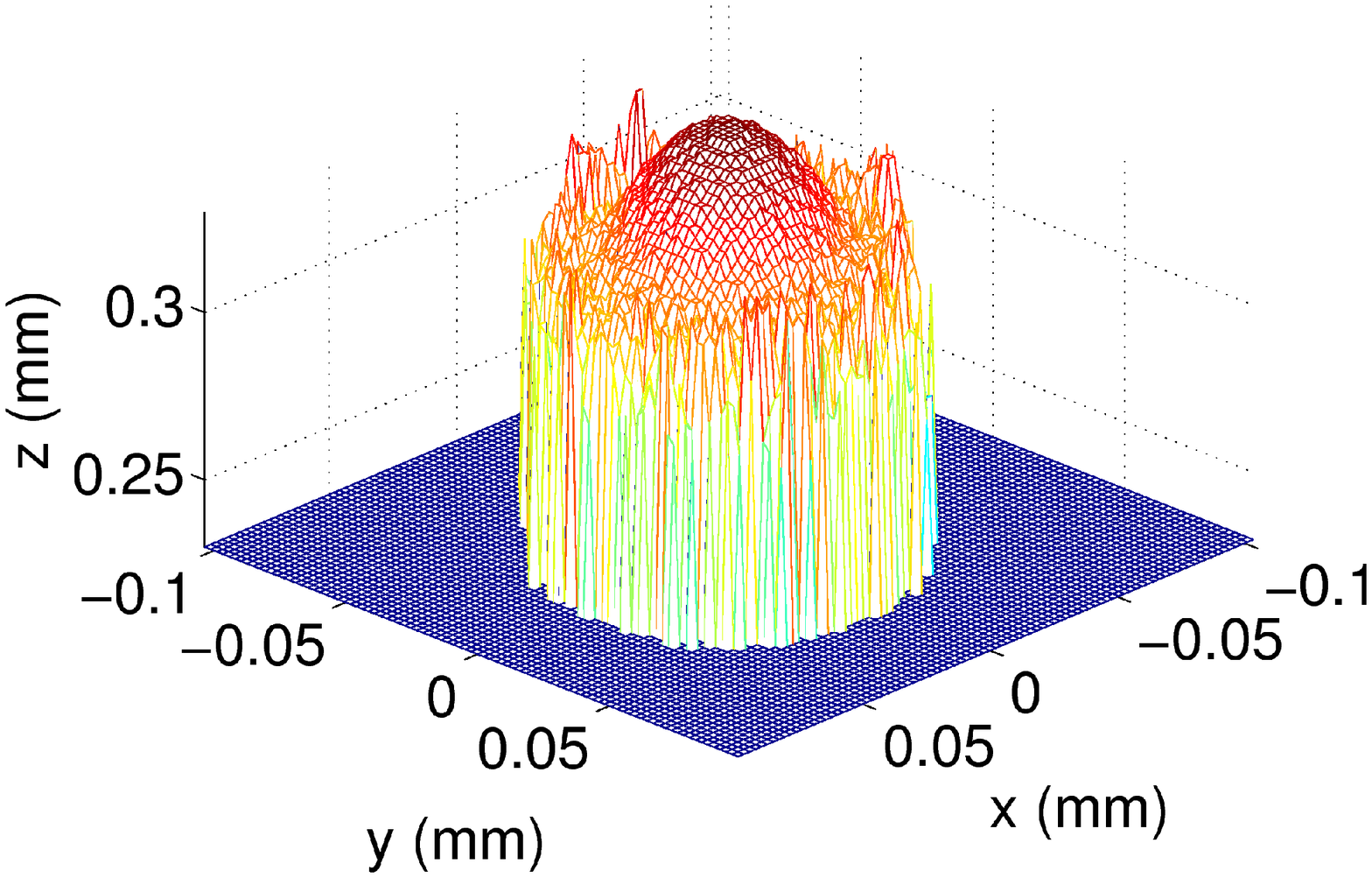}
         }
    \subfigure[$E=0.84~ E_{max}$]{
         \label{fig:where_mi_first_3}
         \includegraphics[width=.30\textwidth]{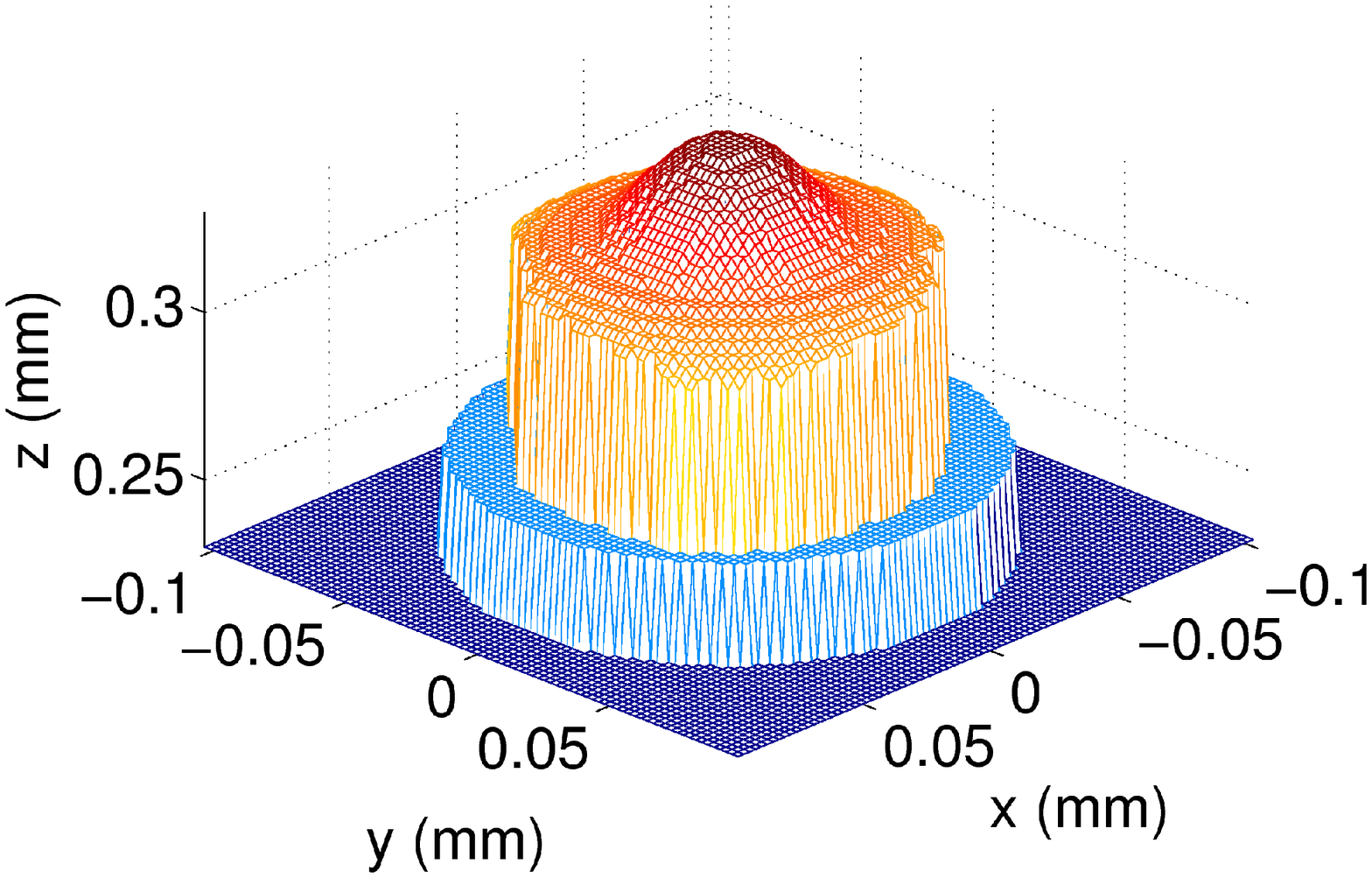}
         }
    \caption{The position of model interfaces for a streamer at t$\approx$0.46. The model interface is at $n_e=n_{e,max}$ (left), $n_e=0.7~n_{e,max}$ (middle) and $E=0.84~ E_{max}$ (right) along each cell-column. Far outside of the streamer where no free electrons and ions exist, the particle model is applied and the model interface is set at $z=0$. When the criterion is $n_e=n_{e,max}$, fluctuations are everywhere. When $n_e=0.7~n_{e,max}$ the interfaces are rather smooth tn the center of the streamer, the fluctuations appear at the side of streamer. When $E=0.84~ E_{max}$, the model interface forms a smooth surface.}
    \label{fig:where_mi_density_07}
\end{figure}

We first try to determine the model interface using the density criterion. But the obtained model interfaces have strong fluctuations, especially along the side of the streamer. In Fig.~\ref{fig:where_mi_first_1} and~\ref{fig:where_mi_first_2}, we show the position of the model interface when it is set at $n_e=n_{e,max}$ (left) or $n_e=0.7~n_{e,max}$ (middle). The fluid model is applied at regions above the model interface and the fluid model is applied beneath. The model interface is set at $n_e=n_{e,max}$ or
$n_e=0.7~n_{e,max}$ only when $n_{e,max}>c$, where $c$ is the density value for one electron per cell.
It means that we split a cell-column into particle and fluid region only if it contains electrons; in the large
area without electrons, the particle model is applied. When the model interfaces are placed at $n_e=n_{e,max}$, the fluctuations are almost everywhere. When the model interfaces are placed at $n_e=0.7~n_{e,max}$, the model interface is a rather smooth surface in the center where $n_{e,max}$ is relatively large, and it fluctuates strongly where $n_{e,max}$ is small.


The density fluctuation is the first problem when we apply the planar front coupling to the cell-columns in 3D. The electron density fluctuations in a planar front can be suppressed by expanding the planar front in the transversal direction, which increases the number of electrons and smoothes the density profile. This method can be applied for the planar front since the particle distributions are uniform in the transversal direction and therefore charged particle densities remain unchanged.
But this numerical trick can not be applied for the cell-column in 3D coupling, while the limited number of electrons or ions within the cell-column gives rise to strong fluctuations in density profiles along the column.

To reduce the computational cost and save memory for high energy electrons at the streamer head, we would like to have the model interfaces of all cell-columns to lie within a smooth surface.
A fluctuating model interface requires extra buffer regions and therefore more electrons to be followed by the particle model, and
it creates a large volume of the model interface in the transversal directions where the motions of all
electrons close to the lateral model interface have to be traced. It is therefore not a good
idea to relate the position of the model interface to the maximum density within a column.

When the interface is placed at smaller densities, e.g., at cells at position $n_e=0.7~n_{e,max}$, it is rather smooth over $z$, at least in the center of the streamer where $n_{e,max}$ is large enough.
For planar fronts in fields between 50 and 200 kV/cm, the particle densities from a hybrid computation with a model interface at $n_e=0.7~n_{e,max}$ deviates from the pure particle simulation results by not more than $4\%$~\cite{Li2010:1}.


To obtain a smooth model interface at the edge of streamer, one possible solution is to find a fitting formula for the interface position near the streamer center and to
extrapolate it to the side to calculate the interface position even if no electron exists in that
cell-column. But besides concerns on the asymmetry of the density over the radius and on the difficulties of
fitting, the interface position at the streamer side would not depend on local properties in this procedure,
but on the fit to the streamer center.


\subsubsection{Field criteria}

In a planar front, there is a clear correlation between the local electric field and the electron
density, and the electric field is always smoother than the electron densities.
Therefore in the planar hybrid calculation, also an interface criterion depending on the field level $E=\xi~ E_{max}$ was tested.
The relation between particle density and electric field in 3D is not as clear as in 1D.
Therefore a field dependent criterion is tried with several values ``$\xi$''; for the particular value of $\xi=0.84$ the positions are given in the Fig.~\ref{fig:where_mi_first_3}. The field criterion $\xi=0.84$ for the model interface agrees very well with the density criterion $\eta=0.7$ in the center of the streamer, while model interfaces are much smoother at the side of streamer.




One problem of the field criterion is that at the side of the streamer where the field varies less than in the center, it becomes $E<0.84~ E_{max}$ everywhere along the cell-columns. Because there are only few electrons which
occasionally fly out of the channel in the sidewards direction, we leave them to the fluid model. Therefore from the point where the
level $E=0.84~ E_{max}$ ceases to exist, the model interface is extended horizontally several cells outwards to include all
particles at the streamer side into the fluid region.

At the streamer side, we also add one or two extra cells for the density diffusion in the fluid calculation.
In the particle model, the electrons are discrete and the electron density is always $N\times c$, where $c$ is the density value for one electron per cell.
But in the fluid calculation, diffusion can
fill the entire fluid region with a small $n_e>0$. To avoid a continuous expansion of the fluid region to
the side, the fluid electron density is set to zero if it drops below $c$.

\subsubsection{After splitting}

\begin{figure}
    \centering
    \subfigure[Electron density (1/cm$^3$) ]{
         \includegraphics[width=.30\textwidth]{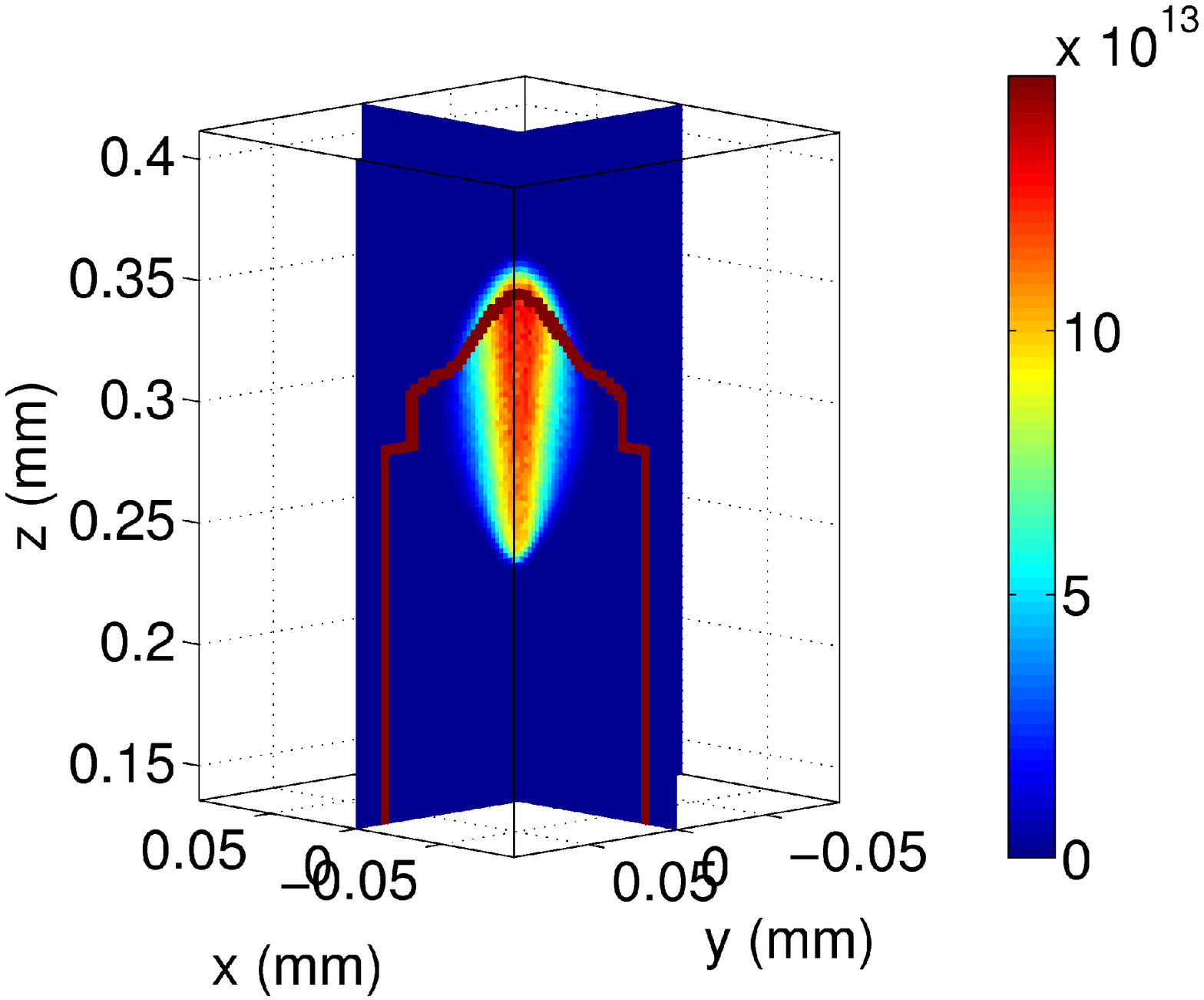}
         }
    \subfigure[Charge density (C/cm$^3$)]{
         \includegraphics[width=.30\textwidth]{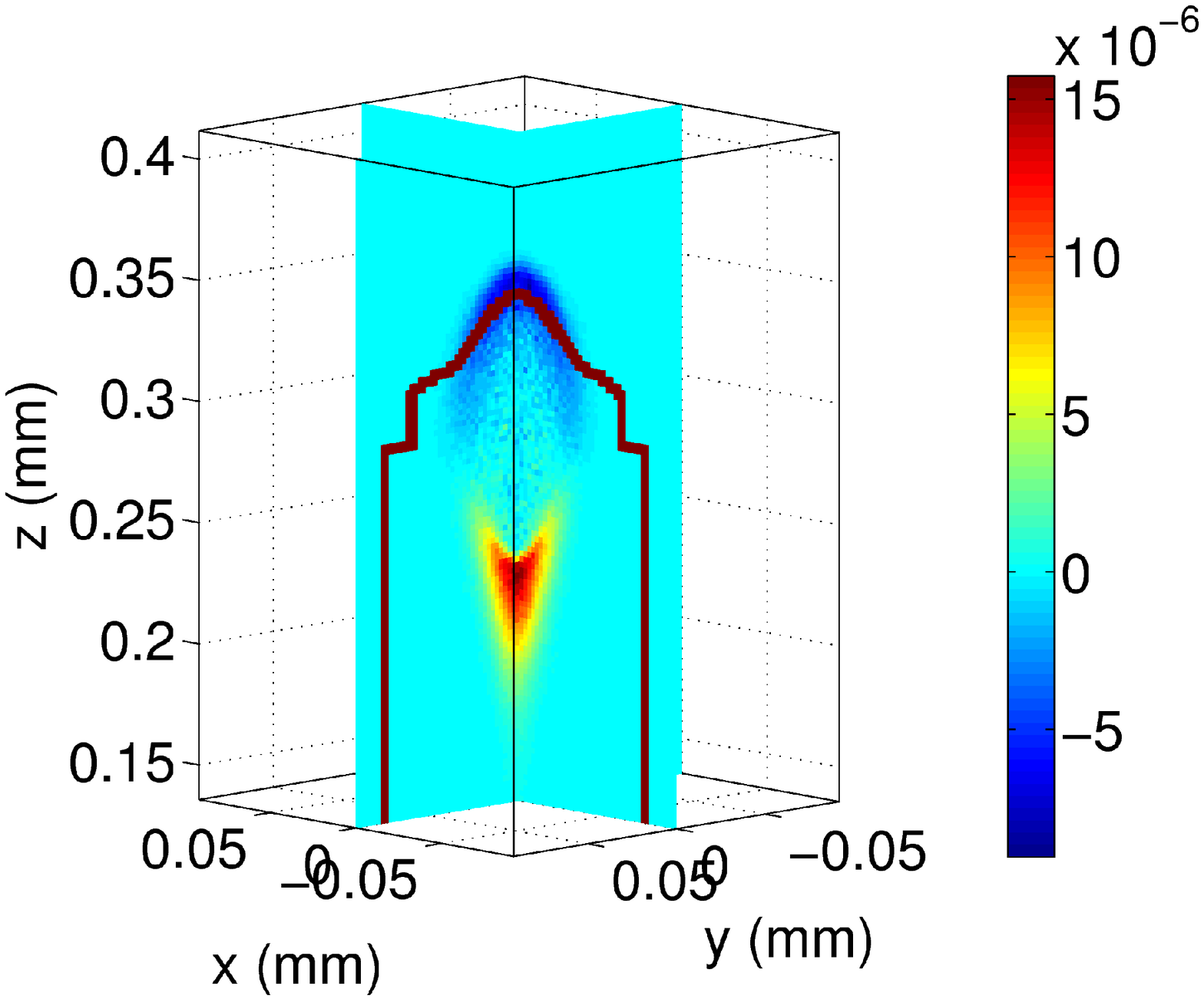}
         }
    \\
    \subfigure[Electric field (kV/cm)]{
          \includegraphics[width=.30\textwidth]{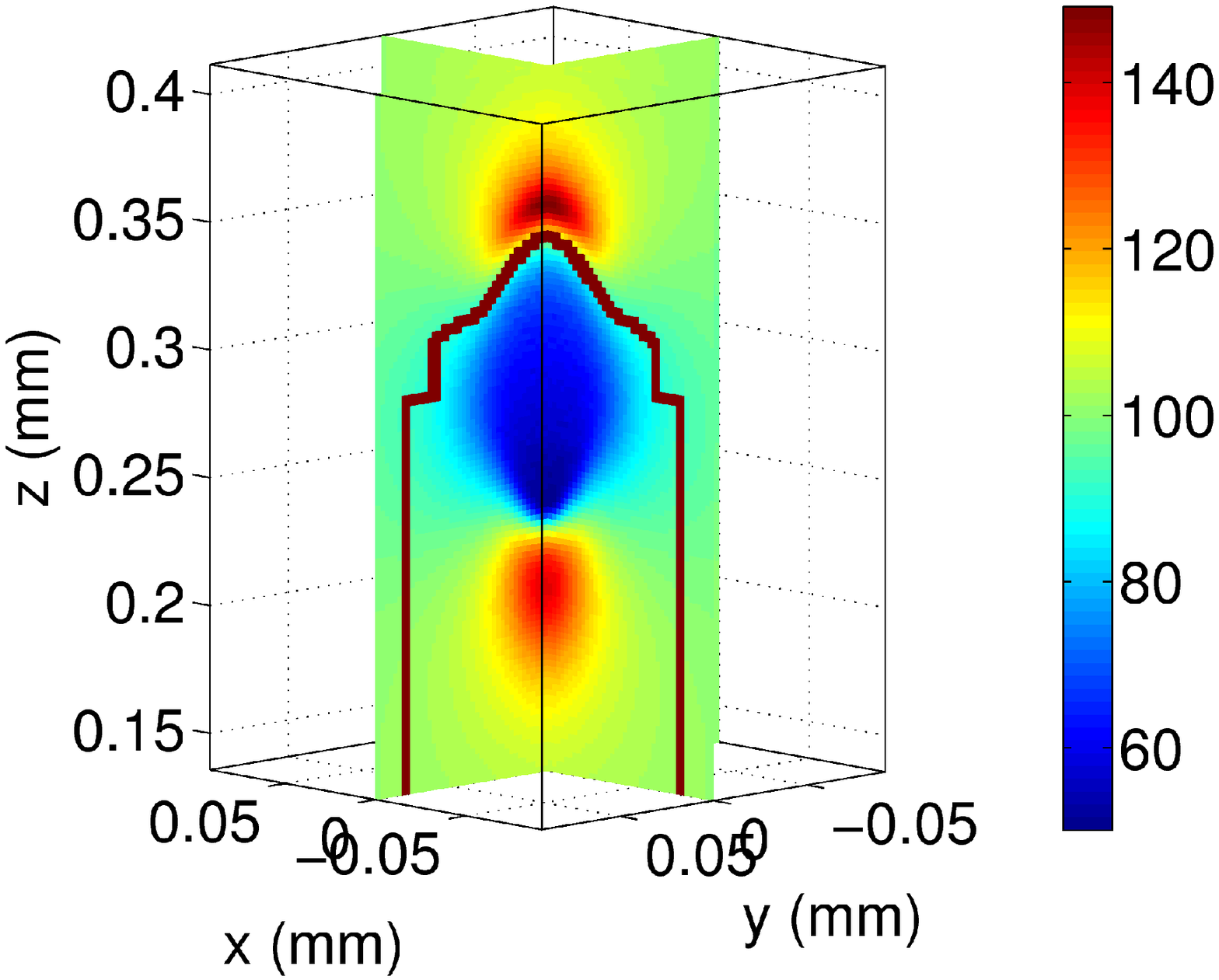}
          }
    \subfigure[Mean electron energy (eV)]{
         \includegraphics[width=.30\textwidth]{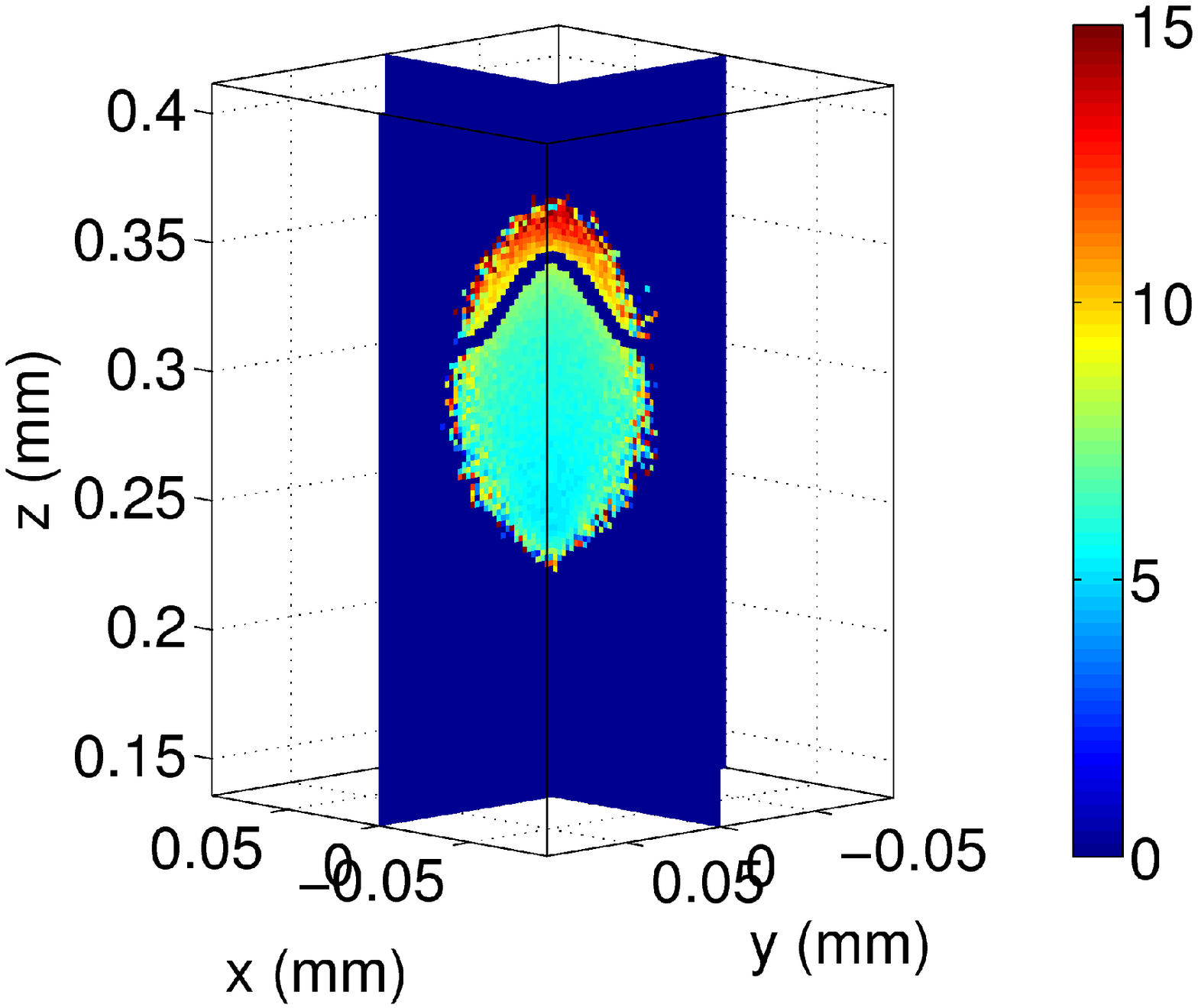}
         }
    \caption{Electron density (upper left), charge density (upper right), electric field (lower left)
    and the electron temperature (lower right) at t$\approx$ 0.46 are presented with the model interface and buffer region marked.}
    \label{fig:3D_hy_start_MI}
\end{figure}

Having described this procedure for locating the model interface, we now present the
splitting results when a 3D simulation is transferred from the particle calculation to the hybrid calculation.
In Fig.~\ref{fig:3D_hy_start_MI}, the electron density (upper left), charge density (upper right),
electric field (lower left) and electron mean energy (lower right) at t$\approx$ 0.46 are shown with the model interface and the buffer
region marked with red lines (blue lines for the electron mean energy). Below the model interface, the fluid model is applied; above the model interface, the particle model is applied. With the cell-column based approach, the particle model focuses on the electrons at the streamer head where the local electric field is most enhanced. And it leaves the large remaining region with high electron densities and low electric field to the fluid model.

The first results of 3D hybrid calculation results based on this approach were reported in the letter~\cite{Li2009} for a negative streamer in N$_2$.
The particle region in this approach focuses on the electrons at the streamer head and leaves the region behind it to the fluid model. However in the situation where i) more than one streamer propagates in the simulation domain and the streamers are not in parallel, ii) a runaway electron runs out of the planar front and creates an avalanche ahead of the streamer, or iii) a double headed streamer emerges, this cell-column based approach is unable to deal with them. To deal with those situations, a more flexible splitting algorithm is needed.

\subsection{Second interface construction: full 3D splitting} \label{sec:2nd_interface}



We have developed another splitting method that is simultaneously based on the electric field and on the electron density.
In contrast to the previous splitting method that was based on local column-based quantities, we now develop a global splitting criterion that operates on the whole 3D domain.

The particle regions are now taken as the regions where the electric fields are higher than a global field threshold {\em and} where electron densities are lower than a global density  threshold, and the fluid regions covers the rest.
Most of the energetic electrons, and electrons with the potential to gain high energies are in the region with the strongest electric field. 

\begin{figure}
\begin{center}
$\begin{array}[c]{p{1.0cm}ccc}
$\xi$=0.7 \hspace{3.0cm}  &
\includegraphics[width=.30\textwidth]{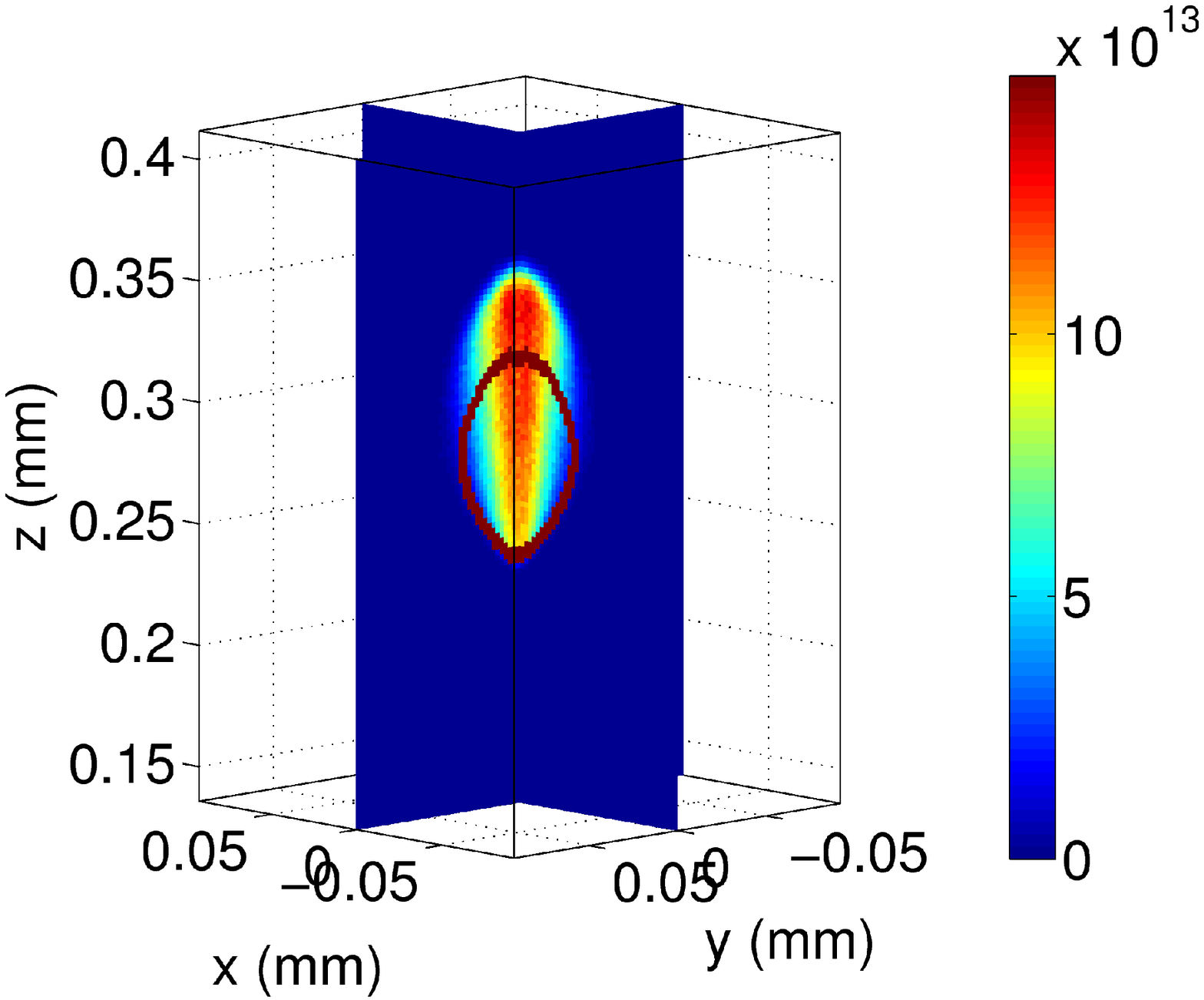}
&
\includegraphics[width=.30\textwidth]{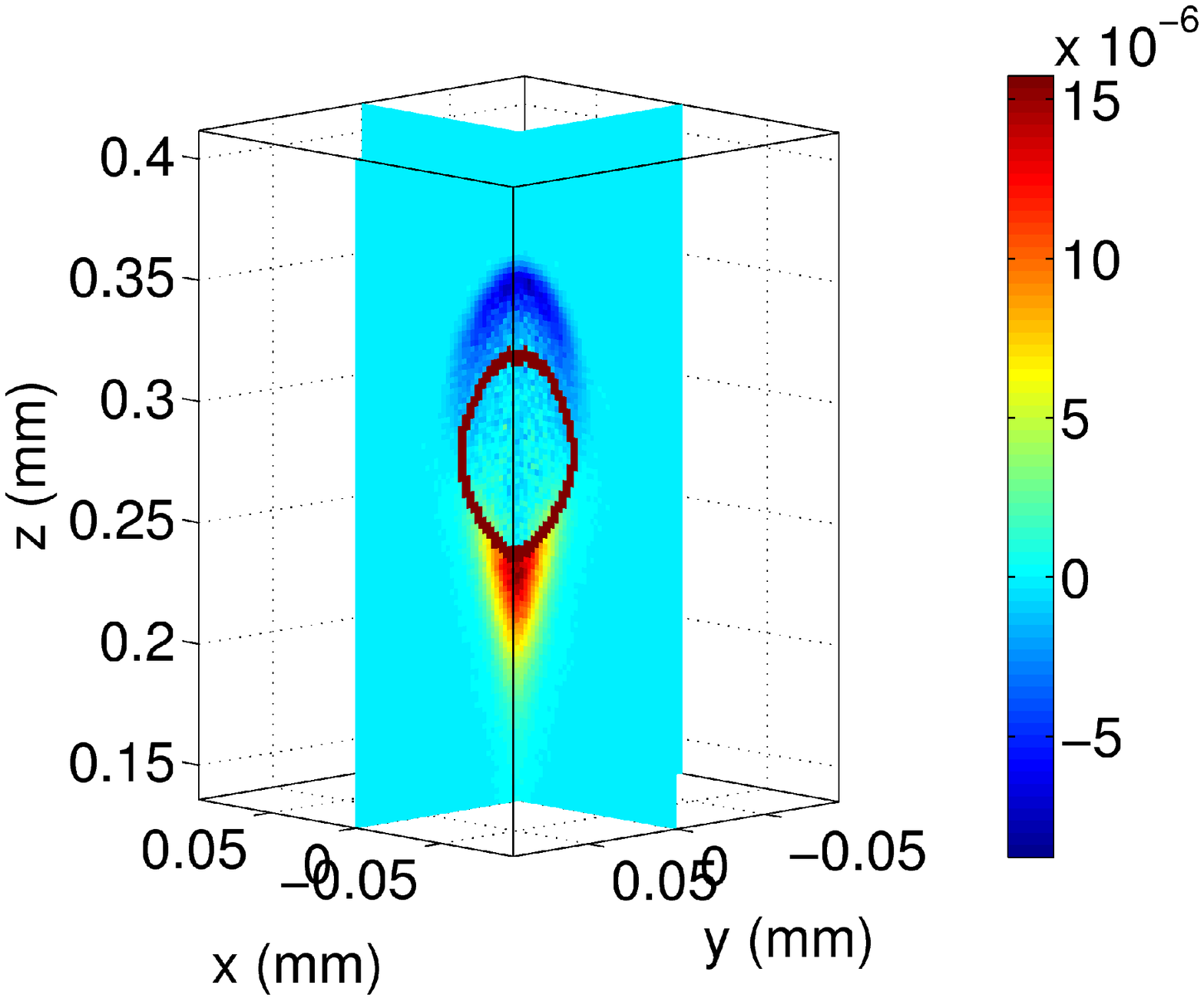}
&
\includegraphics[width=.30\textwidth]{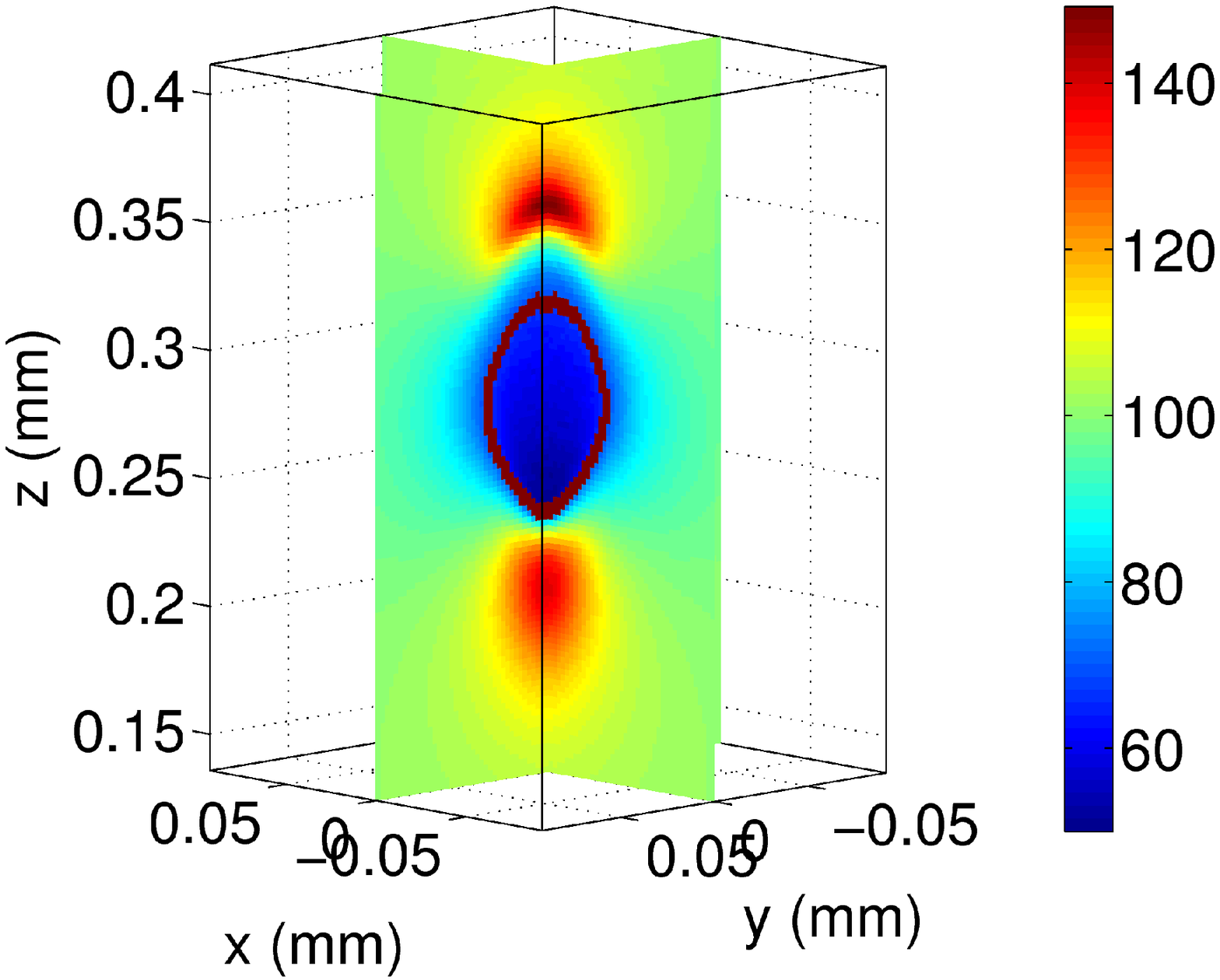}
\\
$\xi$=0.8 &
\includegraphics[width=.30\textwidth]{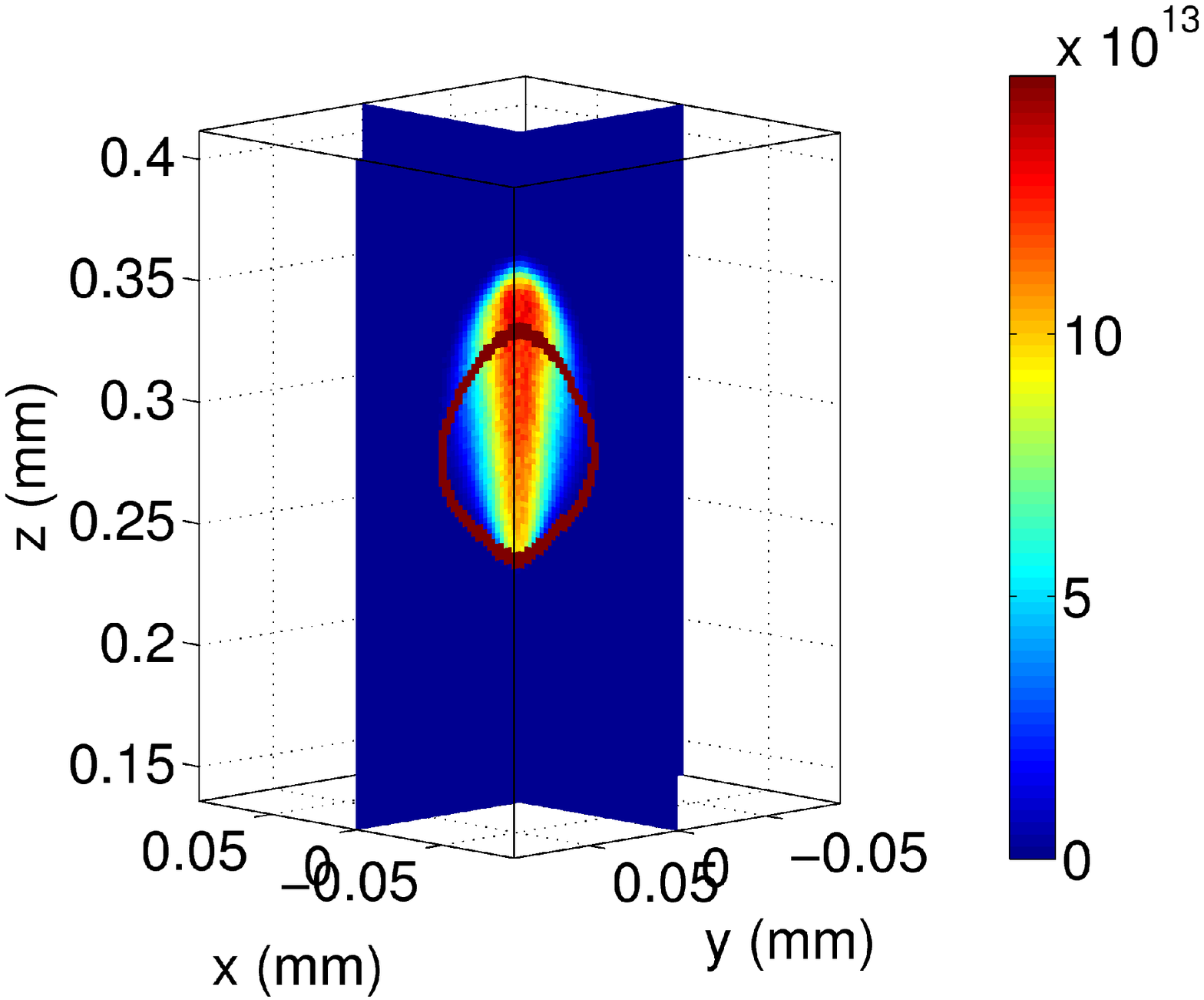}
&
\includegraphics[width=.30\textwidth]{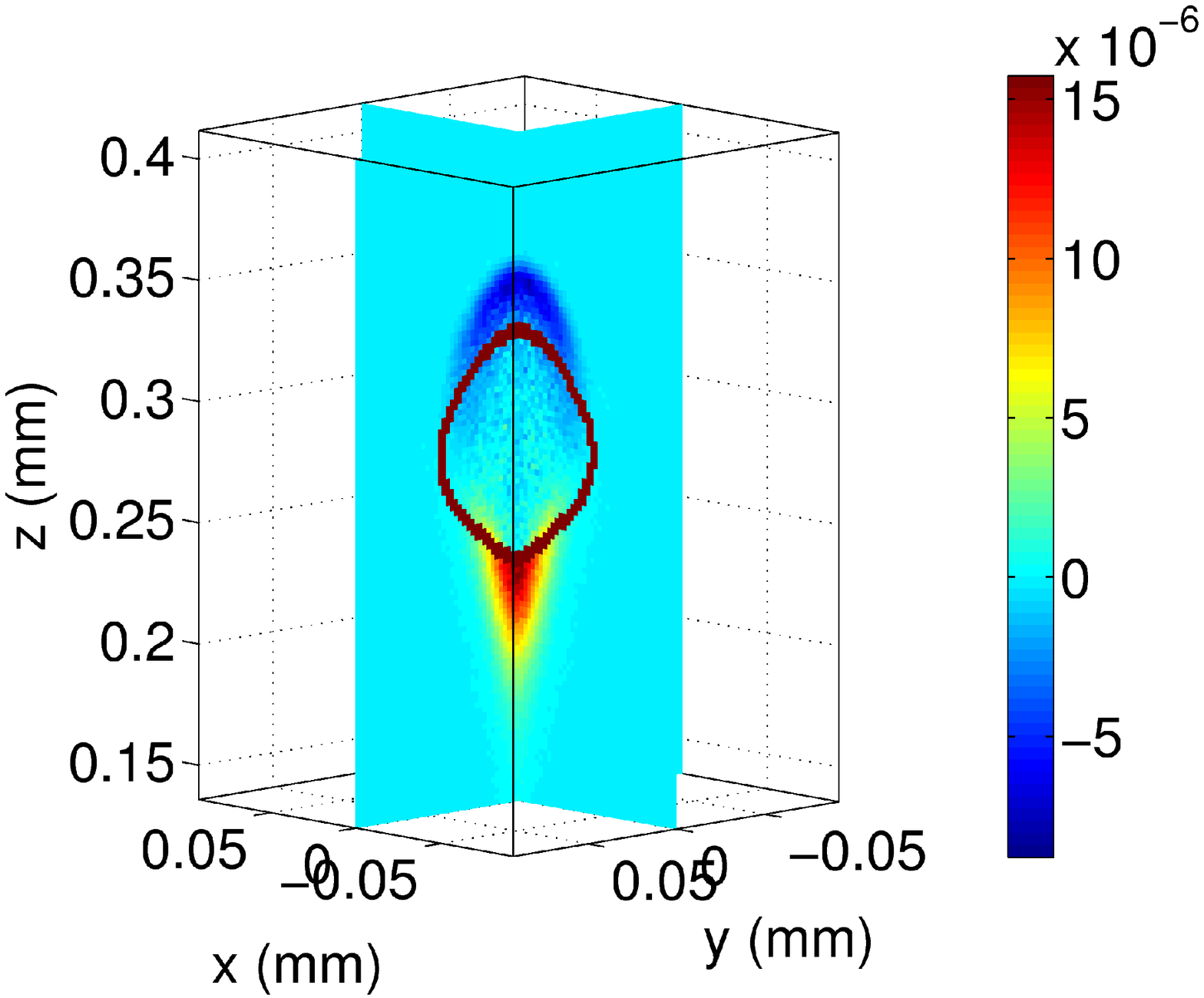}
&
\includegraphics[width=.30\textwidth]{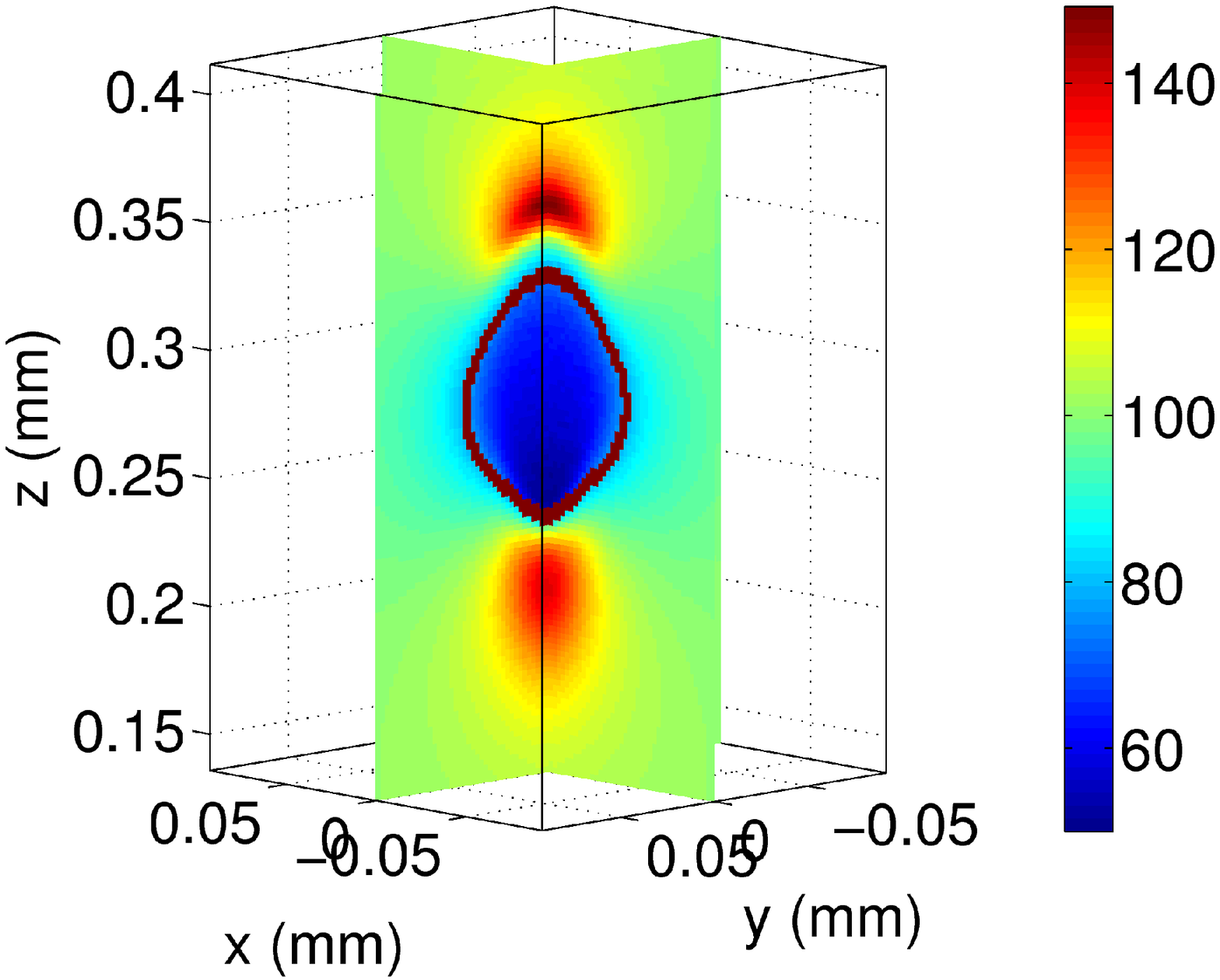}
\\
$\xi$=0.9 &
\includegraphics[width=.30\textwidth]{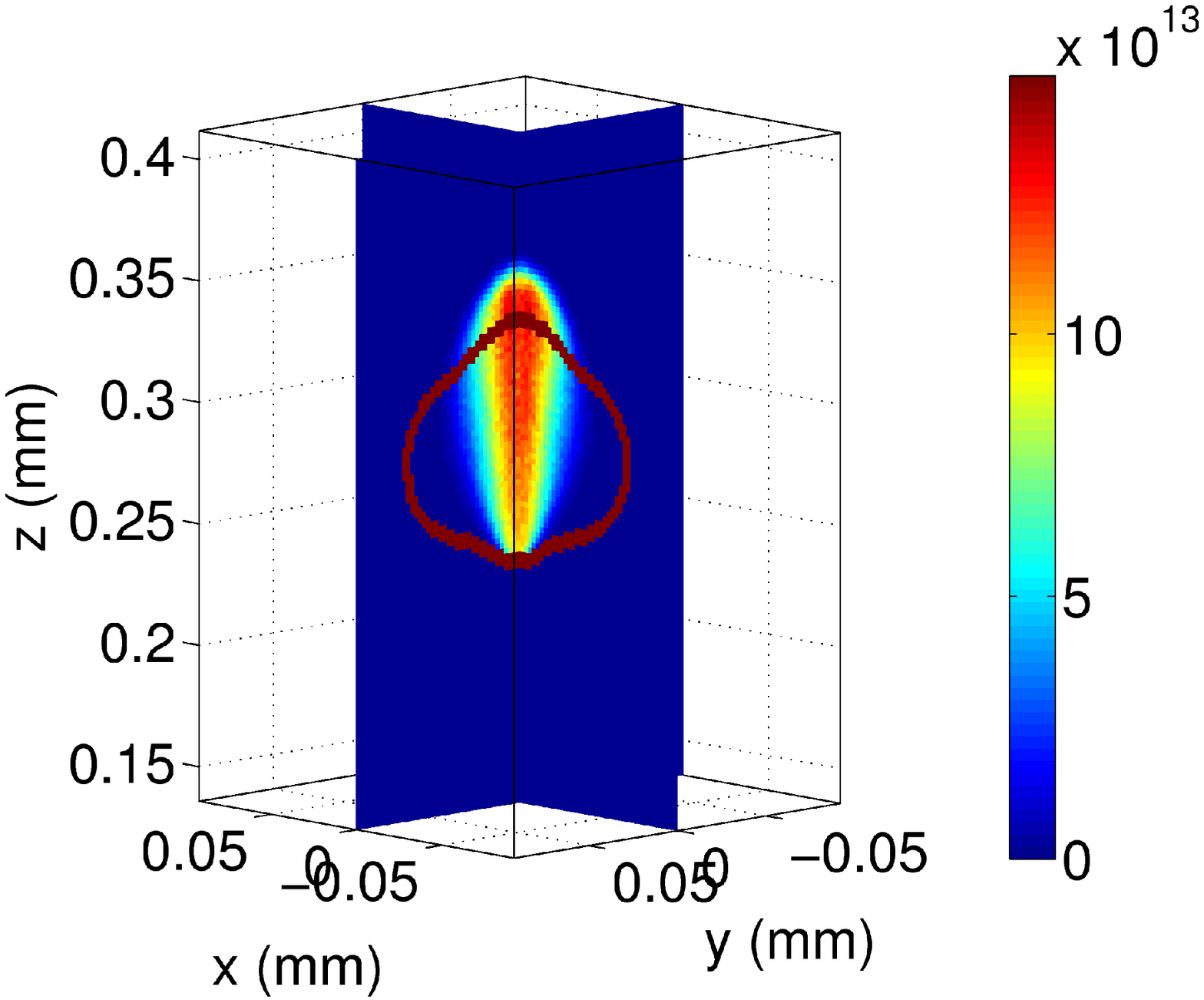}
&
\includegraphics[width=.30\textwidth]{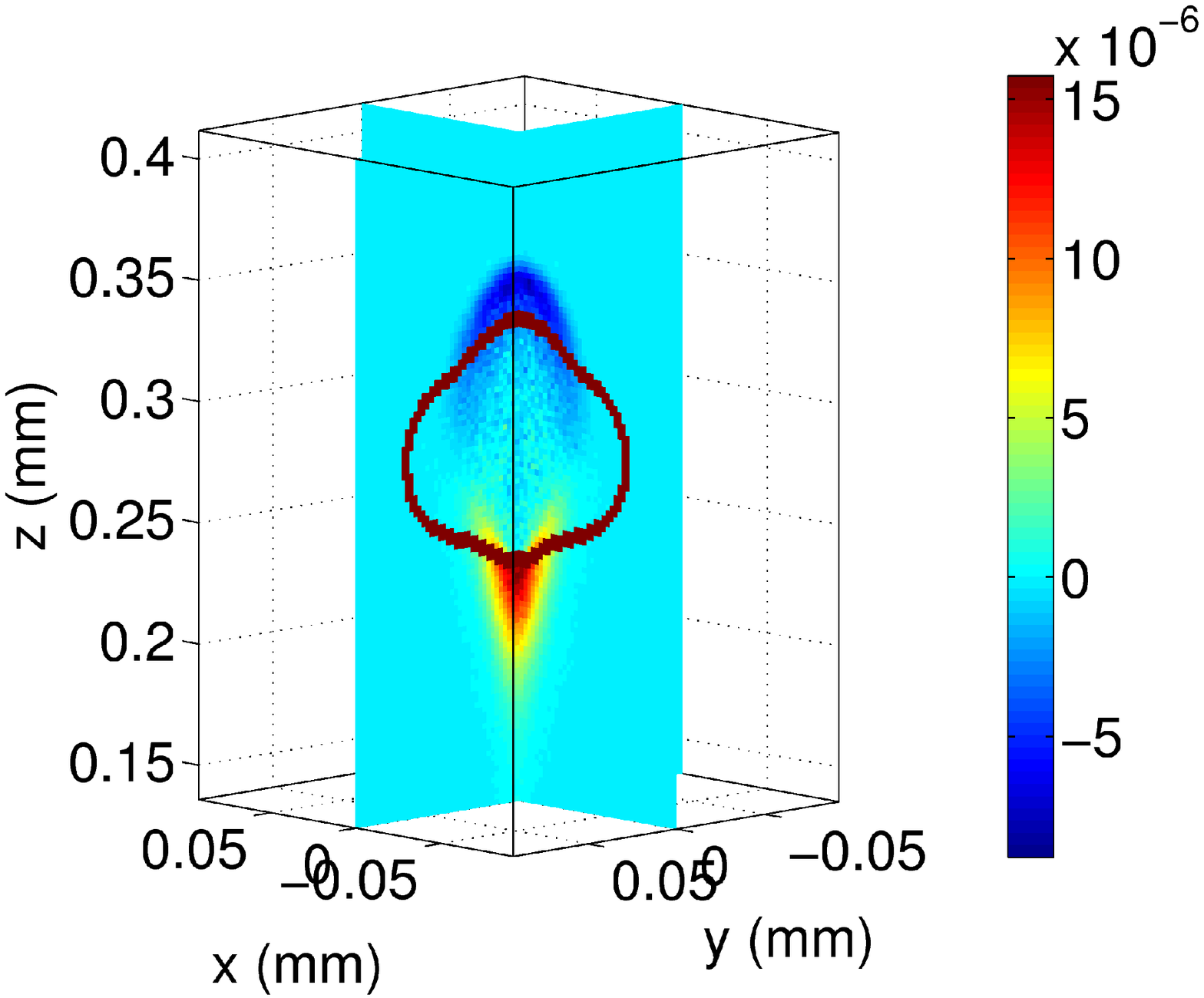}
&
\includegraphics[width=.30\textwidth]{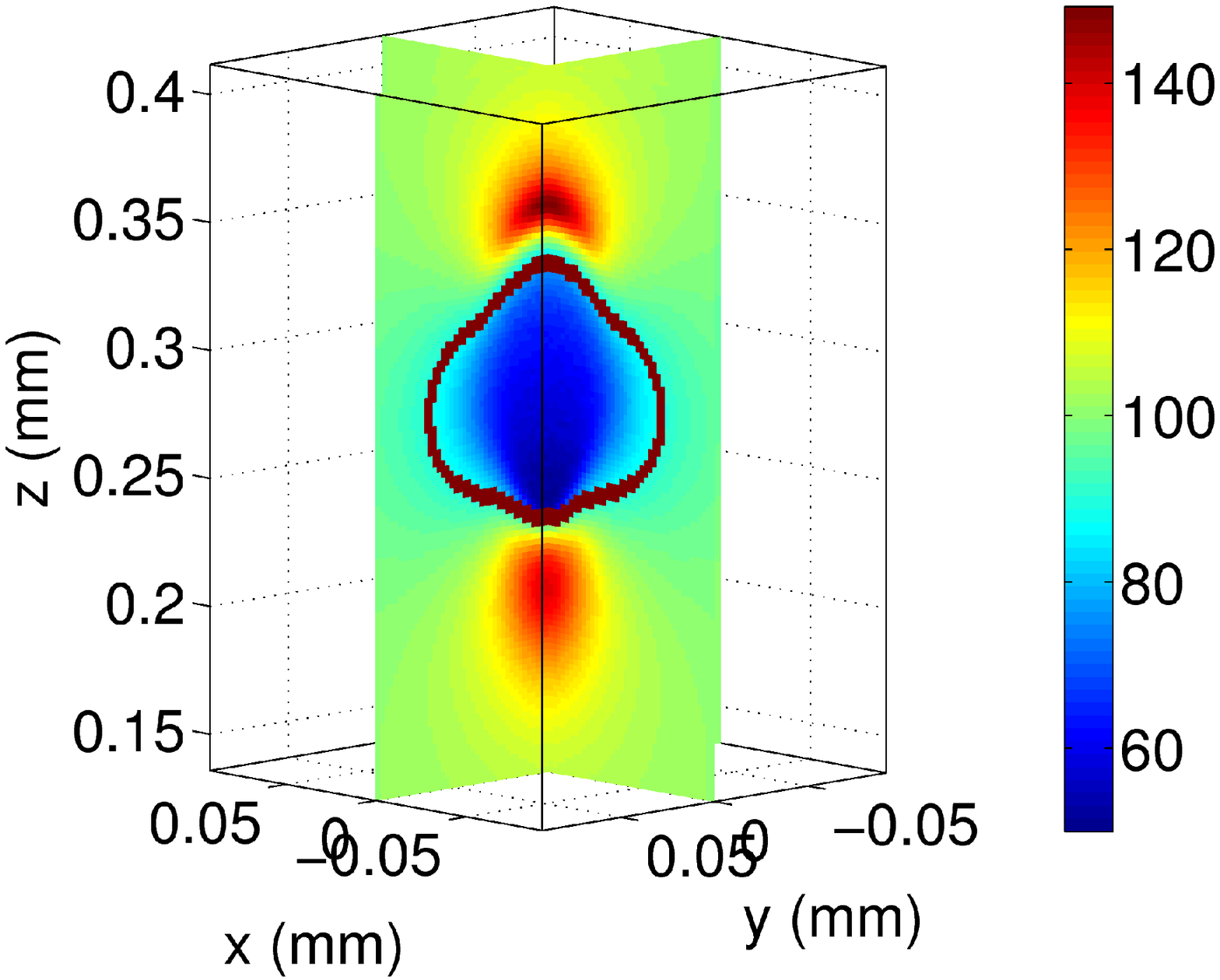}
\\
$\xi$=1.0 &
\includegraphics[width=.30\textwidth]{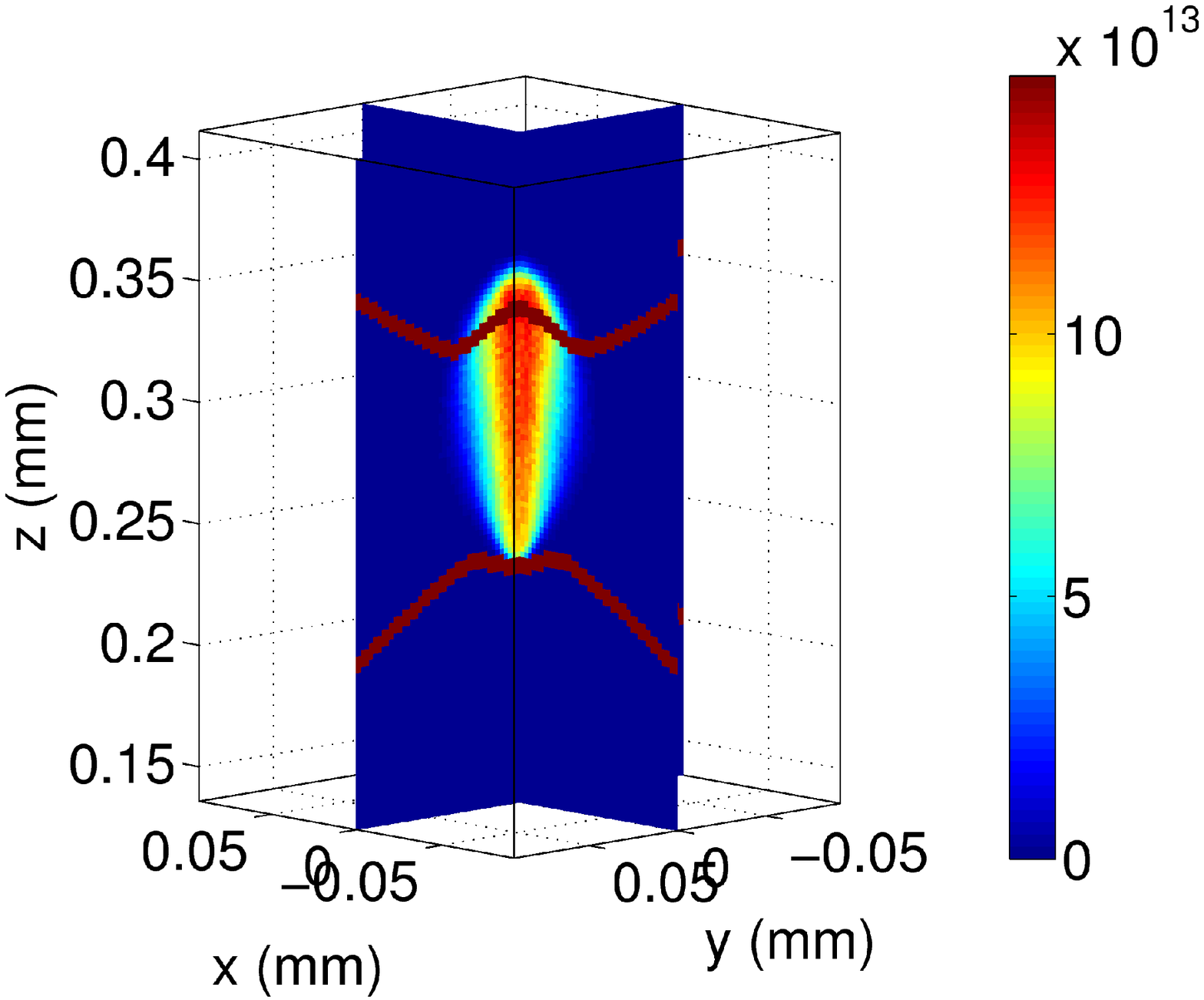}
&
\includegraphics[width=.30\textwidth]{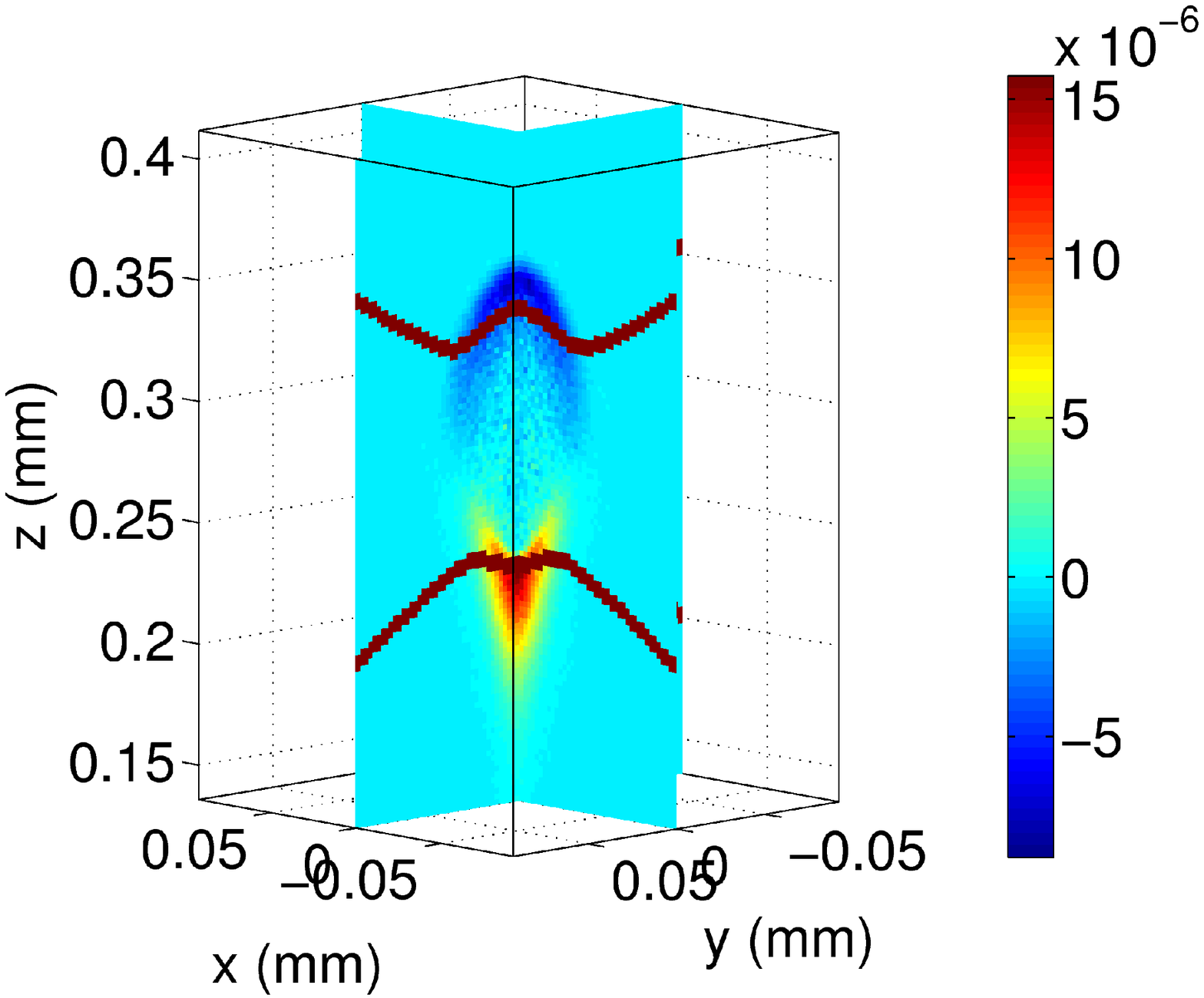}
&
\includegraphics[width=.30\textwidth]{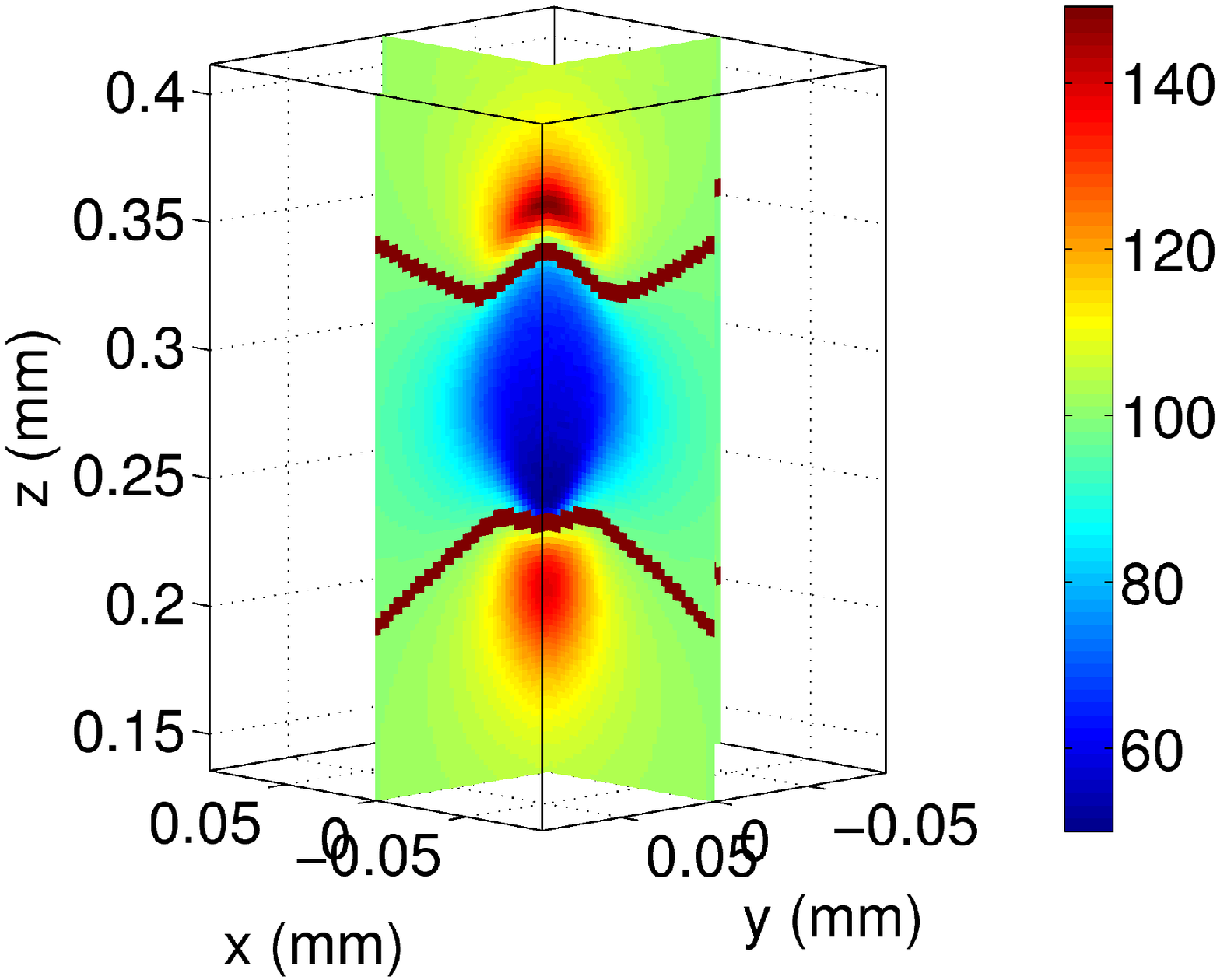}
\\
& \text{electron~density}~(1/\text{cm}^3) & \text{charge~density}~ (\text{C/cm}^3) & \text{E}_z~ \text{(kV/cm)}
\end{array}$
\end{center}
\caption{The model interface based on the electric field only. Electron density (left), charge density (middle), and electric field (right)
     at t$\approx$ 0.46 are presented with the model interface marked. The particle model is applied at the region where the $E > \xi E_b$with $\xi=0.7$ to $1.0$, and the fluid model is applied to the rest. \label{fig:spliting_E}}
\end{figure}

In Fig.~\ref{fig:spliting_E}, we show the model interfaces (marked with red lines) determined only by a field threshold. The presented streamer is from the same simulation at same time t$\approx$ 0.46 as shown in Section~\ref{sec:1st_interface}.
The particle model is applied if the electric field is strong enough, $E>\xi E_b$, in which $E_b$ is the background field 100 kV/cm and $\xi$ is a free parameter. Note that we are not searching a specific model interface anymore, we now search for regions suitable for the particle model.
Fig.~\ref{fig:spliting_E} shows how the particle and the fluid region change when $\xi$ changes from $0.7$ to $1.0$. The fluid region has the smallest volume with $\xi=0.7$ and the largest volume with $\xi=1.0$. From $\xi=0.7$ to $1.0$, the number of electrons in the particle region decreases from $9.6\times10^6$ to $2.5\times10^6$.

Thresholds with $\xi<0.9$ are not appropriate to determine the model interface, not only because there would be too many electrons within the particle region, but also due to the fact that electrons at the side of the streamer can cause a considerable increase of the computational cost when they cross the model interface. In both cases $\xi=0.9$ and $1.0$, electrons at the side of streamer are covered by the fluid regions. But $\xi=1.0$ leaves all the lateral region to the fluid model which makes the model less flexible and computationally more expensive.
The fluid model is computationally more efficient than the particle model only in mesh cells that contain a considerable amount of electrons. Letting the fluid model cover the large open side regions will increase the computational cost due to the non-zero, but sometimes unphysically small densities at side.



\begin{figure}
\begin{center}$
\begin{array}[c]{p{1.0cm}ccc}
$\eta$=0.7 &
\includegraphics[width=.30\textwidth]{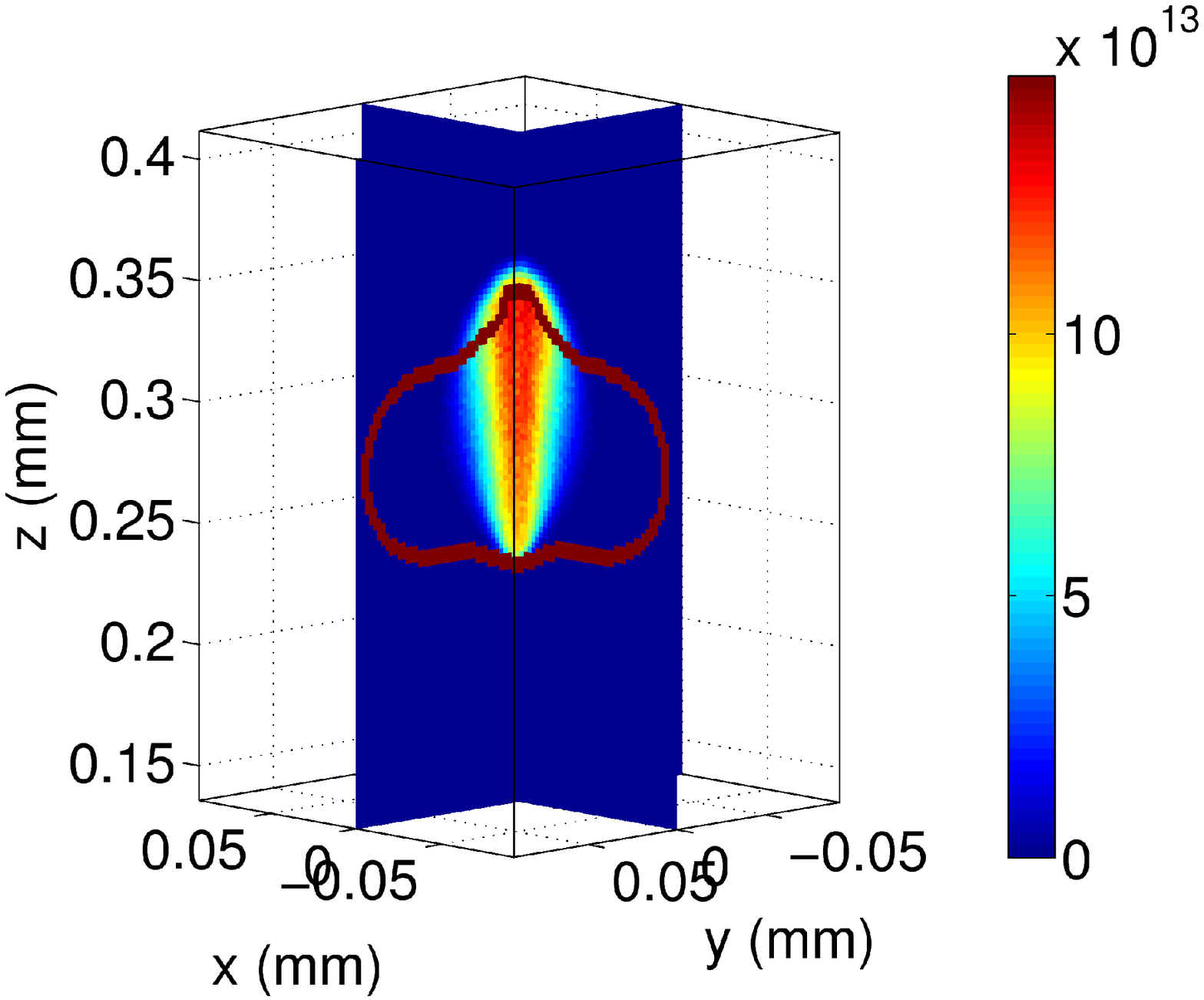}
&
\includegraphics[width=.30\textwidth]{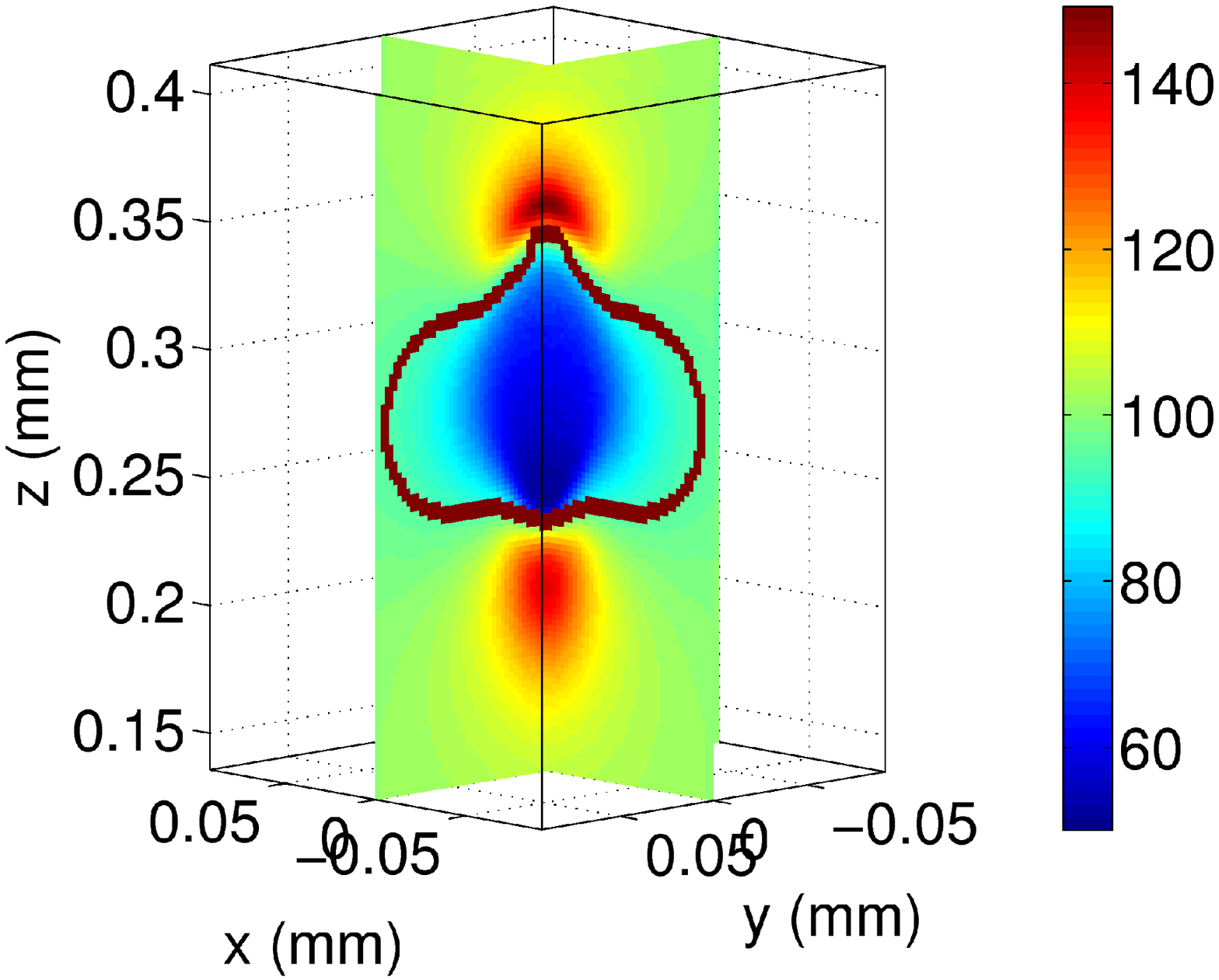}
&
\includegraphics[width=.30\textwidth]{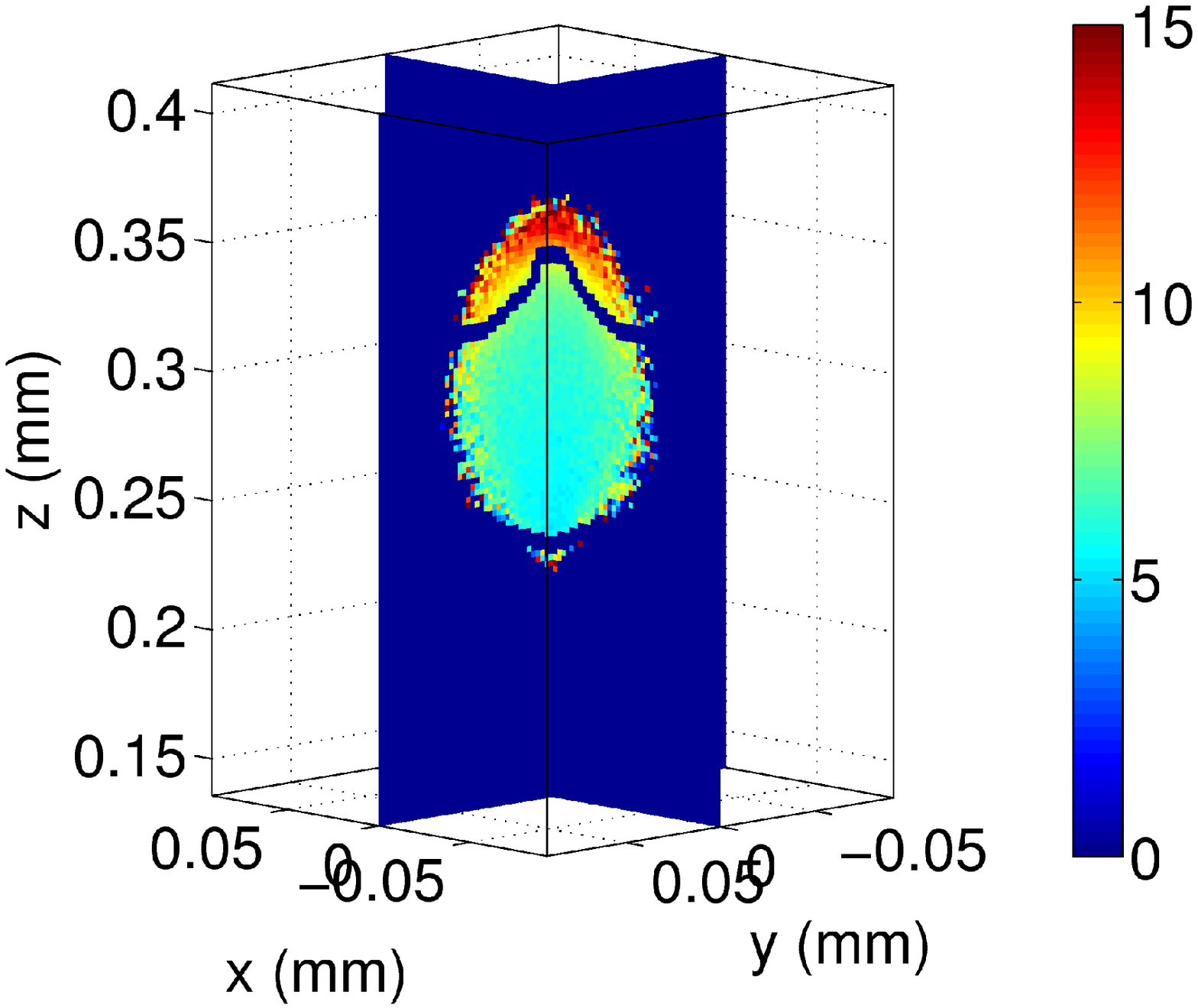}
\\
$\eta$=0.9 &
\includegraphics[width=.30\textwidth]{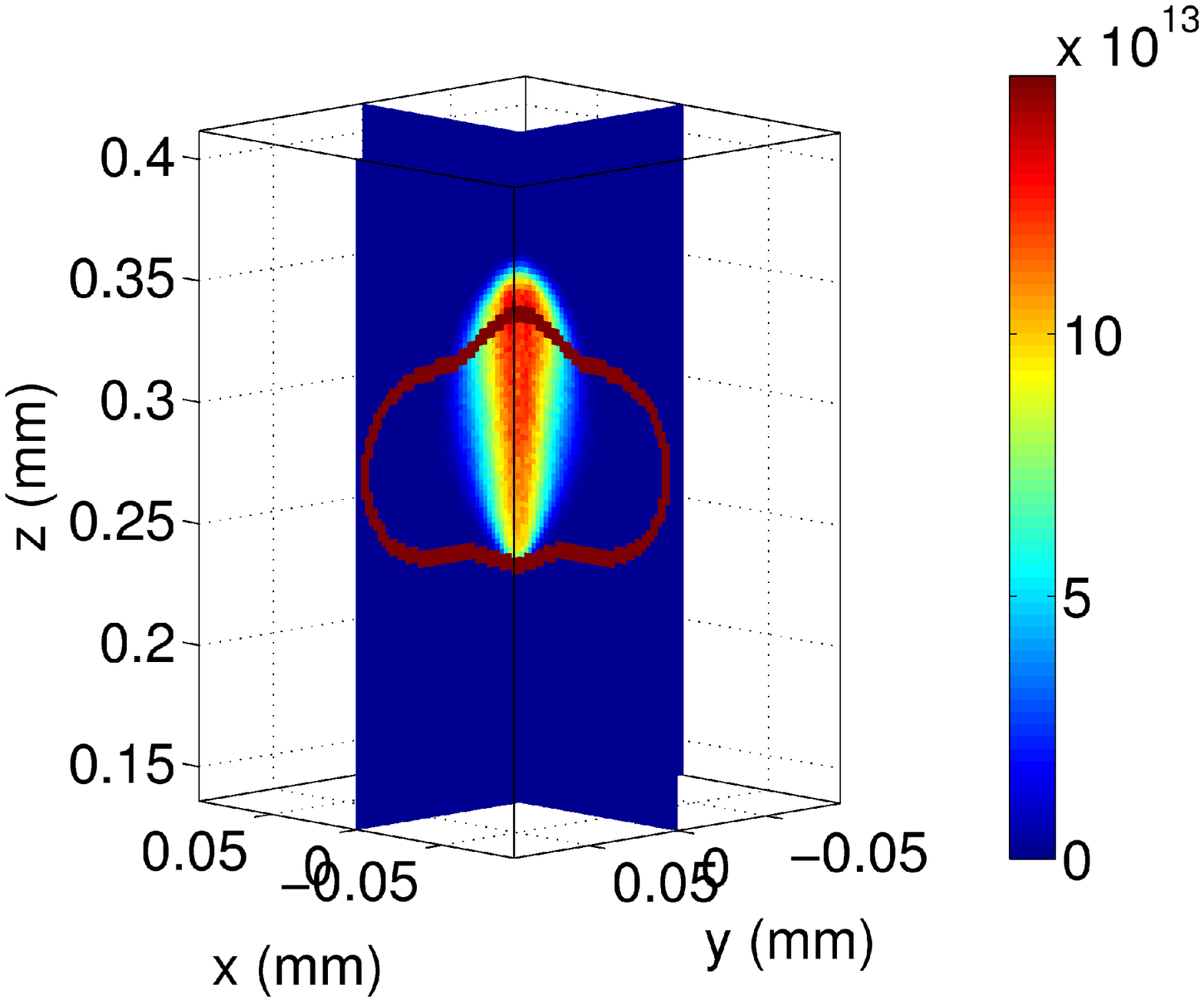}
&
\includegraphics[width=.30\textwidth]{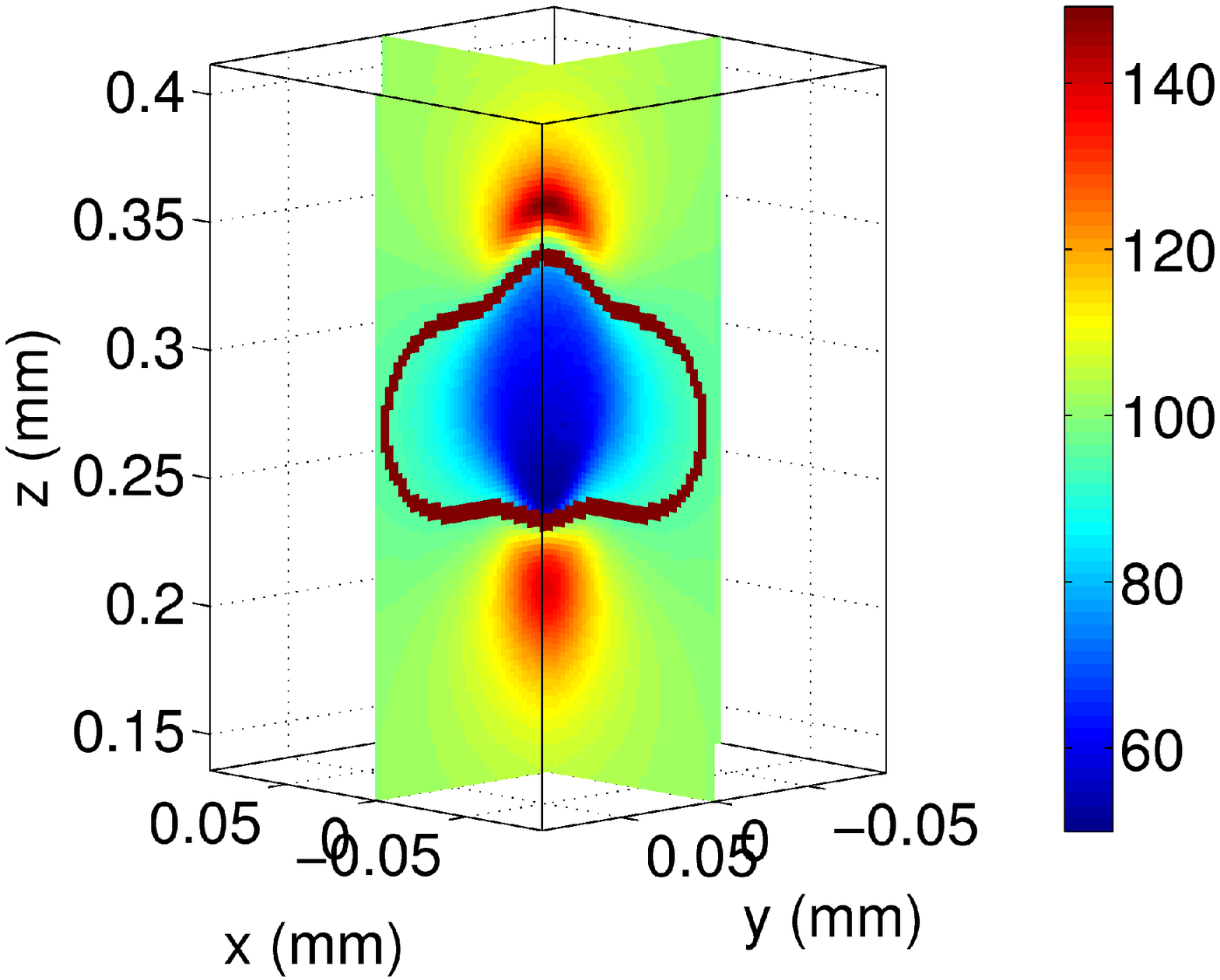}
&
\includegraphics[width=.30\textwidth]{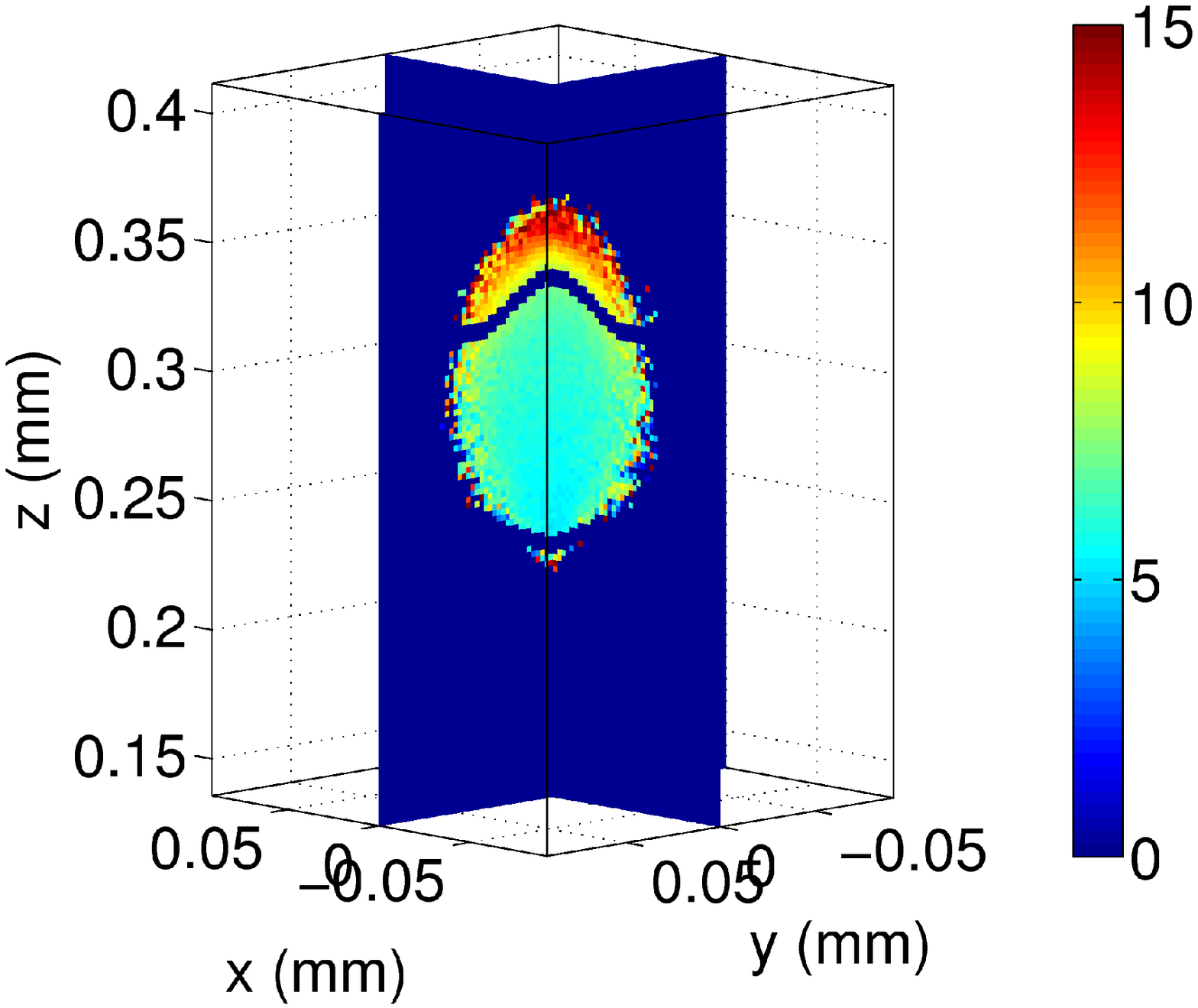}
\\
& \text{electron~density}~ (1/\text{cm}^3) & \text{Electric~field}~(\text{kV/cm}) & \text{Mean~energy}~ (\text{eV})
\end{array}$
\end{center}
\caption{The location of the model interface is simultaneously based on a field and a density threshold. Electron density (left), electric field (middle), and electron mean energy (right) at t$\approx$ 0.46 are presented with the model interface marked. The particle model is applied where the electric field $E > 0.94 E_b$ and the electron density $n_{e} < \eta ~n_{e,max}$ with $\eta=$ $0.7$ (upper panel) and $0.9$ (lower panel), and the fluid model is applied in the remaining region.  }
\label{fig:spliting_E_ne}
\end{figure}

We tested a range of interface positions from $\xi=0.9$ to $\xi=1.0$ and the field criterion $\xi=0.94$ was chosen for the hybrid simulation. With this threshold, the sides of the streamer are fully covered by the fluid model during the period of interest.
Together with the field criterion, a density criterion is added.
The fluid model is applied if the electron density is high enough, $n_{e}>\eta~ n_{e,max}$, where $ n_{e,max}$ is the maximal electron density in the whole simulation domain.
In Fig.~\ref{fig:spliting_E_ne}, we present the electron density, the electric field and the local mean energy of electrons with the model interface indicated with a curved line.
The particle model is applied in the region where the electric field $E>\xi E_b$ {\em and} electron density $n_{e}<\eta ~n_{e,max}$, and the fluid model is applied in the region where the electric field $E <\xi E_b$ {\em or} electron density $n_{e}>\eta ~n_{e,max}$, where $\xi=0.94$ and $\eta=$ $0.7$ (upper panel) and $0.9$ (lower panel).

As shown in Fig.~\ref{fig:spliting_E}, since the side of the streamer has already been allocated to the fluid region, differences only appear at the streamer head due to the choice of $\eta$ in the density criterion.
When the density threshold changes from $\eta=0.9$ to a small value such as $\eta=0.7$, the particle region shrinks and the fluid region expands towards the field enhancement region. 

\begin{figure}
\begin{center}$
\begin{array}[c]{cc}
\it{t}=0.51~ \rm{ns} &
\includegraphics[width=.60\textwidth]{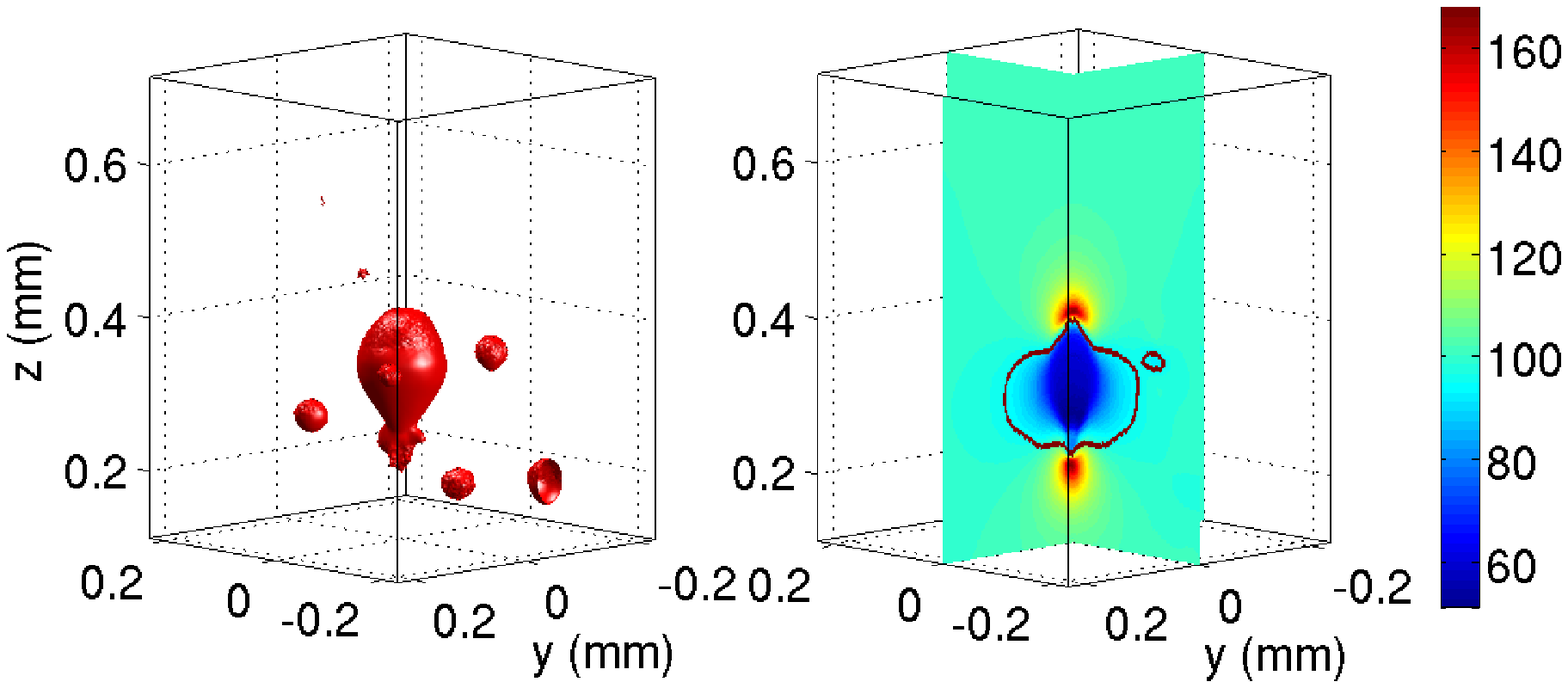}
\\
\it{t}=0.6~ \rm{ns} &
\includegraphics[width=.60\textwidth]{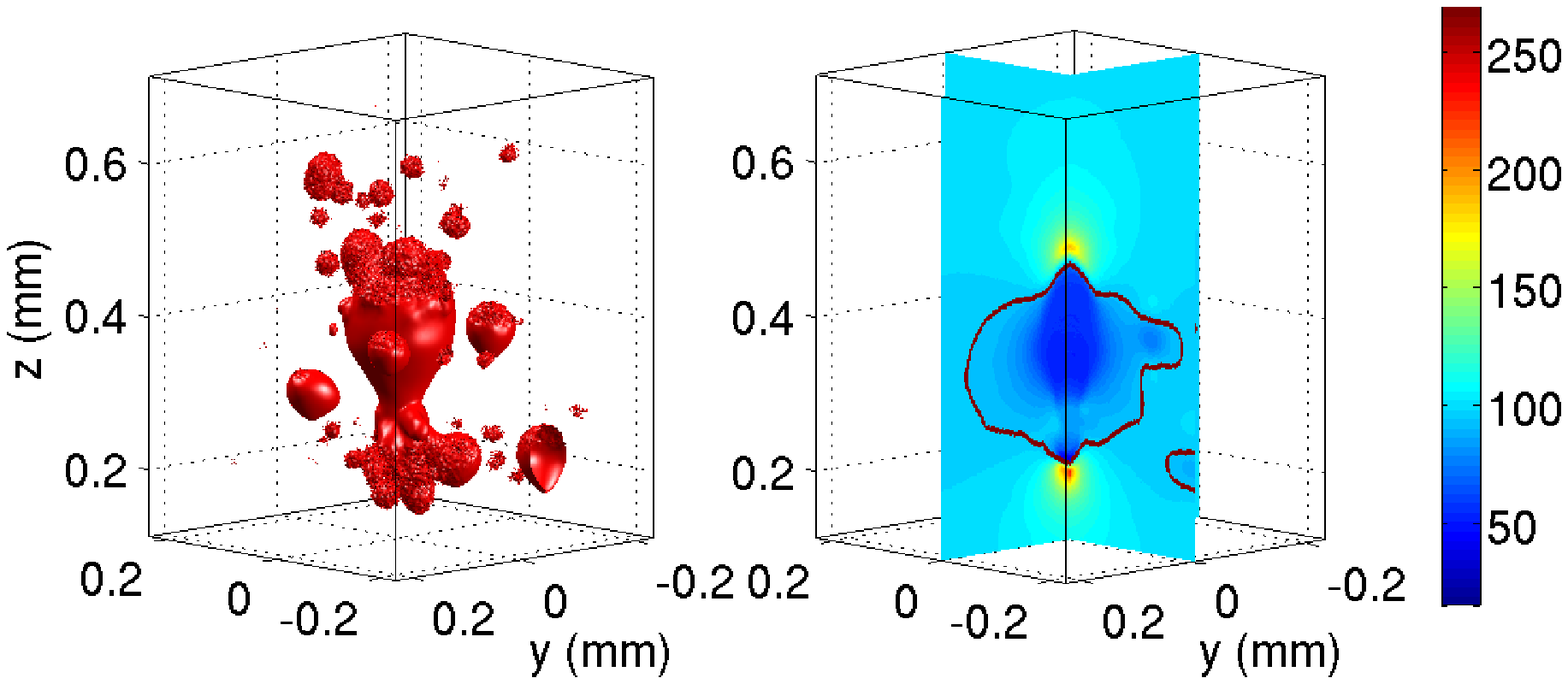}
\end{array}$
\end{center}
\caption{A double headed streamer is developing in air while photo-ionization is included. A surface of equal electron density (left) and the electric field in color coding (right) are presented; the model interface with the buffer region are marked.  }
\label{fig:newMI1}
\end{figure}


In contrast to the column based splitting, which can only follow one streamer head, the full 3D splitting approach can apply the particle model in separate regions. For example, it can follow both the negative and positive ionization fronts when one simulates a double headed streamer, or a second streamer head if another streamer forms.

Hybrid simulations for a double-headed streamer in air with photo-ionization have been carried out in an overvolted gap. The photo-ionization model follows Zheleznyak's model~\cite{Zhe1982}, and the implementation for the particle model is the same as in~\cite{Cha2008,Li2011:IEEE}. Although the development of the double headed streamers, and multi-streamers are not the topic of this paper, two time steps from such simulations are presented in Fig.~\ref{fig:newMI1} to show the flexibility of the full 3D splitting approach.

Fig.~\ref{fig:newMI1} shows a double-headed streamer which is developing in the middle while several small avalanches appear around the streamer. On the left we plot the equal-density surface of the electron bulk with electron density $n_e> 10^{13}$ /cm$^3$. On the right we plot the electric field on two orthogonal planes intersecting the 3D structure, with the model interface and buffer regions marked with red lines.
In the left plot of the upper panel, the larger electron bulk is the main streamer, while some other small avalanches develop around it.
The small avalanches are caused by the electrons created by photo-ionization; i.e., by photons emitted from the streamer head, and several of them have already created a relatively large electron density reaching the density threshold. Separate fluid regions therefore are created in this area. It happens that the $xz$ plane where the electric field is plotted crosses one of these avalanches. A cross section of the separate model interface for this avalanche is therefore also included in the electric field plot.

In the lower panel, we plot the streamers 0.09 ns later. The separate fluid region for the small avalanche has merged with the main fluid region of the main streamer channel, while a new separate fluid region has formed. Note that not only the upper streamer head is followed by the particle model, but that the particle model also follows the field enhanced region of the downward propagating streamer head.

\subsection{Structure of the buffer region}\label{sec:buffer_region}

The interaction of two models is realized through the well know "buffer region" technique which has been employed
in hybrid computations for air flow~\cite{Wad1990,Tal1997,Gar1999,Ale2002,Ale2005,Akt2002,Sun2004}, liquid
flow~\cite{Con1995,Had1999,Del2003}, and also in small scale solid systems~\cite{She1999,Rud2000,Wag2003}.
A similar method that is more suitable for streamer simulations has been implemented and tested for planar fronts~\cite{Li2008:1,Li2010:1}.
Here our implementation in the 3D hybrid model is discussed.

The electrons that cross from the particle region over the model interface will be recorded, since they contribute to the density in the fluid region. We have explained in~\cite{Li2010:1} that in a planar negative front, since the electrons move on average slower than the model interface, we only need to remove the electrons from the particle list if they fall into the fluid region, but we do not need to artificially create new electrons in the buffer region to stabilize the electron flux at the model interface. The same occurs in the (negative) head of a 3D streamer: electrons on average move slower than the model interface in the propagation direction, and we don't need to add new electrons artificially. The side of the streamer is normally attributed to the fluid region in our 3D hybrid model; as no electrons are followed by the particle model here, the interface does not need to be constructed here. At the tail of the streamer, there is no electron flow without photoionization or background ionization, and the model interfaces are more or less stationary. If photo-ionization is added, at the positive head of the streamer, the electrons propagate into the streamer from the particle into the fluid region. Therefore no new electrons need to be created artificially at any part of the model interface.

Here we set the length of the buffer region as 3
cells in the $z$-direction and 2 cells in $x$- and $y$-direction with the cell length $\Delta x= \Delta y=\Delta z=2.3~\mu$m, since the maximum electric field in $z$- direction is normally much higher than in the $x$- or $y$- direction (see Table I in~\cite{Li2010:1}).

It is important in 3D that for any cell face, cell edge and cell corner shared by the particle region and the
fluid region, a buffer region separates them. A direct contact of particle and fluid model
without a buffer region can cause electron leaking, which creates loss of mass and charge.
\begin{figure}
   \centering
 \includegraphics[width=.4\textwidth]{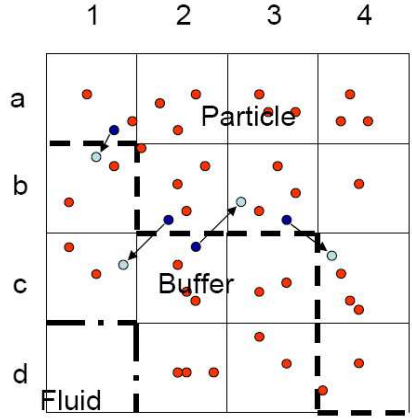}
 \caption{Schematic illustration of the corner problem between the particle and fluid cells around the model interface.}\label{fig:flux_cells}
\end{figure}

The so called ``corner problem''~\cite{Gar1999} is a technical but important issue when calculating the electron flux in the
hybrid model. When an electron passes from the particle region into the fluid region or vice versa, its
contribution to the flux on the cell face is recorded if the cell is in the fluid region, since only the density update in the fluid region needs the electron flux. The cell face is not
necessarily the model interface between particle and fluid region. A 2D cartoon is shown in
Fig.~\ref{fig:flux_cells} to illustrate the problem. The particle region is in the upper right part and the fluid
region is in the lower left corner, the particle region extends into the fluid region by 2 cells in all directions and
creates a buffer region. As illustrated in Fig.~\ref{fig:flux_cells}, when an electron flies from cell "a1"
to "b1", it contributes to the flux on the model interface at cell face "a1$\leftrightarrow$ b1". A more
complicated case occurs when an electron flies from cell "b2" to "c1", it contributes not only to the model
interface at cell face "b2$\leftrightarrow$ c2", but also to the cell face "c1$\leftrightarrow$ c2". On the other hand, when the electron flies from cell "c2" to "b3", it only contributes to the model interface at cell face "b2
$\leftrightarrow$ c2". Finally, consider that the
electron flies from "b3" to "c4"; it flies over two model interfaces, but in the model it contributes to
neither of them since there is no mass change in the fluid region.

\begin{figure}
   \centering
 \includegraphics[width=.5\textwidth]{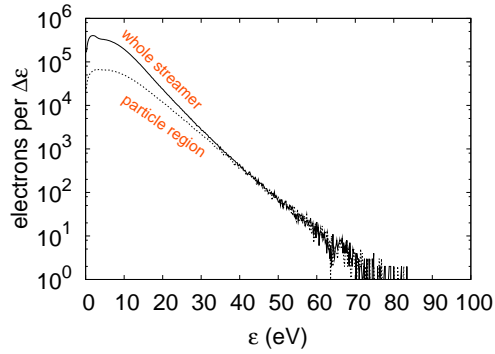} \\
 \caption{Electron energy distribution function (EEDF) of all electrons in the streamer at t$\approx$0.46 ns (solid line) and the part of electrons in the particle and buffer region (dotted line). The EEDF is not normalized and $\Delta \epsilon=0.2 $ eV. The part of electrons which removed from the particle list are mainly the low energy electrons.}\label{fig:eedf_coupling}
\end{figure}

When the simulation reaches a total number of 20 million electrons, at t$\approx$0.46 ns, we split the simulation domain into a particle and a fluid region. If we use the column based splitting with model interface at $E=0.84~ E_{max}$, 12 million electrons which are neither in the particle nor in the buffer region are removed from the particle list and transferred into particle densities in the fluid region, while 5 million electrons remain in particle region and 3 million electrons remain in the buffer region.
If we define the model interface using the full 3D splitting with $E=0.94~ E_b$ and $n_{e}= 0.7 n_{e,max}$, 15.5 million electrons in the fluid region are removed from the particle list, while 2.5 million electrons remain in particle region and 2 million electrons remain in the buffer region.
Both splittings leave most of the ionization front to the particle model. It
is also remarkable that the majority of the high energy electrons remain in the particle list which results in a
good model for the study of runaway electrons. In Fig.~\ref{fig:eedf_coupling}, we show the electron energy distribution function (EEDF) of electrons from the whole streamer and from only the particle and buffer region. The EEDF is not normalized so one can clearly see in which energy range the electrons are removed from the particle list.

\section{Simulation results}\label{sec:sim_result}

Having introduced the new coupling scheme with the numerical details, we now present our hybrid simulation results for streamers in air without photo-ionization.


\begin{figure}
\begin{center}
$\begin{array}[c]{p{1.0cm}c}
t=0.48 &
\includegraphics[width=.80\textwidth]{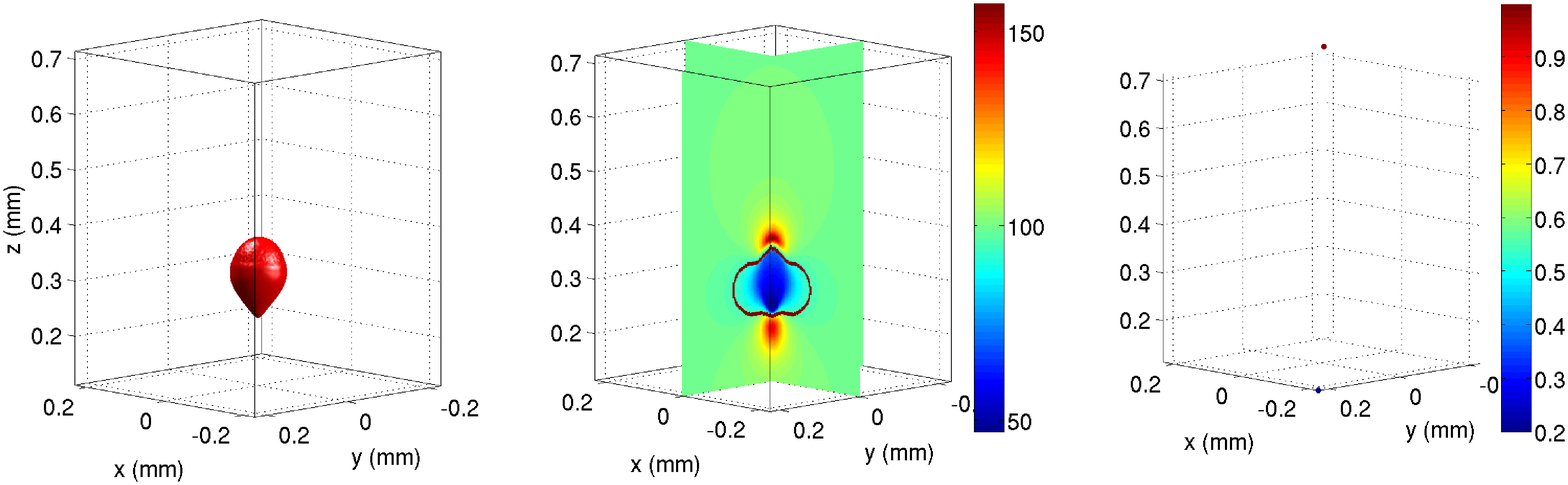}
\\
t=0.56 &
\includegraphics[width=.80\textwidth]{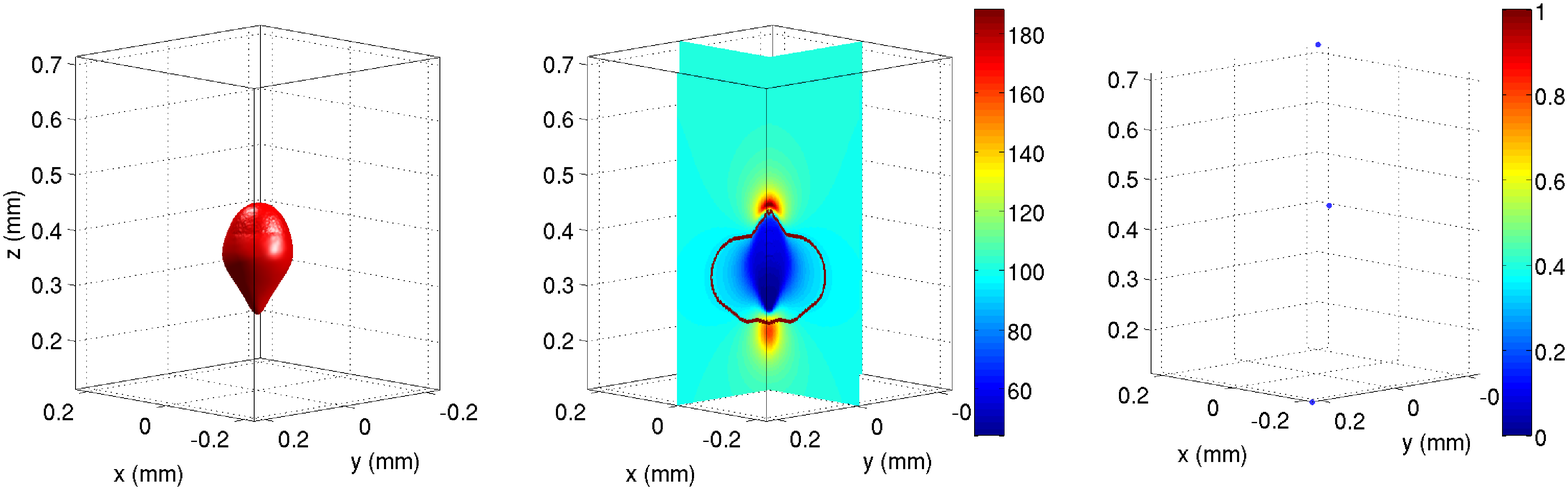}
\\
t=0.72 &
\includegraphics[width=.80\textwidth]{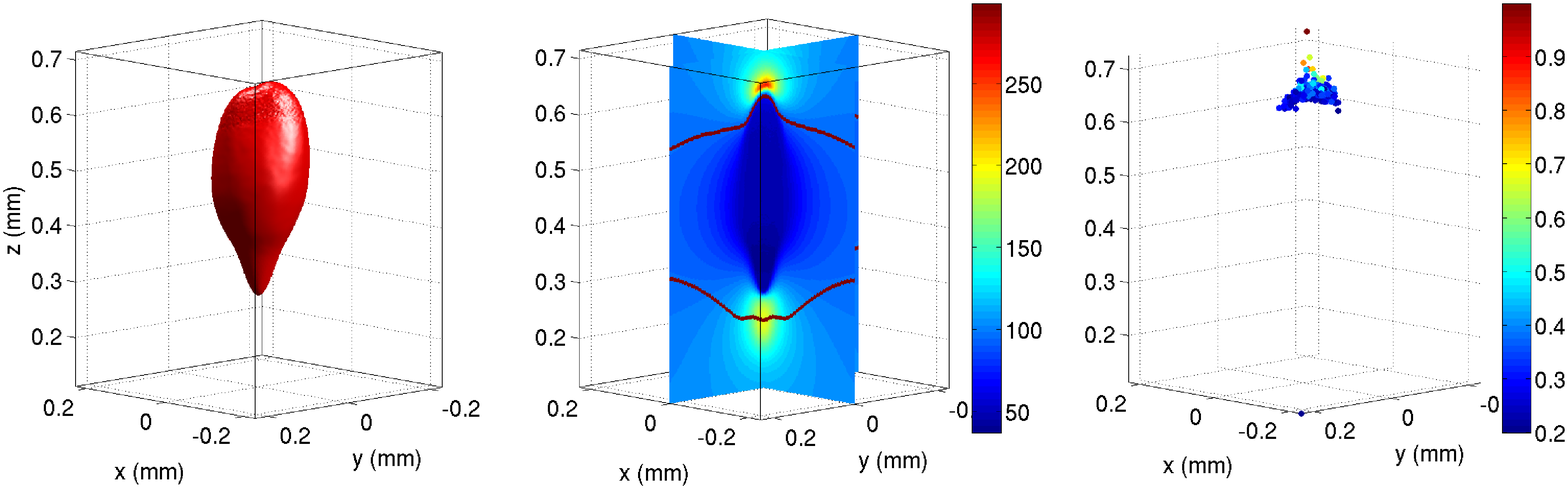}
\\
& \text{electron~density}~(1/\text{cm}^3) \hspace{1cm} \text{Electric~field}~ (\text{kV/cm}) \hspace{1.2cm} \text{High~energy~electrons}~ (\text{eV})
\end{array}$
\end{center}
\caption{3D simulation results of the 3D hybrid model for a negative streamer developing in a background field of -100 kV/cm.  First row: the start of hybrid computation at t = 0.48 ns, second row: streamer at t = 0.56 ns, third row: streamer at t = 0.72 ns. The columns show from left to right: electron density, charge density, and electric field strength. Particle densities and fields are represented on two orthogonal planes that intersect with the 3-dimensional structure. \label{fig:hy_3d} }
\end{figure}

The initial conditions and the configuration are the same as the fluid simulation in Section~\ref{sec:fluid_results}. The model interface is determined by the full 3D splitting approach with the field criterion $E_{\xi}=0.94 E_b$ and density criterion $n_{e,\eta}= 0.7 n_{e,max}$. In Fig.~\ref{fig:hy_3d}, we show a negative streamer followed by the hybrid model at different times. At time 0.48 ns (upper panel), after the hybrid calculation started 0.02 ns, there are $2.8\times 10^7$ electrons in total, where $6.0\times10^6$ electrons are followed by the particle model in the particle and buffer region, and the maximal electric field at that time is around 150 kV/cm.

At time 0.56 ns (middle panel), the electric field reach 180 kV/cm, there are $7.5\times 10^7$ electrons in total, in which $8\times10^6$ electrons followed by the particle model.

In the bottom panel of Fig.~\ref{fig:hy_3d}, we show the streamer at t=0.72 ns. The maximal electric field at that time is around 300 kV/cm. There are $4.5\times10^8$ electrons in total, and $2.0\times10^7$ electrons are followed by the particle model.

At time 0.56 ns, electrons with energies above 200 eV start to appear at the streamer head. But electrons with such energies can not hold their energy long, they lose their energies very quickly and disappear in the next time step. As the streamer propagates, the maximal electric field keeps increasing, there are more electrons with energies above 200 eV generated and the maximal electron energy also increases. At t= 0.72 ns, the maximal electron energy is around 1 keV.


\section{Conclusion}\label{sec:conclusion}



\subsection{Numerical technique}

The new splitting method, the full 3D splitting, has overall advantages over old column-based splitting. It adaptively applies the particle model in the most field-enhanced region and the fluid model in regions where the density is high enough. It is flexible in complicated situations where spatial coupling is required. Moreover, it is easier to add new features such as local grid refinement in the full 3D splitting than in the column based splitting.

The accurate definition and the evaluation of the electron transport coefficients and reaction rates from the particle model for the fluid model are essential for the subsequent successful coupling of the models. This is the reason why a large part of this paper and also our previous papers~\cite{Li2007,Li2010:1} are devoted to the correct calculation of these coefficients. These coefficients are required for the successful implementation of the hybrid model.

Compared to simulations with the pure fluid or pure particle model, the 3D hybrid model combines the advantages of both models. It describes the dynamics of a streamer channel in a very efficient way while being able to follow the movement of each single electron in the most active region of the streamer head. The 3D hybrid model is therefore a powerful tool to study the kinetics of electrons in the important regions of streamers.

\subsection{Physical predictions}

The 3D hybrid streamer model is a major step forward for several physical problems. First, it correctly traces the high electron energies in space and therefore can supply correct densities for the many different excited states of molecules; these distributions determine the further chemical reactions and end products at later times. Second, the model traces both electron density fluctuations and electron energy fluctuations, and therefore it can correctly trace the influence of fluctuations, e.g., on streamer branching. These two aspects have not been evaluated yet from the simulations. Third, as shown in~\cite{Li2009}, it can predict electron run-away from streamers. We now elaborate on this aspect.

Electrons with energies above 200 eV are shown in Fig.~\ref{fig:hy_3d} and similarly in~\cite{Li2009}; they appear in the regions with high local field enhancement, and some of them are accelerated up to 2.5 keV before they disappear into the anode. The energy of 200 eV is the threshold value for electron runaway in air. As shown in Fig.~\ref{fig:fre_total}, above 200 eV, the electron collision frequency decreases, and the electrons are accelerated further more easily.

The development of run-away electrons has been studied in a range of constant uniform electric fields with particle swarm experiments~\cite{Kun1986:3,Kun1988:2,Bak2000,Bab2001,Vrh1992,Vrh2001}, in agreement with~\cite{Gur1961,Mos2006}, where it was shown that thermal electrons run away when electric field exceeds a critical strength of E$\approx$ 260 kV/cm at standard temperature and pressure, and the runaway rate increases as the field increases. This is quite reasonable if one considers that in such a field an electron gains 200 eV within less than 8$\mu$m, which is the range of the mean free ionization length.

However, much lower background fields together with the field enhancement at the streamer tip are sufficient to let electrons run away. In our simulation, the background field is 100 kV/cm and the streamers are quite short (as the simulations do not have adaptive grid refinement yet). The field in these simulations is eventually enhanced by a factor of 3, but even when it just exceeds 180 kV/cm locally, electron run-away sets in. Chanrion {\it et al.}\ even observed electrons with 1 keV when the locally enhanced field reaches 160 kV/cm $N/N_0$ for a negative streamer in air at 70 km altitude~\cite{Cha2010} (where $N_0$ is air density at ground level and $N$ at 70 km altitude). The discrepancy between our and Chanrion's data is probably due to the fact that the total number of electrons in a streamer scales as $1/N$~\cite{Ebert2010/JGR}, and therefore streamers at high altitudes contain more electrons, and therefore the rate of run-away electrons is larger. But how can the different run-away thresholds of 260 kV/cm in homogeneous fields and of 160 or 180 kV/cm in the streamer head be explained?

The answer is two-fold. First, 260 kV/cm is the threshold where the majority of electrons experience more acceleration than friction and run away, while single electrons in the high energy tail of the distribution can run away earlier. Second, the electrons in the streamer head are not in equilibrium to the local electric field. Even in a constant field, the electrons in the tip of a swarm on average have higher energies than at the back end~\cite{Li2007, Li2010:1}. It is these high energy electrons at the tip of the ionization front that run into the highly enhanced electric field at the streamer tip and are accelerated further.

But if the electrons run away, they run into the region with lower field ahead of the streamer. For example, electrons with an energy of 200 eV run $\sim$6 times faster than the streamer when its maximally enhanced electric field is 250 kV/cm. Depending on the spatial profile of the electric field and on the energy of the electrons, they will either predominantly loose their energy ahead of the streamer or accelerate further (if no anode is in their way).

For fully understanding the hard radiation from sparks~\cite{Ngu2008,Dwy2008,Rep2008} and corona streamer discharges~\cite{Ngu2010}, two ingredients are necessary: the acceleration of thermal electrons to energies above 200 eV in streamers or next to pointed electrodes; and the profile of the electric field ahead of a streamer corona or next to another strongly curved electrode that supports the further acceleration of the energetic electrons.

\subsection{Outlook}

The size of the system is kept small to obtain sufficient numerical accuracy, while accurate solutions are not required everywhere. In the future local grid refinement will be incorporated in the hybrid model. It will allow us to study streamer propagation in a large system, and the generation rate of the runaway electrons can therefore be obtained even for lower background fields.

We have presented preliminary results for a double headed streamer in air with photo-ionization in Section~\ref{sec:2nd_interface}. Photo-ionization is a three-step process where first a nitrogen molecule is excited by electron impact, then it emits a photon, and the photon ionizes oxygen. In our particle model the nitrogen impact excitation is already included. Therefore it will be possible to develop the widely-used photo-ionization model further to take the lifetime of the excited species into account. It may change the position of the photon source and may consequently influence the streamer propagation.

{\bf Acknowledgment:} The authors wish to thank L. Pitchford for helpful discussions on the momentum transfer cross sections and also S. Dujkov for the suggestions of flux and bulk diffusion coefficients.
The main part of this paper is from the thesis of C.Li, which was supported by the Dutch National Program BSIK, in the ICT project BRICKS, theme MSV1. C.Li also acknowledges recent support through STW-project 10118 of the Netherlands Organization for Scientific Research NWO, and the EPG group of Eindhoven University of Technology.

\appendix
\section{Comparing transport and reaction parameters of the fluid model for air, nitrogen, oxygen, and argon} \label{app:parameters}

The electron transport coefficients, attachment and the ionization rate in air have been presented in section~\ref{sec:3d_particle}. The simulated air is composed by 78.12\% nitrogen, 20.946\% oxygen , and 0.934\% argon. 
In this section the generated transport coefficients and reaction rate are compared with the {\sc bolsig}+ package~\cite{Siglo}. {\sc bolsig}+ is a Boltzmann solver to calculate electron transport coefficients in gases or gas mixtures.
It is based on the two-term Legendre expansion solution of the Boltzmann equations~\cite{Pit1981,Hag2005}.
And the {\sc bolsig}+ calculated transport coefficients have been widely used as an input of the fluid simulation for various plasma applications.

To compare with each other, both our particle model and the {\sc bolsig}+ package use the same cross section data ({\sc siglo} database~\cite{Siglo}) with the same energy splitting mode in ionization collisions. The energy splitting between the primary and the secondary electrons during an ionization has two different modes in {\sc bolsig}+, "equal sharing" or "primary electron takes all". In both {\sc bolsig}+ and our particle model, we used  "equal sharing".

Both the bulk coefficients and the flux coefficients are calculated from our particle model. For each of them, the particle swarm experiments are carried out both with isotropic and with anisotropic scattering. Since the momentum transfer cross section is fixed, as discussed in Section~\ref{sec:Data_DCS}, the electron swarms shall be similar in both cases.

\subsection{Air}

\begin{figure}
\centering
   \subfigure[\label{fig:coe_mu} ~Mobility $\mu(E)$ ]{
         \includegraphics[width=.40\textwidth]{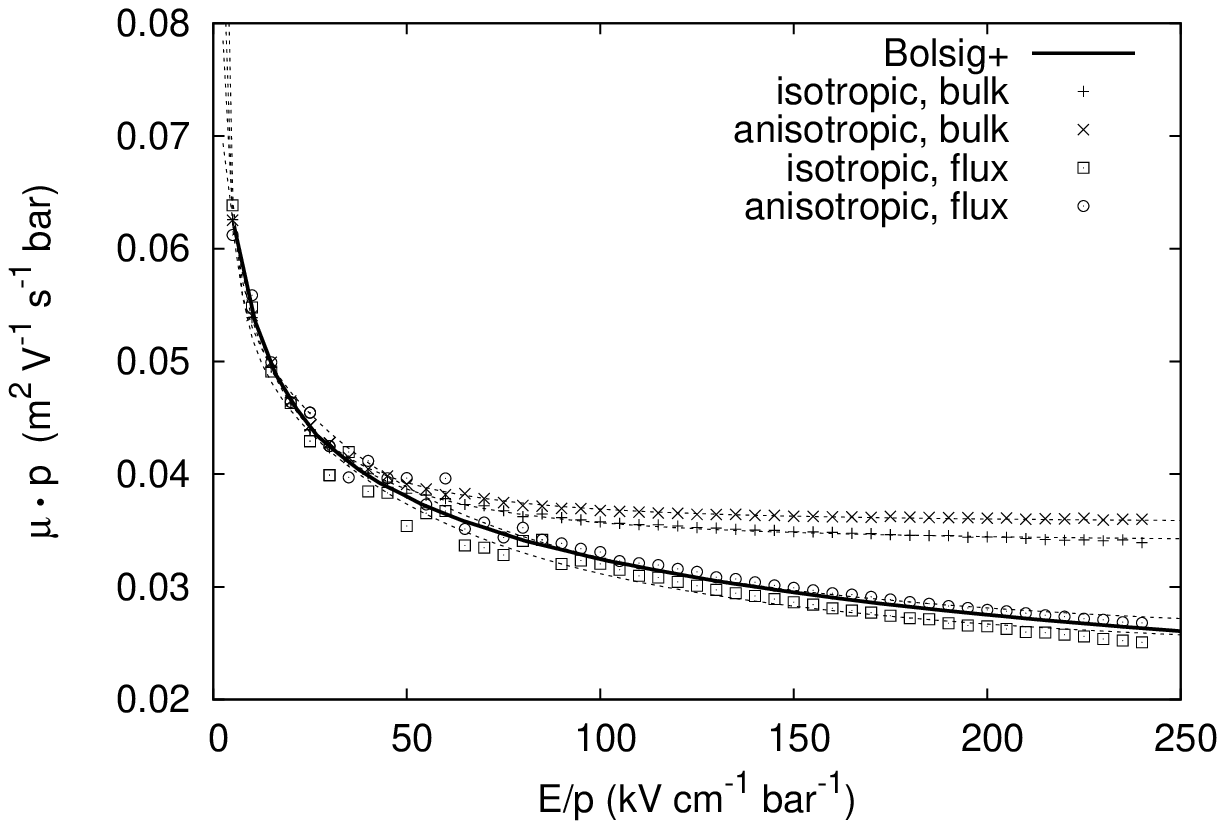}
}
   \subfigure[\label{fig:coe_al} ~Ionization and attachment rates $\alpha(E)$ ]{
         \includegraphics[width=.40\textwidth]{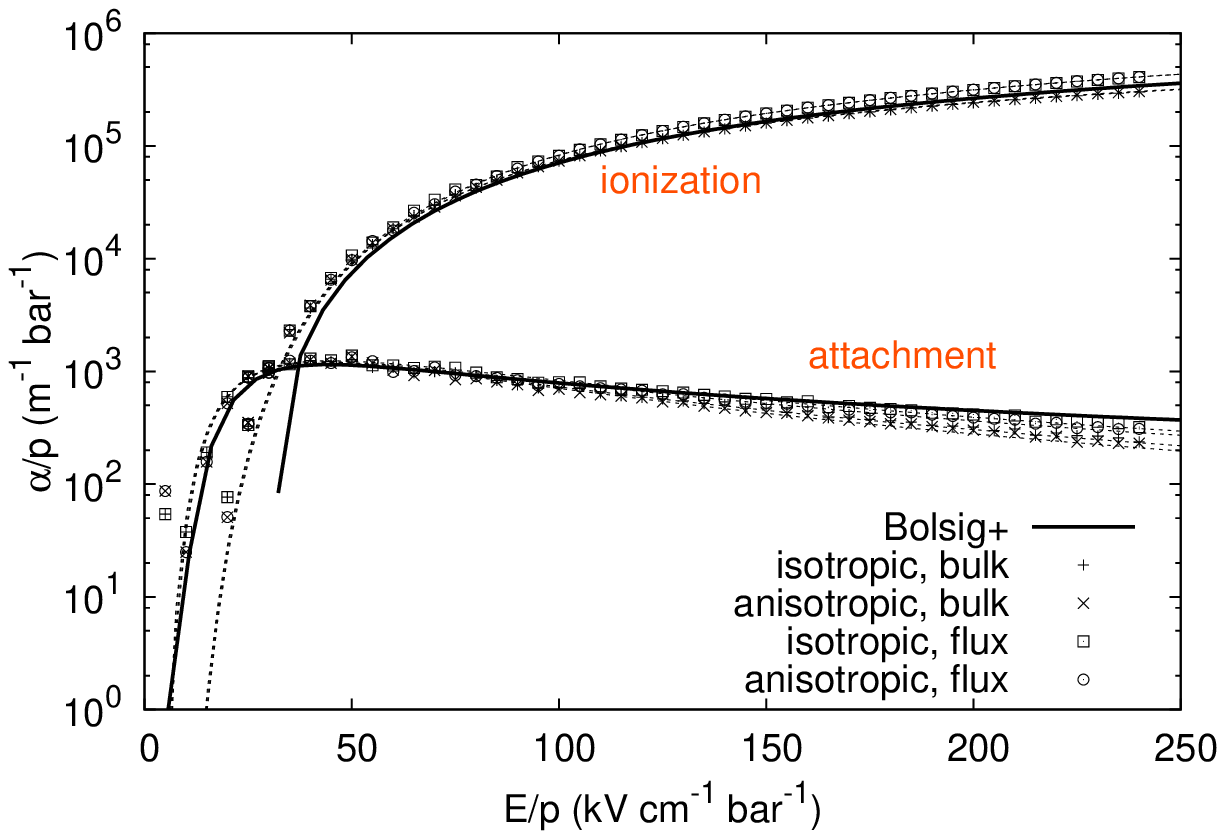}
}
   \subfigure[\label{fig:coe_dr} ~Transversal diffusion $D_T(E)$]{
         \includegraphics[width=.40\textwidth]{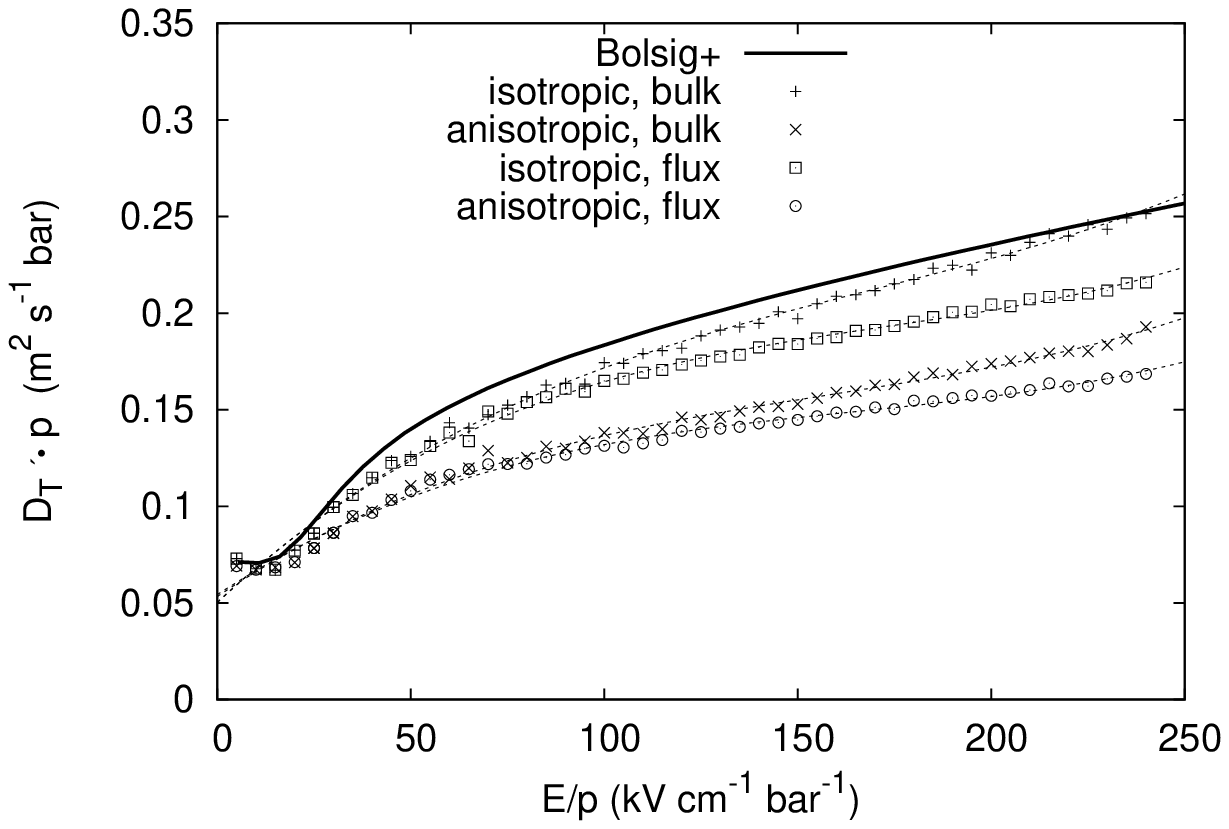}
}
    \subfigure[\label{fig:coe_dl} ~Longitudinal diffusion $D_L(E)$ ]{
         \includegraphics[width=.40\textwidth]{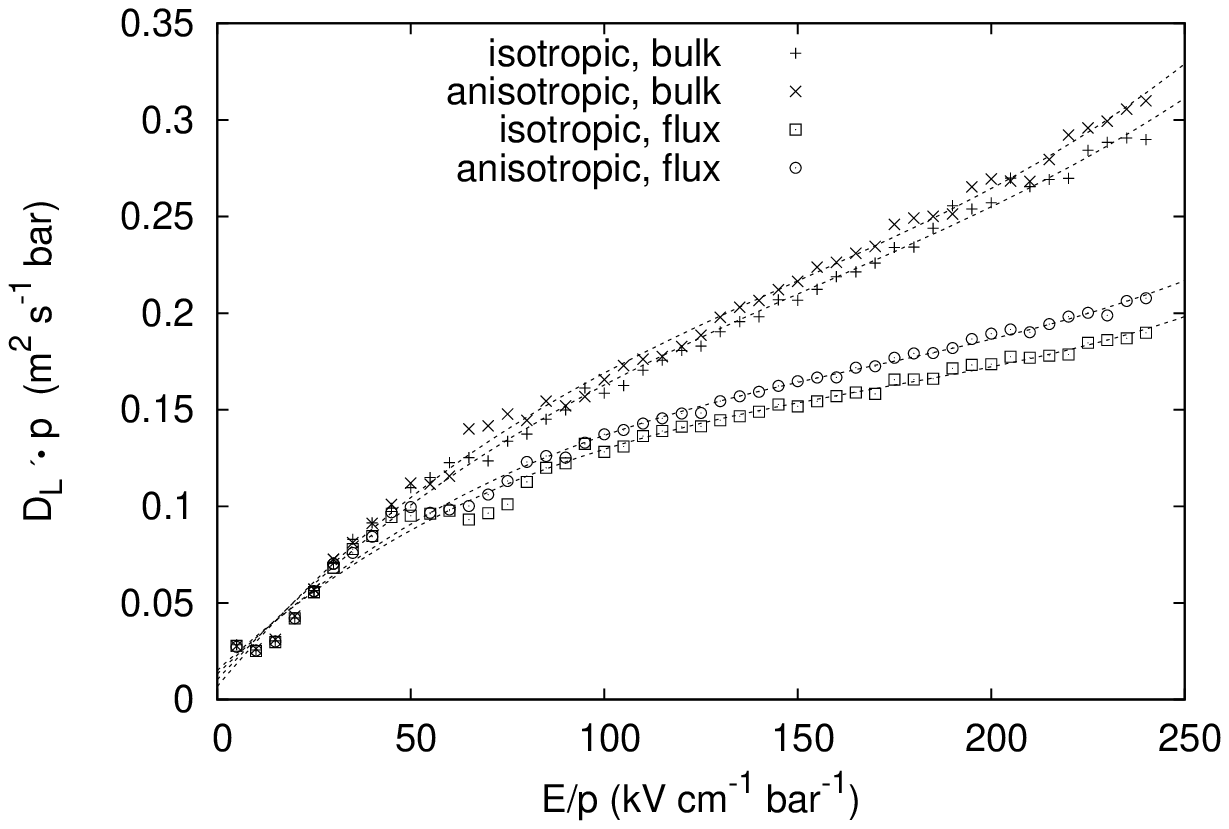}
}
   \subfigure[\label{fig:coe_en} ~Average energy $\varepsilon(E)$]{
         \includegraphics[width=.40\textwidth]{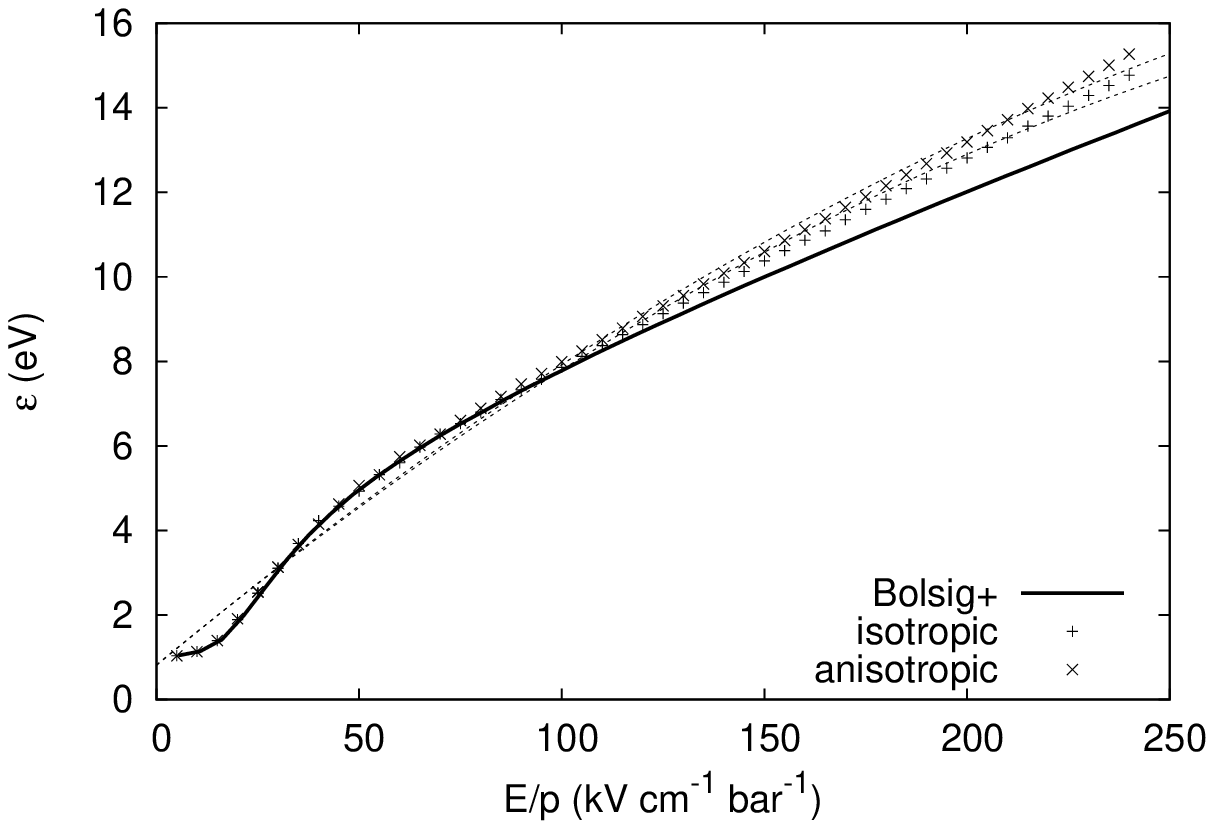}
}
   \subfigure[\label{fig:coe_k1} ~Coefficient k$_1$(E)]{
         \includegraphics[width=.40\textwidth]{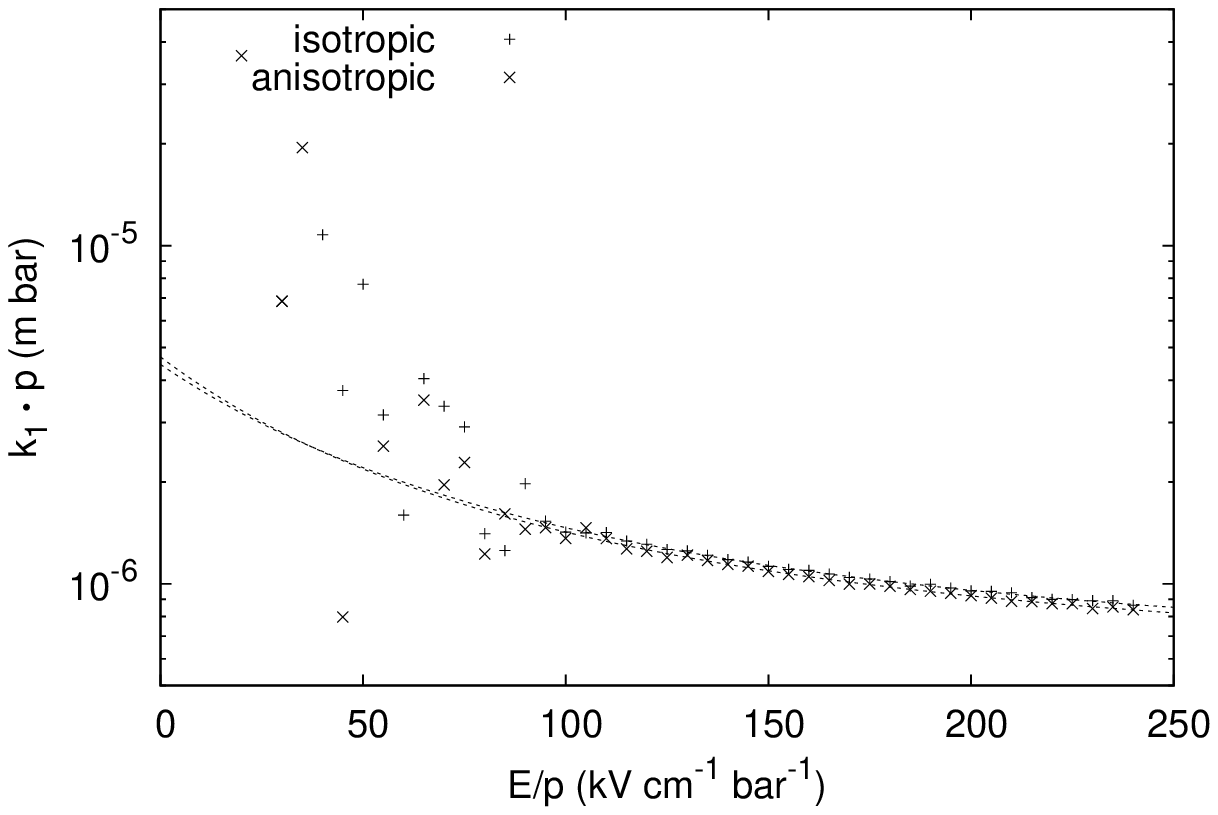}
}
\caption{Shown are the transport coefficients, ionization and attachment rate, and the mean energy of electrons in air. The presented coefficients are the bulk coefficients and the flux coefficients calculated from the particle swarm experiments with both isotropic and anisotropic scattering, and with equal energy sharing in the ionization collision. The results are compared with results of the Boltzmann solver {\sc Bolsig}+~\cite{Siglo,Hag2005}.}
\label{fig:Com_with_Bolsig_Air}
\end{figure}

In Fig.~\ref{fig:Com_with_Bolsig_Air}, the transport parameters: electron mobility $\mu$, electron diffusion rate ${\bf D}$, mean energies $\bar{\epsilon}$, and the ionization $\alpha_i$ together with the attachment rate $\alpha_{att}$, and the nonlocal parameter $k_1$ of electron ensembles, are presented.

They are generated from {\sc bolsig}+ (solid line), the flux coefficients from our particle simulation with isotropic scattering (marked with ``$\Box$''),  flux coefficients with anisotropic scattering (marked with ``o''), the bulk coefficients with isotropic scattering (marked with ``$+$''), and bulk coefficients with anisotropic scattering (marked with ``x''). The presented transport parameters from {\sc bolsig}+ have the same definition with the flux transport coefficients. We also made the empirical fittings for both flux and bulk coefficients from our particle swarm experiments and they are plotted with dashed lines.

As shown in Fig.~\ref{fig:Com_with_Bolsig_Air}, the coefficients with isotropic or anisotropic scattering have no difference for low electric field ($E$ $<$ 50 kV/cm). In the high field range ($E$ $>$ 100 kV/cm), the electron swarm simulated with anisotropic scattering has slightly larger mean energies and slightly higher mobilities than that of isotropic scattering;
 the transversal diffusion rate is larger with isotropic scattering and the longitudinal diffusion rate is larger with anisotropic scattering;
 the ionization rates are higher with anisotropic scattering but almost undistinguishable.
In general, the transport coefficients obtained with isotropic scattering are very close to the one with anisotropic scattering.
The flux coefficients from our particle swarm simulations is undistinguishable to the {\sc Bolsig}+ calculated coefficients for the low field, while small differences remain at strong electric fields.

The fitting functions for the bulk coefficients with the anisotropic scattering are
\ba
\mu(\bar{E}) ~/ {\rm m^2V^{-1}s^{-1}} & = & \exp\left[{-3.02-6.22\times10^{-2}\cdot \ln{\bar{E}}+2.19/{\bar{E}} - (1.67 /{\bar{E}})^2} \right] \nonumber \\
\alpha_i(\bar{E})~ /{\rm m^{-1}}  & = & \exp\left[1.13\times10+3.81\times10^{-1}\cdot \ln{\bar{E}}-1.85\times10^{2}/{\bar{E}}  \right] \nonumber \\
\alpha_{att}(\bar{E})~ /{\rm m^{-1}}  & = & \exp\left[1.63\times10-1.94\cdot \ln{\bar{E}}-8.24\times10/{\bar{E}}  \right] \nonumber \\
D_T(\bar{E})~/ {\rm m^2s^{-1}} & = &   {5.34\times10^{-2}+  1.35\times10^{-3}\cdot {\bar{E}}-6.52\times10^{-6} \cdot \bar{E}^2 + 1.38\times10^{-8} \cdot \bar{E}^3} \nonumber \\
D_L(\bar{E}) ~/ {\rm m^2s^{-1}} & = &   {6.83\times10^{-3}+  2.40\times10^{-3}\cdot {\bar{E}}-1.00\times10^{-5} \cdot \bar{E}^2 + 2.23\times10^{-8} \cdot \bar{E}^3}  \nonumber
\ea
and for the flux coefficients:
\ba
\mu^*(\bar{E}) ~/ {\rm m^2V^{-1}s^{-1}} & = & \exp\left[   {-2.32-2.36\times10^{-1}\cdot \ln{\bar{E}}-4.83\times10^{-1} /{\bar{E}} + (2.94\times10^{-1}/{\bar{E}})^2} \right] \nonumber \\
\alpha^*_i(\bar{E})~ /{\rm m^{-1}}  & = & \exp\left[1.03\times10+6.25\times10^{-1}\cdot \ln{\bar{E}}-1.80\times10^{2}/{\bar{E}}  \right] \nonumber \\
\alpha^*_{att}(\bar{E})~ /{\rm m^{-1}}  & = & \exp\left[1.50\times10-1.65\cdot \ln{\bar{E}}-7.35\times10/{\bar{E}}  \right] \nonumber \\
D_T^*(\bar{E})~/ {\rm m^2s^{-1}} & = &   {5.43\times10^{-2}+  1.31\times10^{-3}\cdot {\bar{E}}-6.69\times10^{-6} \cdot \bar{E}^2 + 6.85\times10^{-8} \cdot \bar{E}^3} \nonumber \\
D_L^*(\bar{E})~/ {\rm m^2s^{-1}} & = &  {1.33\times10^{-2}+  1.95\times10^{-3}\cdot {\bar{E}}-1.17\times10^{-5} \cdot \bar{E}^2 + 2.76\times10^{-9} \cdot \bar{E}^3}   \nonumber \\
k_1(\bar{E})  ~/ {\rm m} & = & 5.58\times10^{-7}+2.94\times10^{-2}/(\bar{E}+8.44\times10).
\ea
We recall that the fitting functions for air in Section.~\ref{sec:swarm_tran_rea} are from particle swarm simulations with Opal's formula applied to distribute the electron energies between two out-comming electrons in an ionization, while the functions presented here using an even spliting between two out-coming electrons in ionization.

\begin{figure}
    \centering
          \includegraphics[width=.6\textwidth]{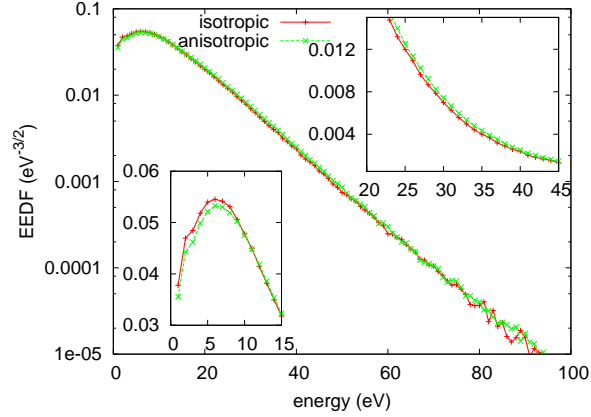}
    \caption{
The electron spectrum in the swarm experiments at 200 kV/cm. The ``$+$'' marked curve shows the particle simulation results with anisotropic scattering. The ``x'' marked curve is obtained from the particle simulation with isotropic scattering. Two subplots zoom into the $0$-$15$ eV and $20$-$45$ eV of the EEDF. The zoomings show the influences of the different scattering methods: more low energy electrons appear when isotropic scattering is used, while more high energy electrons are generated when the anisotropic scattering is used.
}
    \label{fig:en_distribution}
\end{figure}

Although more or less the same swarm and transport parameters can be obtained for the different scattering method, the coefficients from the anisotropic scattering are preferred in our hybrid model. Not only because anisotropic scattering is closer to physical reality,
but also because the small variance in the electron mean energy at relatively high fields can make a difference for the presence of high energy electrons. The choice of the differential cross sections has direct influence on the electron energy distribution function (EEDF). In Fig.~\ref{fig:en_distribution}, we show the EEDF in a particle simulation with both isotropic (curve marked with ``$+$'') and anisotropic (curve marked with ``*'') scattering. The two EEDFs in general agrees with each other very well. However, when we zooming into the low energy part $0$-$15$ eV and slightly higher energy part $20$-$45$ eV of the EEDF, the comparison clearly shows that there are more low energy electrons with isotropic scattering and high energy electrons are easier to produce with anisotropic scattering. That is, the scattering method influences the generation rate of the high energy electrons. Since one goal of the hybrid model is to study the generation of runaway electrons, the anisotropic scattering is used in our particle model, and the parameters from the anisotropic scattering are used in our fluid model.

\subsection{Nitrogen}

\begin{figure}
\centering
   \subfigure[\label{fig:coe_N2mu} ~Mobility $\mu(E)$ ]{
         \includegraphics[width=.40\textwidth]{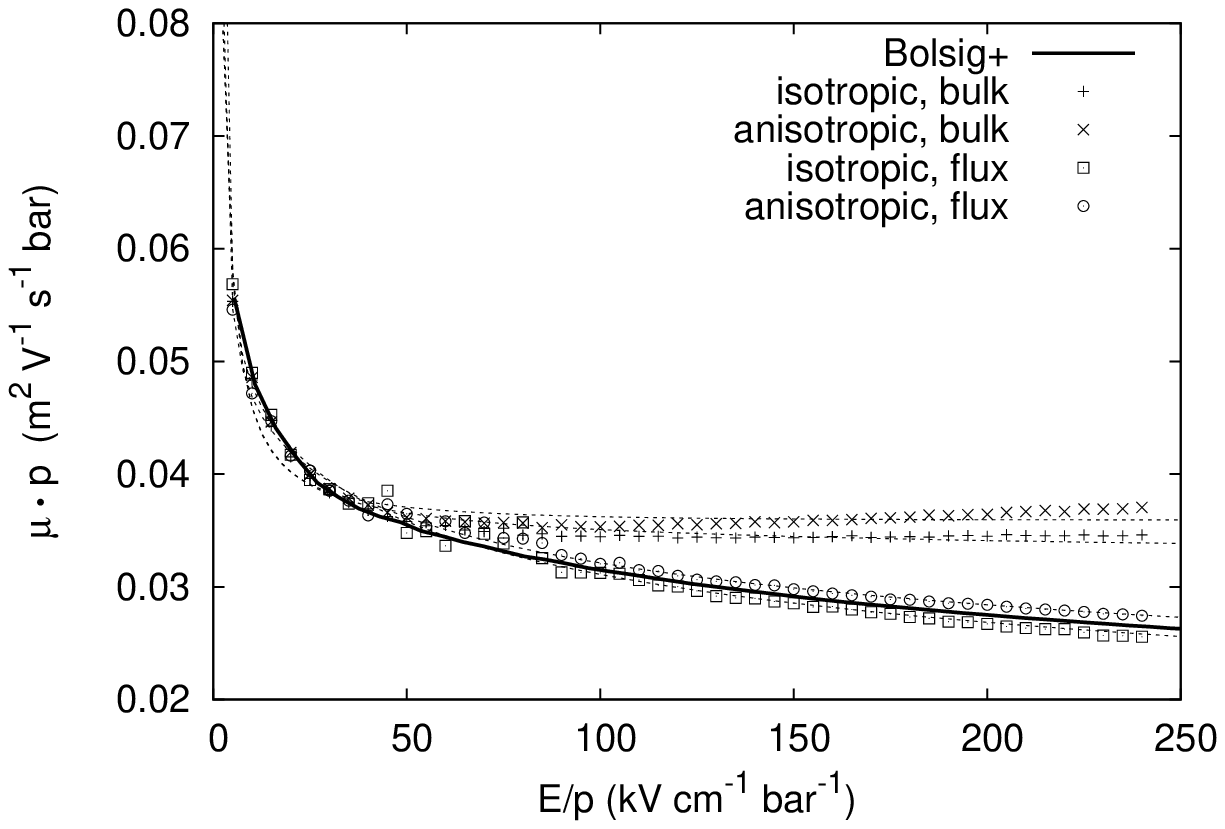}
}
   \subfigure[\label{fig:coe_N2al} ~Ionization rate $\alpha(E)$ ]{
         \includegraphics[width=.40\textwidth]{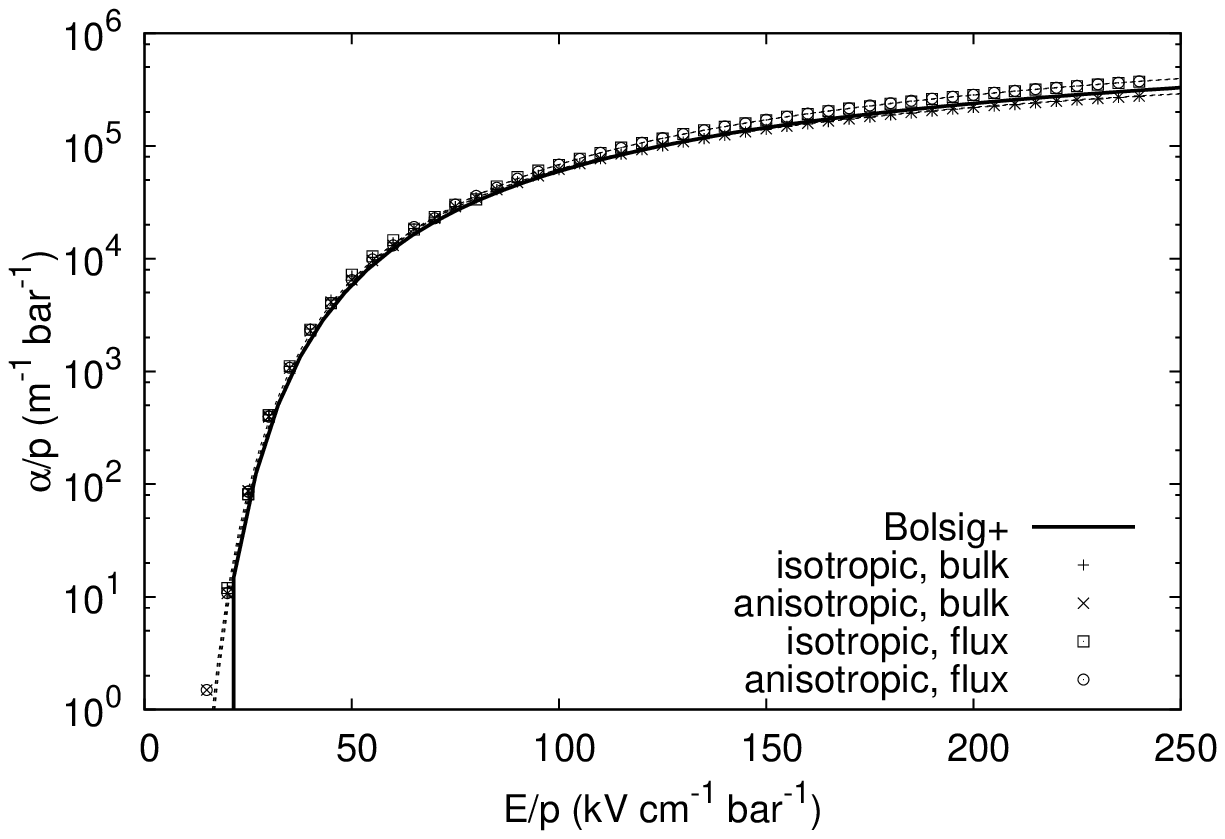}
}
   \subfigure[\label{fig:coe_N2dr} ~Transversal diffusion $D_T(E)$]{
         \includegraphics[width=.40\textwidth]{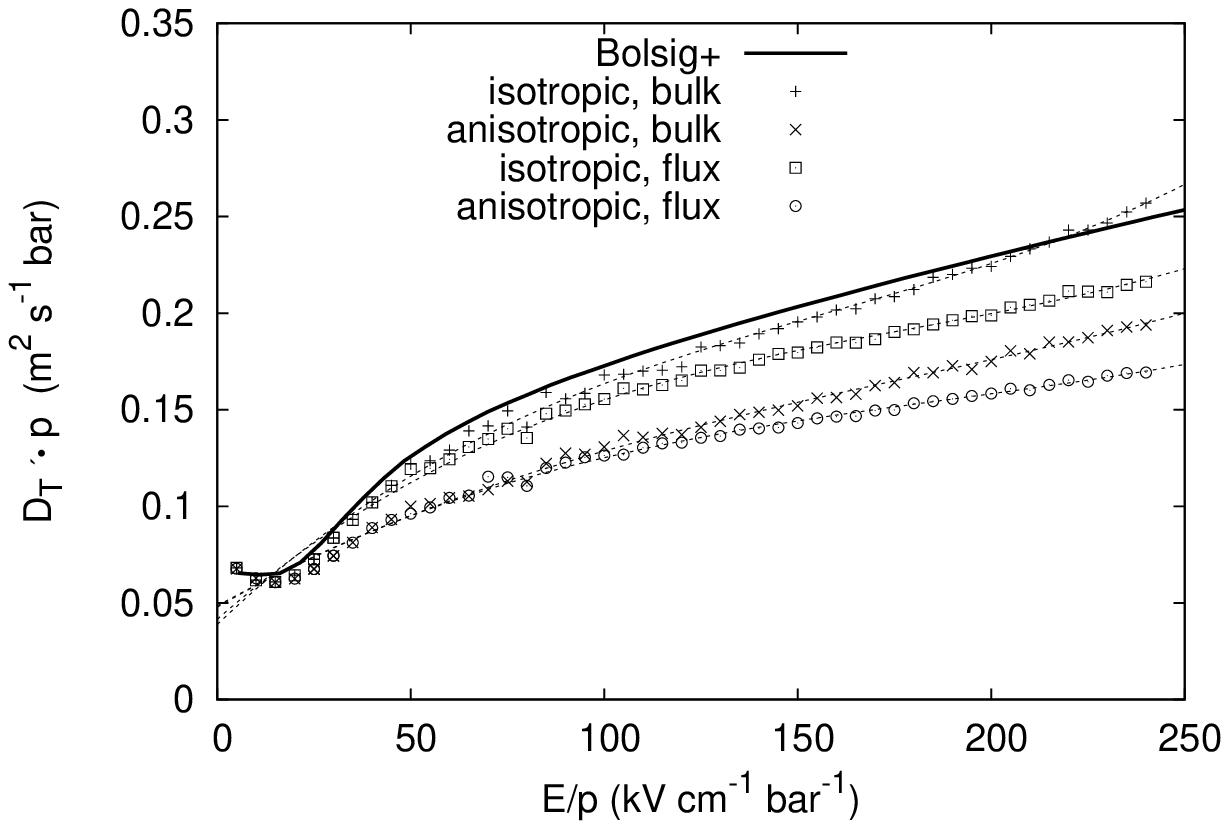}
}
    \subfigure[\label{fig:coe_N2dl} ~Longitudinal diffusion $D_L(E)$ ]{
         \includegraphics[width=.40\textwidth]{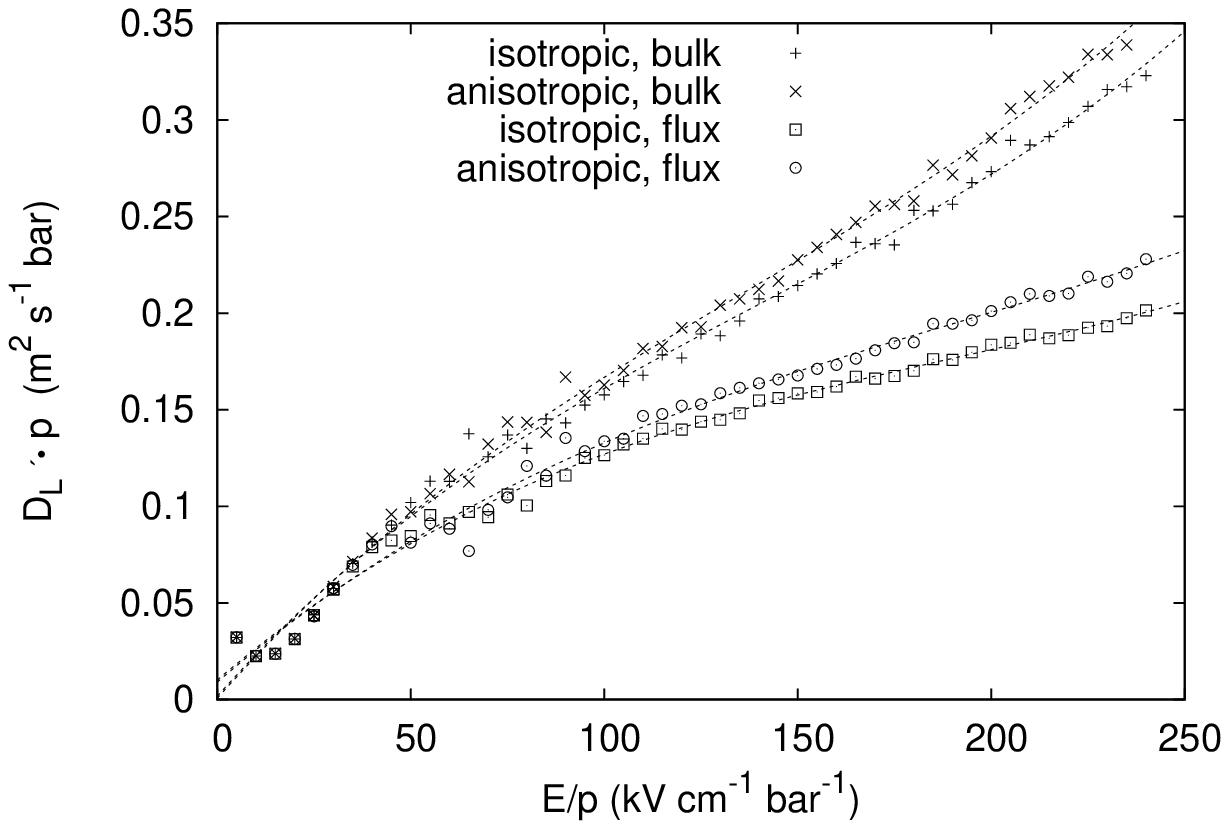}
}
   \subfigure[\label{fig:coe_N2en} ~Average energy $\varepsilon(E)$]{

         \includegraphics[width=.40\textwidth]{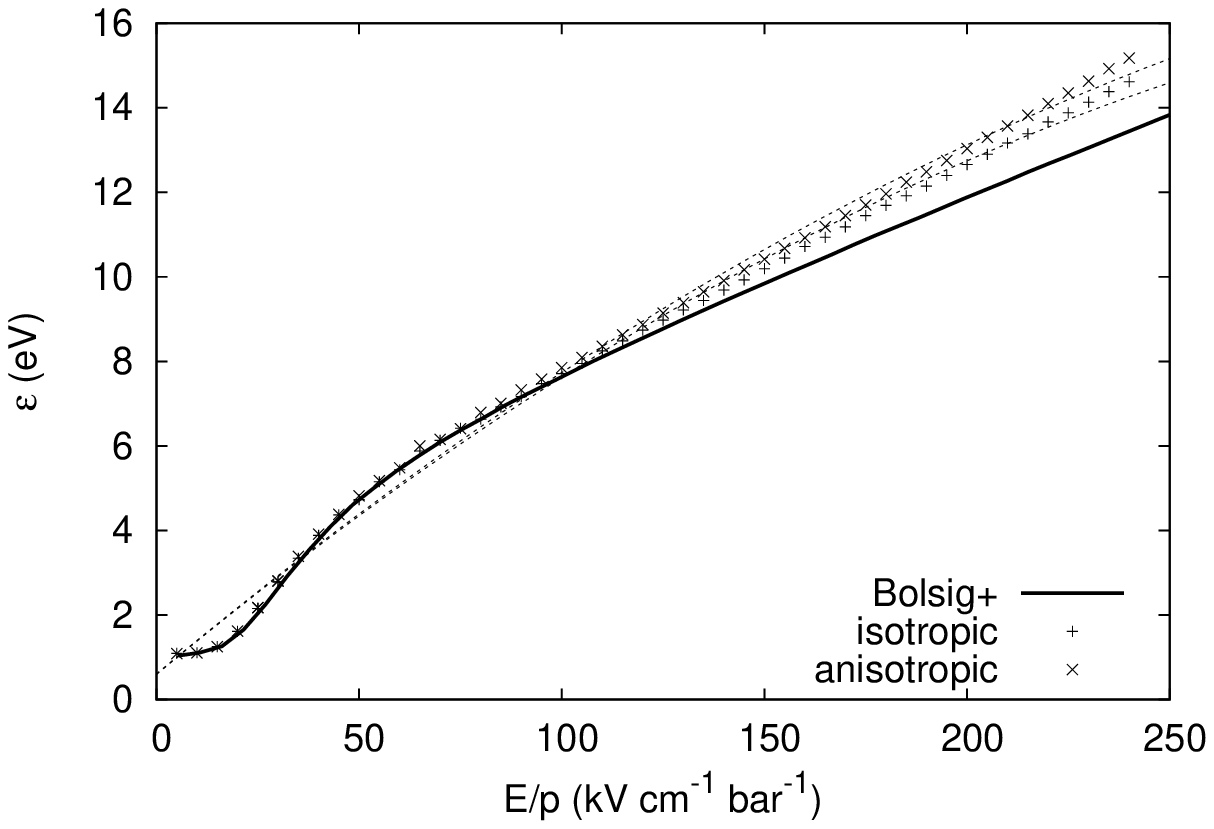}
}
   \subfigure[\label{fig:coe_N2k1} ~Coefficient k$_1(E)$]{
         \includegraphics[width=.40\textwidth]{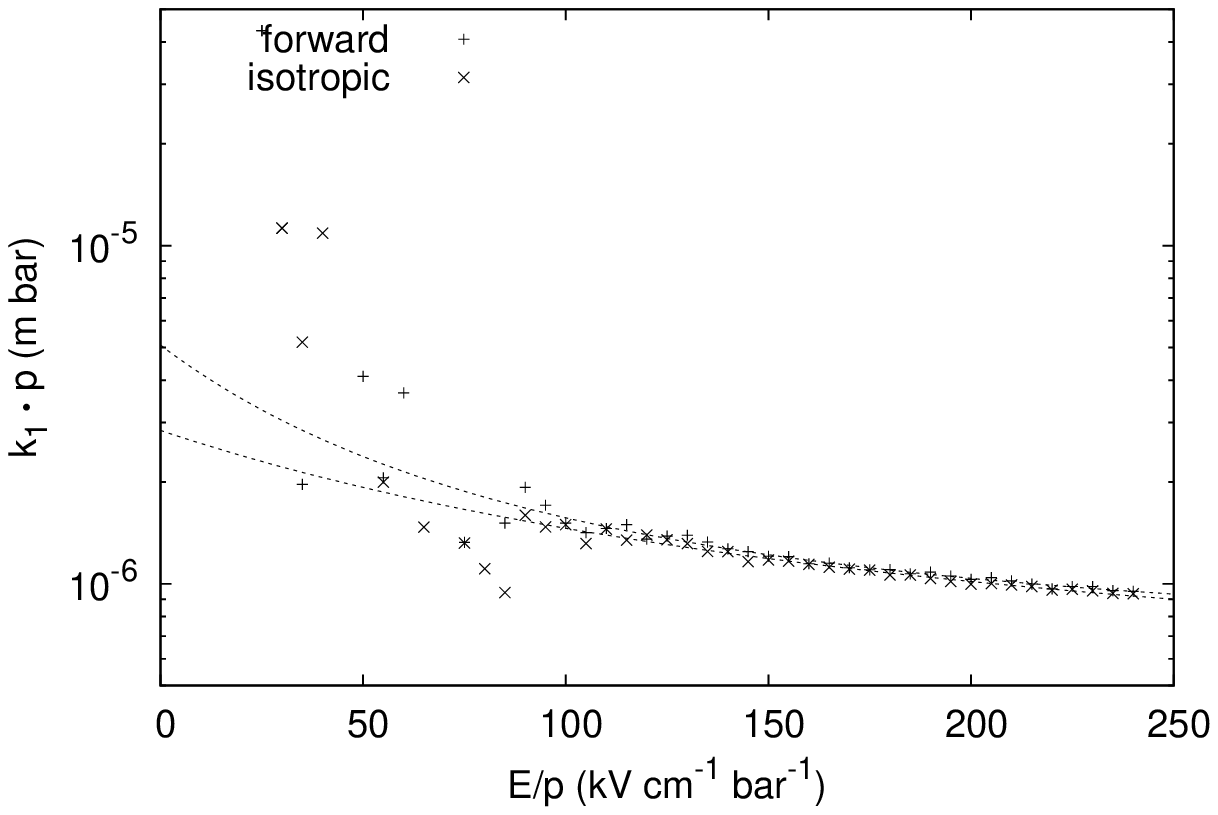}
}
\caption{Nitrogen: transport coefficients, ionization rate, the mean energy of electrons, and $k_1$ presented as in Fig.~\ref{fig:Com_with_Bolsig_Air} for air. }
\label{fig:Com_with_Bolsig_N2}
\end{figure}

The electron transport coefficients and the ionization rates in nitrogen have been reported in~\cite{Li2007}, and the parameter $k_1$ has been reported in~\cite{Li2010:1}. However, the previous particle swarm calculation fixed the total elastic cross sections and didn't take into account the corresponding changes when different scattering methods are applied. The previously calculated parameters with isotropic scattering and anisotropic scattering therefore have noticeable differences, especially at electric fields above 150 kV/cm. Now the momentum elastic cross section is fixed, therefore we here present a new set of the transport parameters and reaction rates for nitrogen.

In Fig.~\ref{fig:Com_with_Bolsig_N2}, the transport coefficients $\mu$, {\bf D}, ionization rate $\alpha_i$, nonlocal parameter k$_1$, and the mean energy of electrons $\bar{\epsilon}$ are presented for nitrogen, in which the $\mu$, $\alpha_i$, $D_T$, and $\bar{\epsilon}$ are also compared with the solution of {\sc bolsig}+. As for air, we were also able to find empirical fittings for the parameters in nitrogen. 

The fitting functions for the bulk coefficients are
\ba
\mu(\bar{E}) ~/ {\rm m^2V^{-1}s^{-1}} & = & \exp\left[{-3.79+8.25\times10^{-2}\cdot \ln{\bar{E}}+8.24/{\bar{E}} - (2.21\times10 /{\bar{E}})^2} \right] \nonumber \\
\alpha_i(\bar{E}) ~ /{\rm m^{-1}} & = & \exp\left[1.17\times10+3.06\times10^{-1}\cdot \ln{\bar{E}}-2.11\times10^{2}/{\bar{E}}  \right] \nonumber \\
D_T(\bar{E})~/ {\rm m^2s^{-1}}  & = &   {4.85\times10^{-2}+  1.10\times10^{-3}\cdot {\bar{E}}-3.67\times10^{-6} \cdot \bar{E}^2 + 6.78\times10^{-9} \cdot \bar{E}^3} \nonumber \\
D_L(\bar{E})~/ {\rm m^2s^{-1}}  & = &   {7.56\times10^{-4}+  2.24\times10^{-3}\cdot {\bar{E}}-7.73\times10^{-6} \cdot \bar{E}^2 + 1.89\times10^{-8} \cdot \bar{E}^3}  \nonumber
\ea
and for the flux coefficients:
\ba
\mu^*(\bar{E}) ~/ {\rm m^2V^{-1}s^{-1}} & = & \exp\left[   {-2.58-1.84\times10^{-1}\cdot \ln{\bar{E}}-9.60\times10^{-1}/{\bar{E}} + (4.11/{\bar{E}})^2} \right] \nonumber \\
\alpha^*_i(\bar{E}) ~ /{\rm m^{-1}} & = & \exp\left[1.05\times10+5.83\times10^{-1}\cdot \ln{\bar{E}}-2.05\times10^{2}/{\bar{E}}  \right] \nonumber \\
D_T^*(\bar{E})~/ {\rm m^2s^{-1}}  & = &   {4.79\times10^{-2}+  1.16\times10^{-3}\cdot {\bar{E}}-4.65\times10^{-6} \cdot \bar{E}^2 + 8.12\times10^{-9} \cdot \bar{E}^3} \nonumber \\
D_L^*(\bar{E}) ~/ {\rm m^2s^{-1}} & = &  {9.08\times10^{-3}+  1.72\times10^{-3}\cdot {\bar{E}}-5.90\times10^{-6} \cdot \bar{E}^2 + 1.03\times10^{-8} \cdot \bar{E}^3}   \nonumber \\
k_1(\bar{E}) ~/ {\rm m}  & = & 4.90\times10^{-7}+7.58\times10^{-2}/(\bar{E}+1.80\times10^{2})^2.
\ea

\subsection{Oxygen}

\begin{figure}
\centering
   \subfigure[\label{fig:coe_O2mu} ~Mobility $\mu(E)$ ]{
         \includegraphics[width=.40\textwidth]{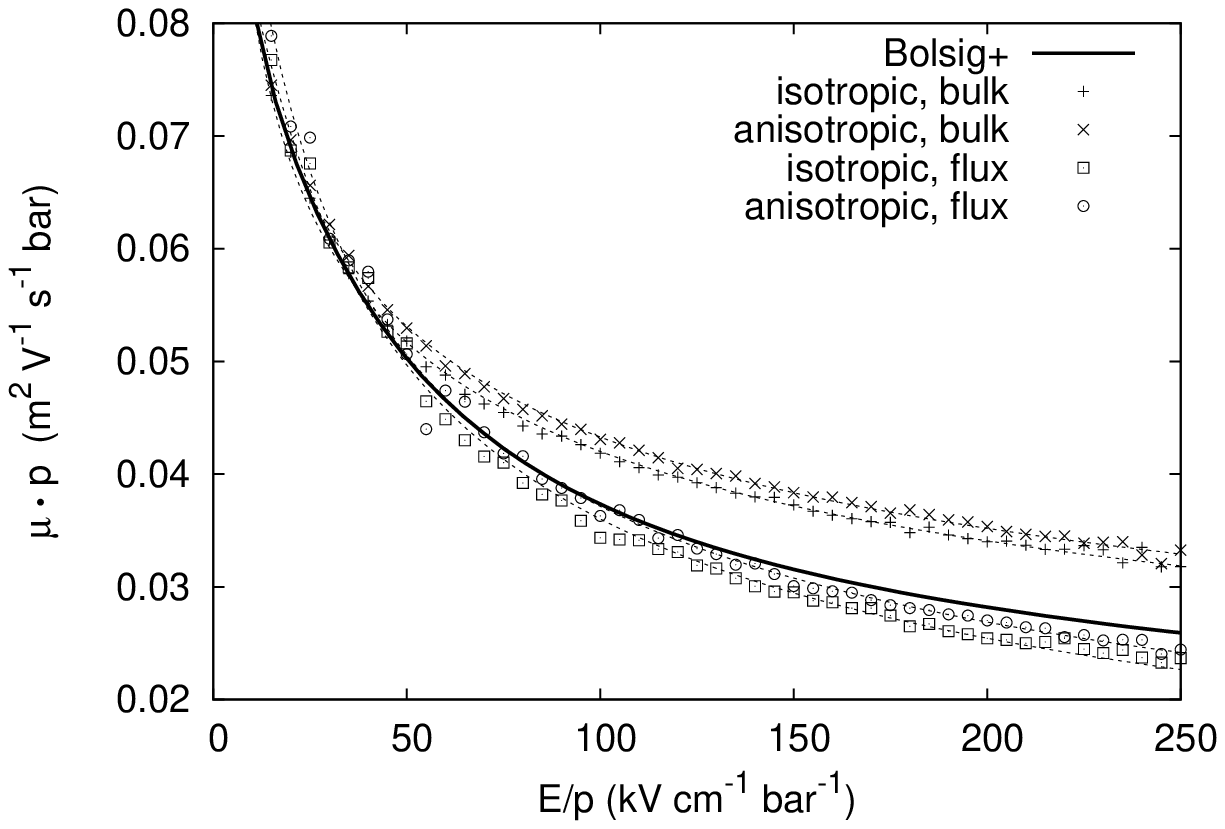}
}
   \subfigure[\label{fig:coe_O2al} ~Ionization and attachment rates $\alpha(E)$ ]{
         \includegraphics[width=.40\textwidth]{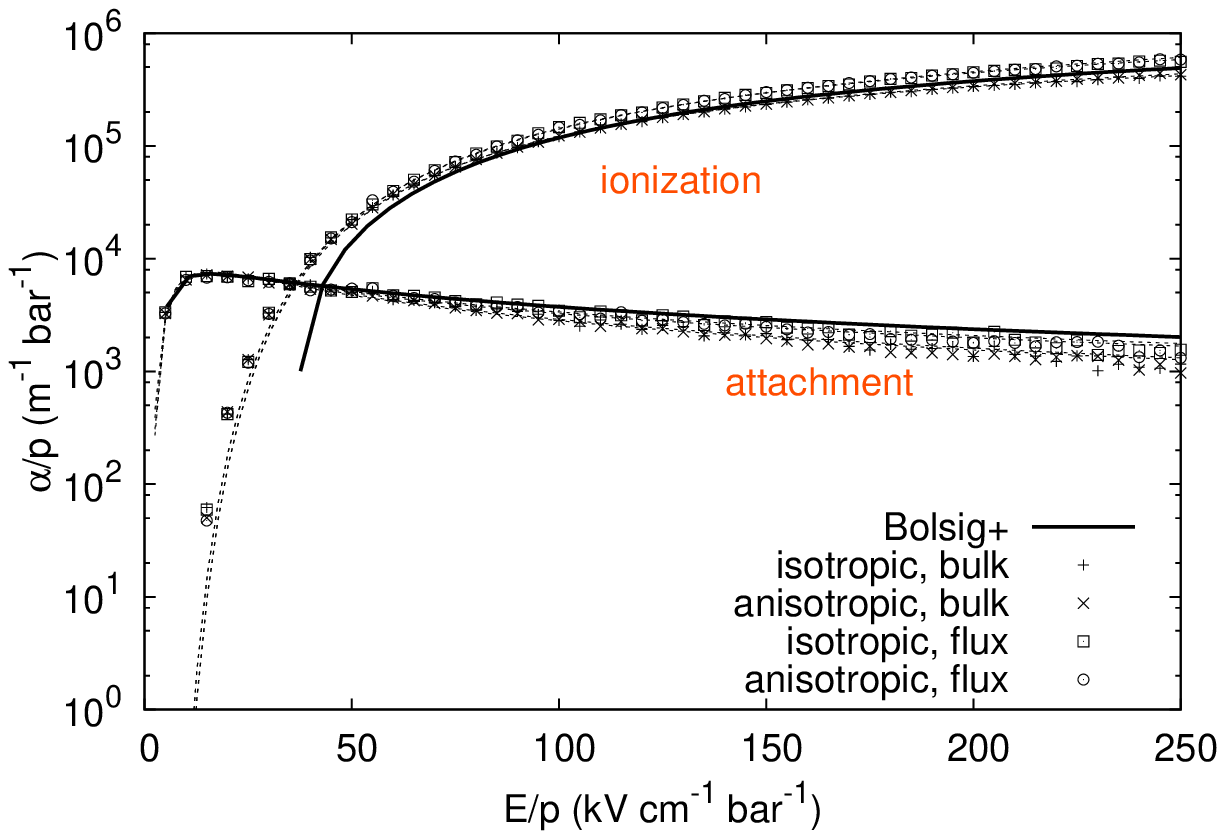}
}
   \subfigure[\label{fig:coe_O2dr} ~Transversal diffusion $D_T(E)$]{
         \includegraphics[width=.40\textwidth]{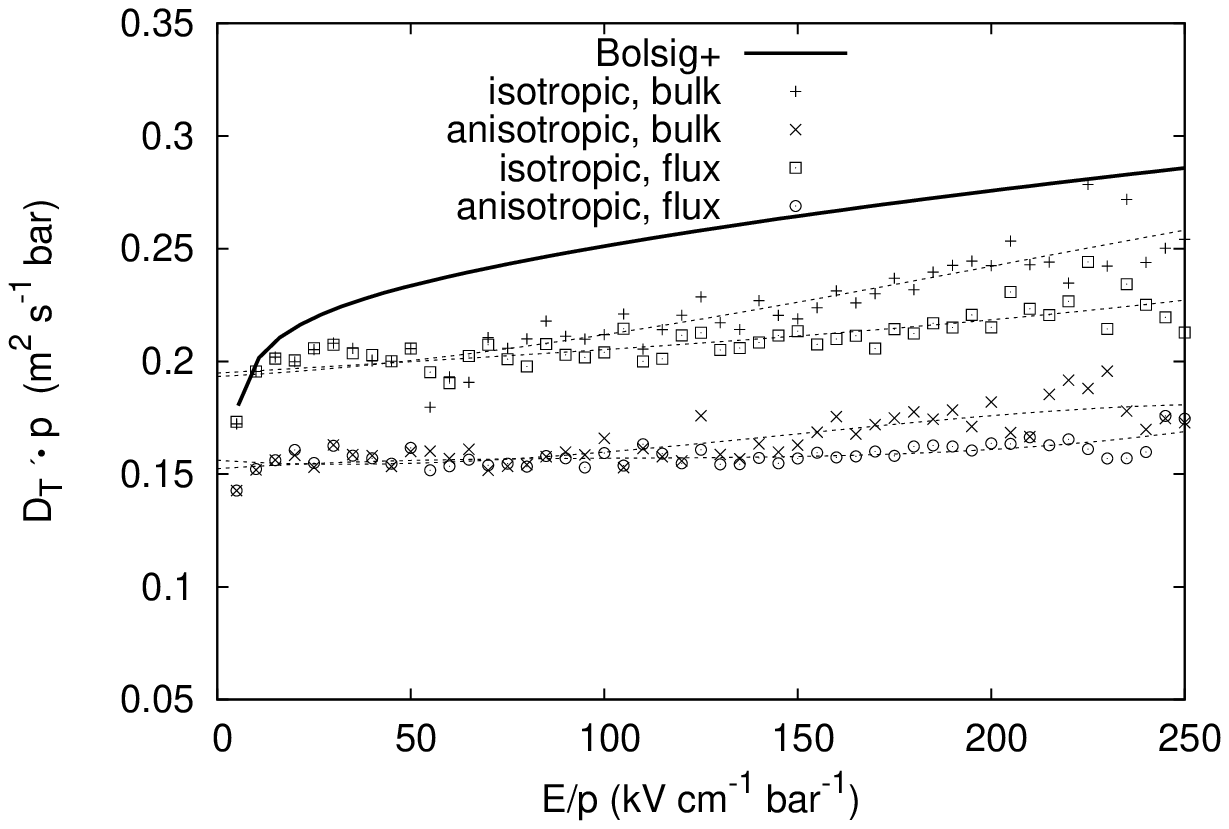}
}
    \subfigure[\label{fig:coe_O2dl} ~Longitudinal diffusion $D_L(E)$ ]{
         \includegraphics[width=.40\textwidth]{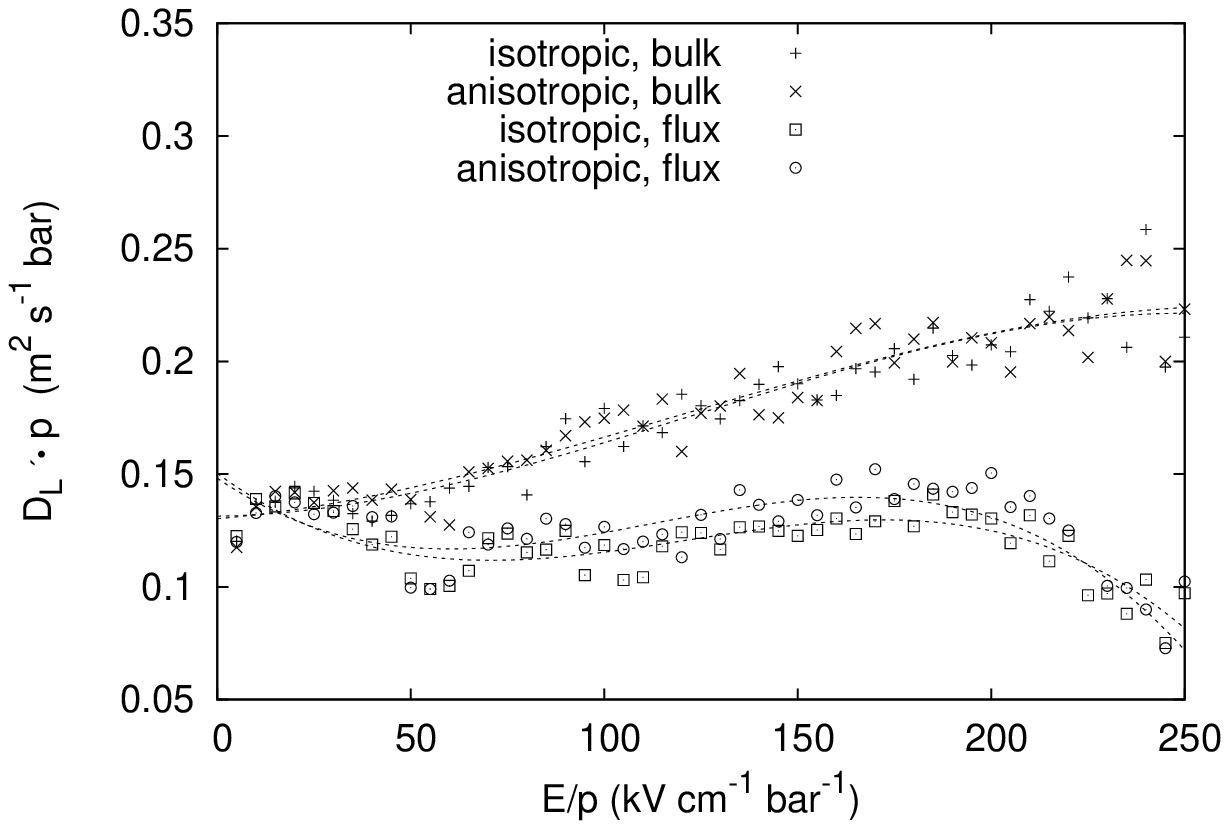}
}
   \subfigure[\label{fig:coe_O2en} ~Average energy $\varepsilon(E)$]{
         \includegraphics[width=.40\textwidth]{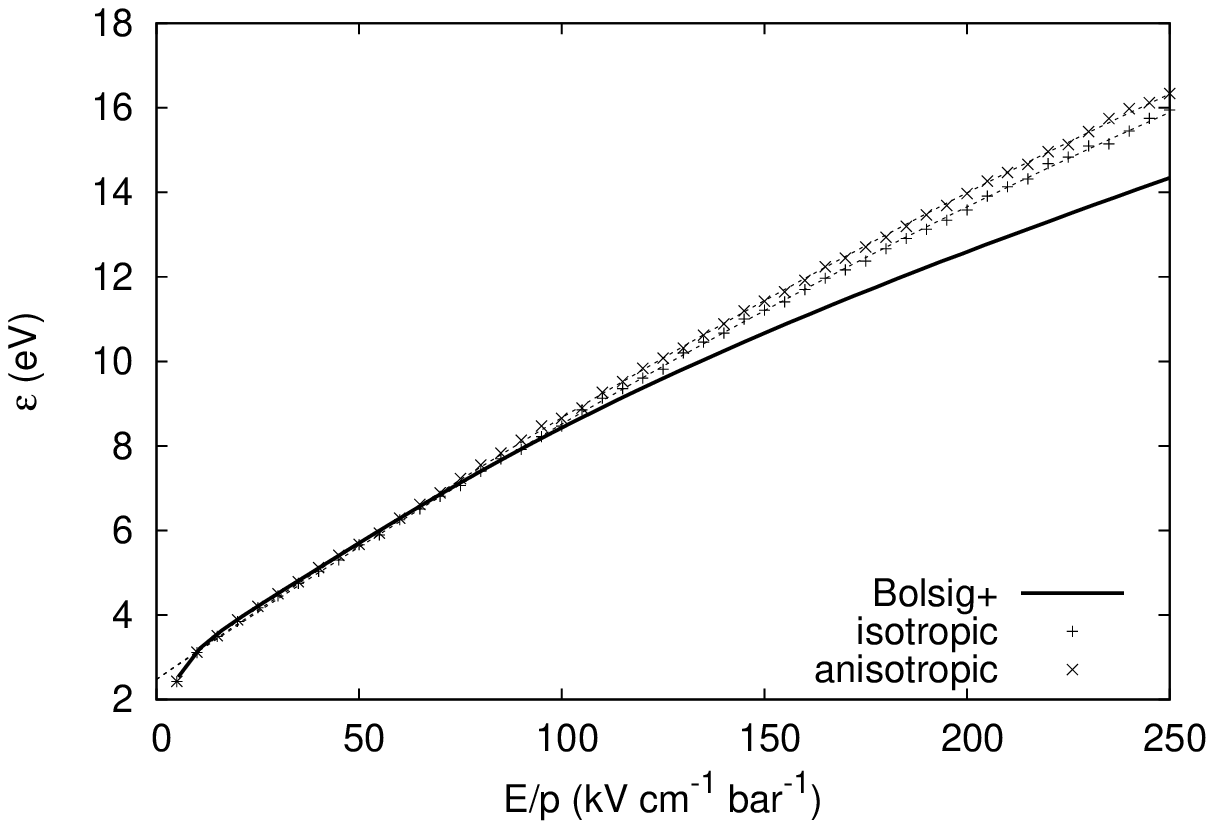}
}
   \subfigure[\label{fig:coe_O2k1} ~Coefficient k$_1(E)$]{
         \includegraphics[width=.40\textwidth]{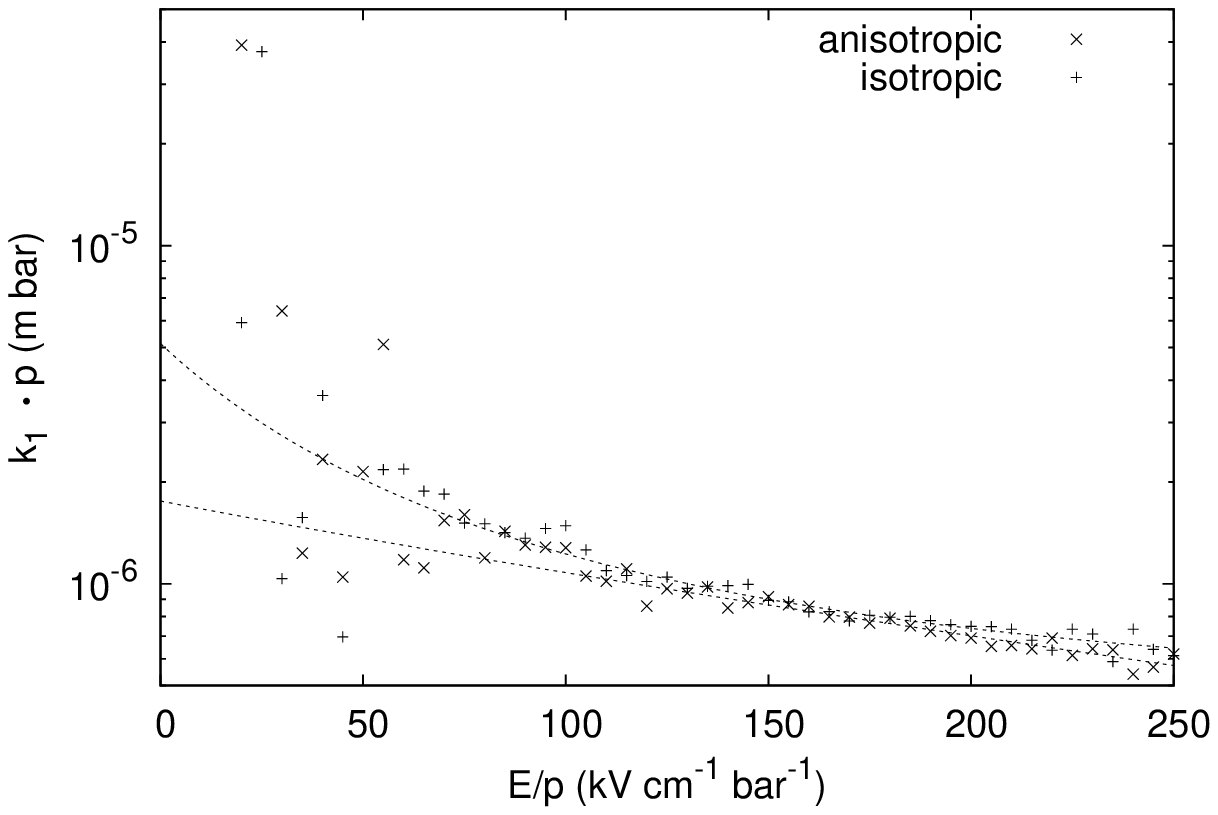}
}
\caption{Oxygen: transport coefficients, ionization and attachment rates, and the mean energy of electrons presented as in Fig.~\ref{fig:Com_with_Bolsig_Air} for air. }
\label{fig:Com_with_Bolsig_O2}
\end{figure}

The electron transport coefficients in oxygen are shown in Fig.~\ref{fig:Com_with_Bolsig_O2}. The electron mobilities $\mu$, ionization rate $\alpha_{i}$ and attachment rates $\alpha_{att}$, transversal diffusion $D_T$, longitudinal diffusion $D_L$, mean energy $\bar{\epsilon}$, and the coefficient $k_1$ from Eq.~\eqref{equ:source_extended} are presented in Fig.~\ref{fig:coe_O2mu}, Fig.~\ref{fig:coe_O2al}, Fig.~\ref{fig:coe_O2dr}, Fig.~\ref{fig:coe_O2dl}, Fig.~\ref{fig:coe_O2en}, Fig.~\ref{fig:coe_O2k1} respectively. As for the air and nitrogen, we compared our coefficients with the calculation results from {\sc bolsig}+ for mobilities, reaction rate, transversal diffusion and the mean energies, and a rather good agreement have been obtained between the flux coefficients calculated from our particle swarm simulation and the {\sc bolsig}+ except the diffusion rates.

The difference between transversal diffusion rate $D_T$ with the isotropic and anisotropic scattering are already remarkable at the very low electric fields. Similar behaviors have been also observed in  $D_T$ in argon and it is not clear the reasons causing this. 
Empirical fittings for the fluid parameters are given in the following. The fitting functions for the bulk coefficients are
\ba
\mu(\bar{E})  ~/ {\rm m^2V^{-1}s^{-1}} & = & \exp\left[   {-1.75-3.02\times10^{-1}\cdot \ln{\bar{E}}-4.34\times10^{-1} /{\bar{E}} - (8.68\times10^{-1} /{\bar{E}})^2} \right] \nonumber \\
\alpha_i(\bar{E}) ~ /{\rm m^{-1}} & = & \exp\left[1.11\times10+4.52\times10^{-1}\cdot \ln{\bar{E}}-1.44\times10^{2}/{\bar{E}}  \right] \nonumber \\
\alpha_{att}(\bar{E})~ /{\rm m^{-1}}  & = & \exp\left[1.26\times10-9.71\times10^{-1}\cdot \ln{\bar{E}}-1.53\times10/{\bar{E}}  \right] \nonumber \\
D_T(\bar{E})~/ {\rm m^2s^{-1}}  & = &   {1.56\times10^{-1}-  1.12\times10^{-4}\cdot {\bar{E}}+1.90\times10^{-6} \cdot \bar{E}^2 - 4.24\times10^{-9} \cdot \bar{E}^3} \nonumber \\
D_L(\bar{E})~/ {\rm m^2s^{-1}}  & = &   {1.30\times10^{-1}+  1.46\times10^{-4}\cdot {\bar{E}}+2.99\times10^{-6} \cdot \bar{E}^2 -8.29\times10^{-9} \cdot \bar{E}^3}  \nonumber
\ea
and for the flux coefficients:
\ba
\mu^*(\bar{E}) ~/ {\rm m^2V^{-1}s^{-1}}  & = & \exp\left[{-1.01-4.89\times10^{-1}\cdot \ln{\bar{E}}-3.16/{\bar{E}} + (3.26 /{\bar{E}})^2} \right] \nonumber \\
\alpha^*_i(\bar{E}) ~ /{\rm m^{-1}} & = & \exp\left[1.01\times10+6.87\times10^{-1}\cdot \ln{\bar{E}}-1.39\times10^{2}/{\bar{E}}  \right] \nonumber \\
\alpha^*_{att}(\bar{E})~ /{\rm m^{-1}}  & = & \exp\left[1.19\times10-8.10\times10^{-1}\cdot \ln{\bar{E}}-1.32\times10/{\bar{E}}  \right] \nonumber \\
D_T^*(\bar{E}) ~/ {\rm m^2s^{-1}} & = &   {1.52\times10^{-1}+  1.17\times10^{-4}\cdot {\bar{E}}-1.04\times10^{-6} \cdot \bar{E}^2 + 3.35\times10^{-9} \cdot \bar{E}^3} \nonumber \\
D_L^*(\bar{E}) ~/ {\rm m^2s^{-1}} & = &  {1.44\times10^{-1}-  1.18\times10^{-4}\cdot {\bar{E}}+1.34\times10^{-5} \cdot \bar{E}^2 -3.94\times10^{-8} \cdot \bar{E}^3}   \nonumber \\
k_1(\bar{E}) ~/ {\rm m}  & = & 4.17\times10^{-7}+2.36\times10^{-2}/(\bar{E}+7.07\times10)^2.
\ea

\subsection{Argon}

\begin{figure}
\centering
   \subfigure[\label{fig:coe_Armu} ~Mobility $\mu(E)$ ]{
         \includegraphics[width=.40\textwidth]{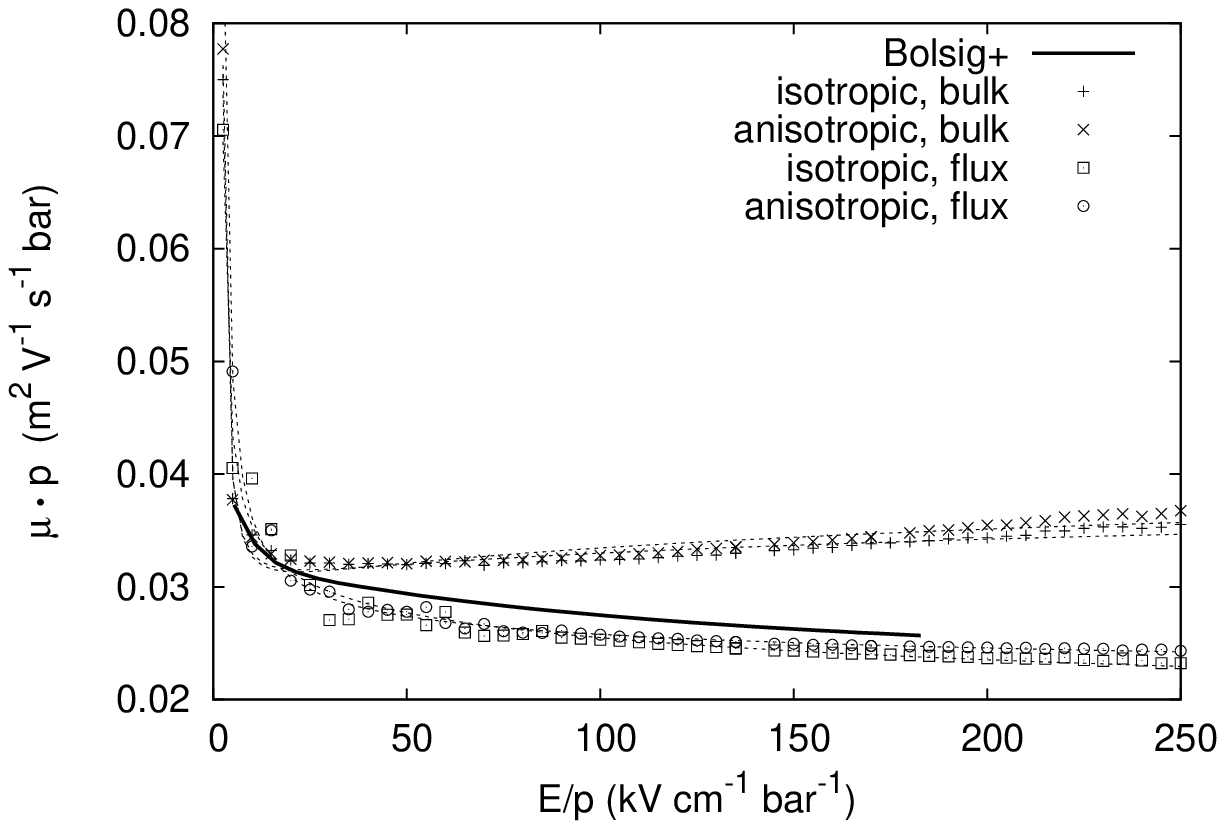}
}
   \subfigure[\label{fig:coe_Aral} ~Ionization rate $\alpha(E)$ ]{
         \includegraphics[width=.40\textwidth]{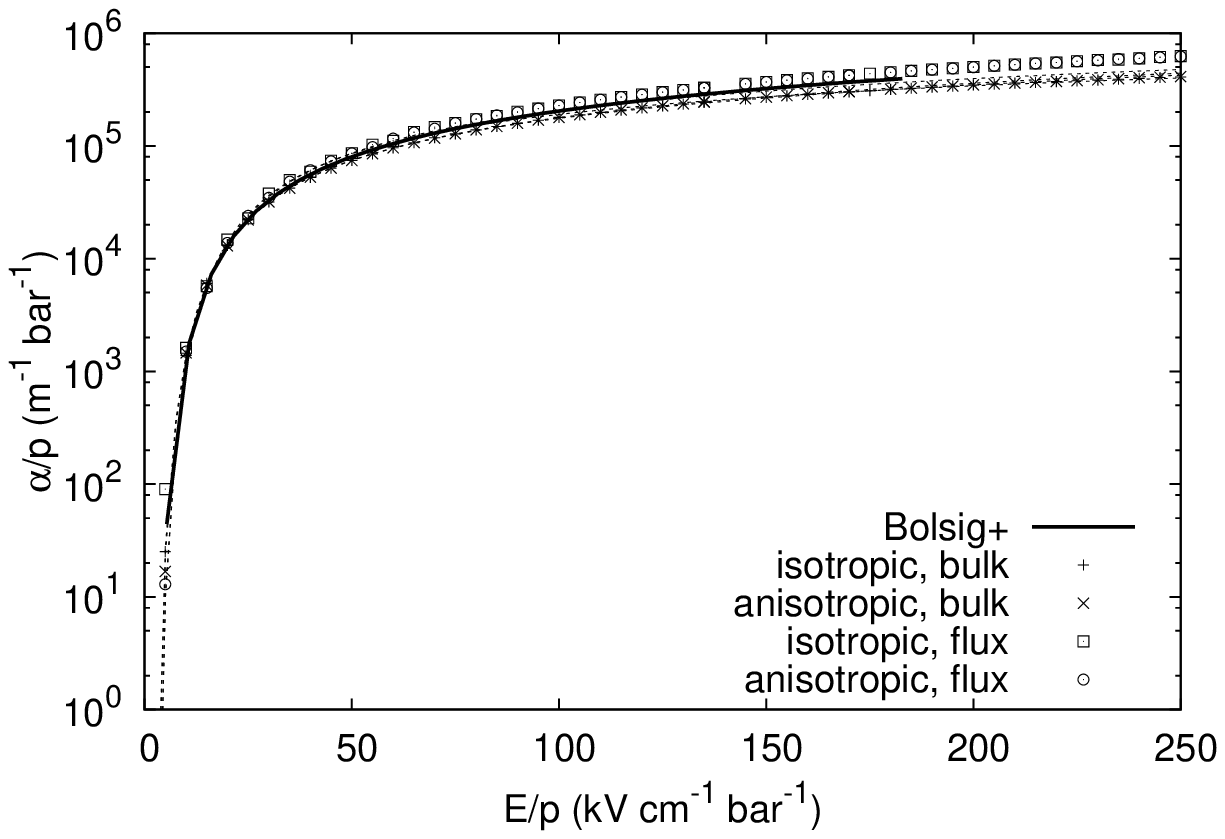}
}
   \subfigure[\label{fig:coe_Ardr} ~Transversal diffusion $D_T(E)$]{
         \includegraphics[width=.40\textwidth]{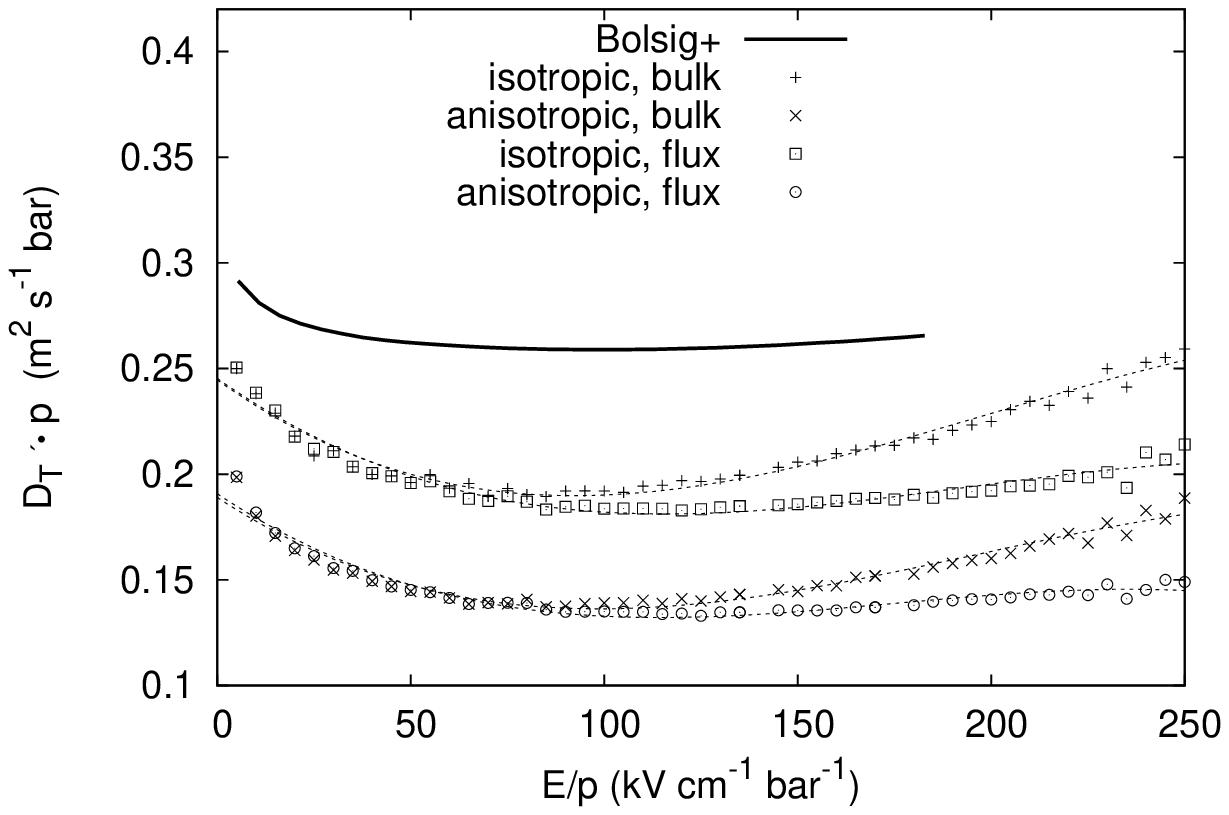}
}
    \subfigure[\label{fig:coe_Ardl} ~Longitudinal diffusion $D_L(E)$ ]{
         \includegraphics[width=.40\textwidth]{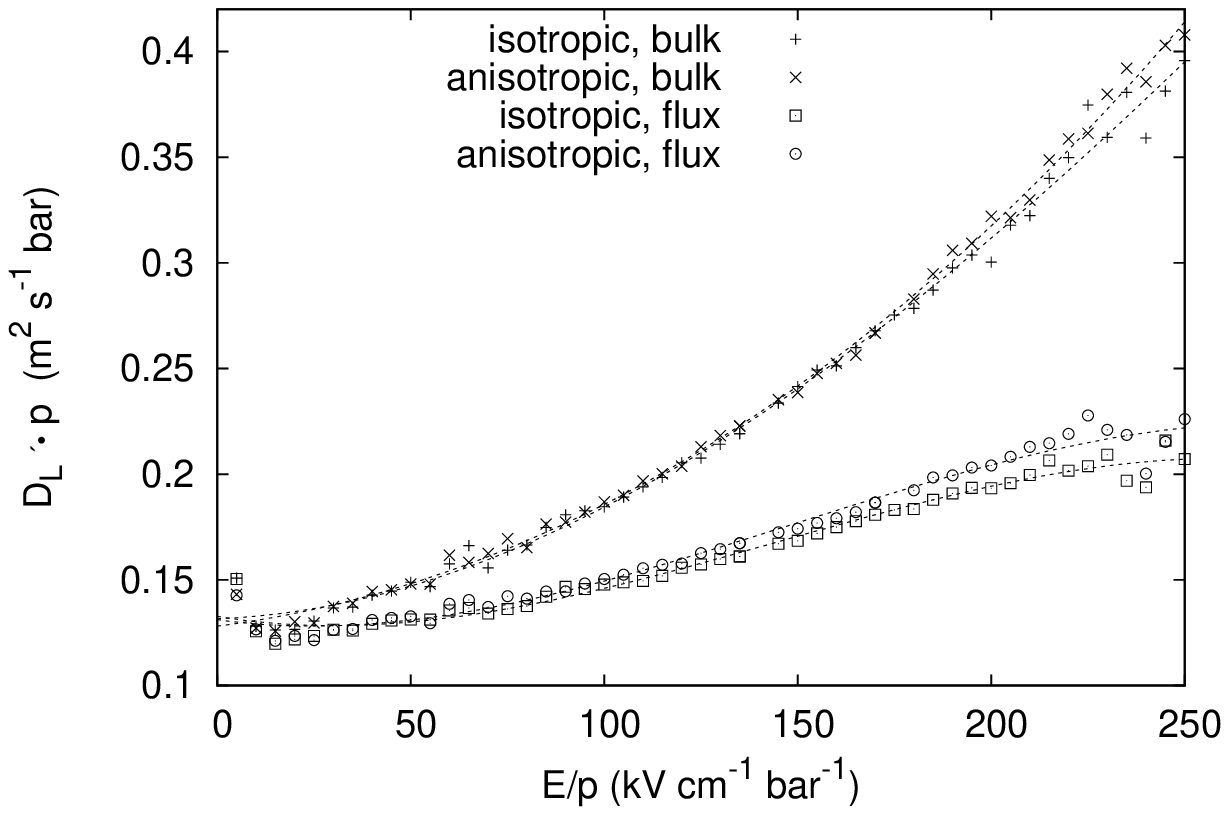}
}
   \subfigure[\label{fig:coe_Aren} ~Average energy $\varepsilon(E)$]{
         \includegraphics[width=.40\textwidth]{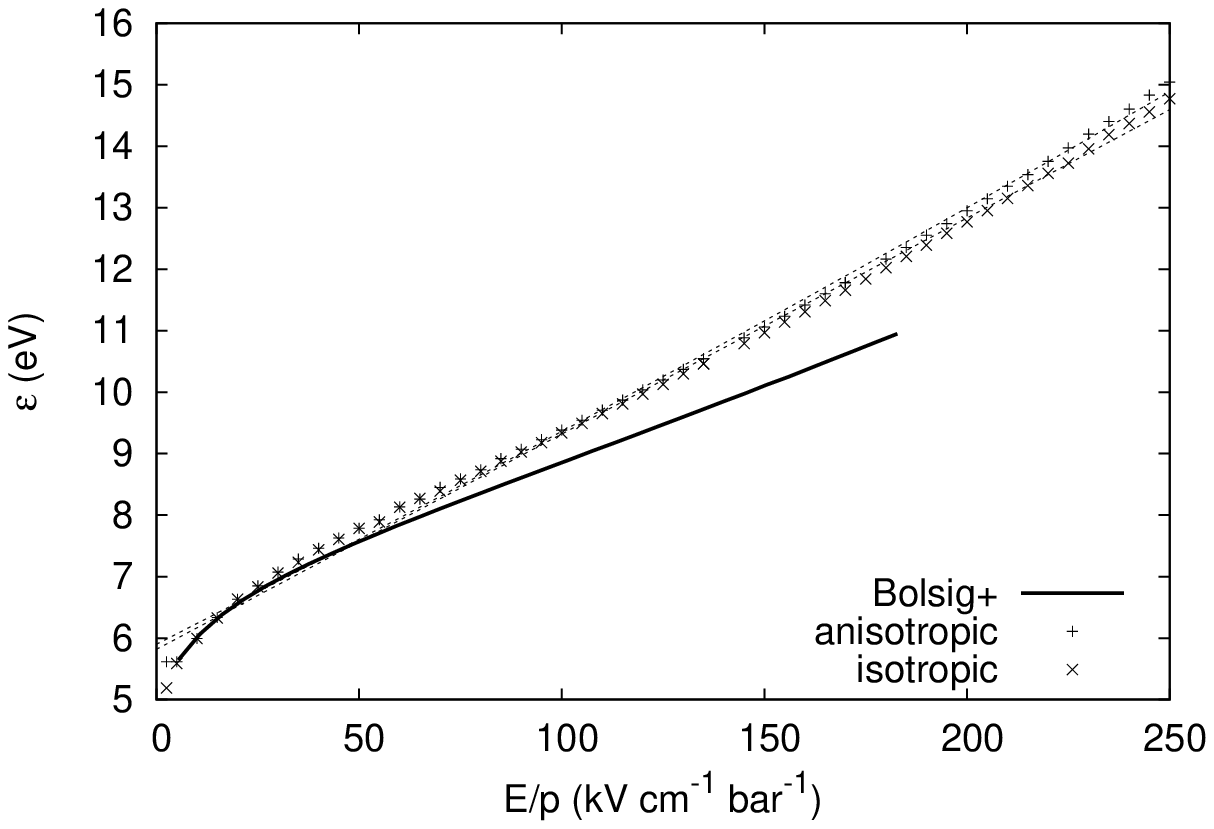}
}
   \subfigure[\label{fig:coe_Ark1} ~Coefficient k$_1(E)$]{
         \includegraphics[width=.40\textwidth]{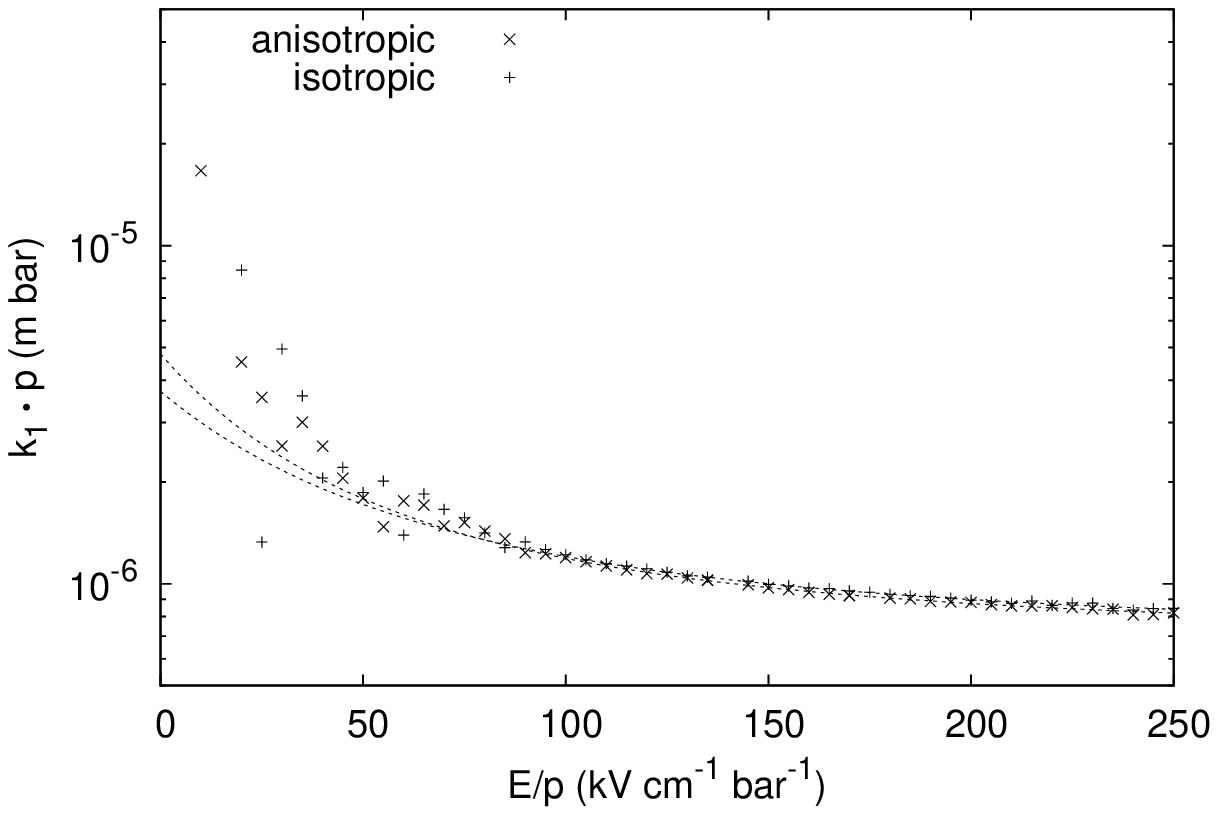}
}
\caption{Argon:  transport coefficients, ionization and attachment rates, and the mean energy of electrons presented as in Fig.~\ref{fig:Com_with_Bolsig_Air} for air.}
\label{fig:Com_with_Bolsig_Ar}
\end{figure}

The same parameters and comparisons for argon can also be found in Fig.~\ref{fig:Com_with_Bolsig_Ar}, and their empirical fittings are listed here. For the bulk coefficients, we have
\ba
\mu(\bar{E})  ~/ {\rm m^2V^{-1}s^{-1}} & = & \exp\left[   {-3.87+7.69\times10^{-2}\cdot \ln{\bar{E}}+2.42/{\bar{E}} - (1.66/{\bar{E}})^2} \right] \nonumber \\
\alpha_i(\bar{E})~ /{\rm m^{-1}} & = & \exp\left[9.07+7.29\times10^{-1}\cdot \ln{\bar{E}}-3.56\times10/{\bar{E}}  \right] \nonumber \\
D_T(\bar{E})  ~/ {\rm m^2s^{-1}} & = &   {1.98\times10^{-1}-  1.45\times10^{-3}\cdot {\bar{E}}+1.01\times10^{-5} \cdot \bar{E}^2 -1.83 \times10^{-8} \cdot \bar{E}^3} \nonumber \\
D_L(\bar{E})  ~/ {\rm m^2s^{-1}} & = &   {1.33\times10^{-1}+  8.87\times10^{-5}\cdot {\bar{E}}+4.40\times10^{-6} \cdot \bar{E}^2 + 1.10\times10^{-9} \cdot \bar{E}^3}  \nonumber
\ea
and for the flux coefficients:
\ba
\mu^*(\bar{E})  ~/ {\rm m^2V^{-1}s^{-1}} & = & \exp\left[{-3.52-3.86\times10^{-2}\cdot \ln{\bar{E}}+3.26\times10^{-1}/{\bar{E}} - (2.57/{\bar{E}})^2} \right] \nonumber \\
\alpha^*_i(\bar{E})~ /{\rm m^{-1}} & = & \exp\left[9.67+6.46\times10^{-1}\cdot \ln{\bar{E}}-4.18\times10/{\bar{E}}  \right] \nonumber \\
D_T^*(\bar{E}) ~/ {\rm m^2s^{-1}}  & = &  {1.96\times10^{-1}-  1.38\times10^{-3}\cdot {\bar{E}}+9.19\times10^{-6} \cdot \bar{E}^2 -1.80\times10^{-8} \cdot \bar{E}^3} \nonumber \\
D_L^*(\bar{E}) ~/ {\rm m^2s^{-1}}  & = &  {1.37\times10^{-1}-  4.20\times10^{-4}\cdot {\bar{E}}+6.93\times10^{-6} \cdot \bar{E}^2 -1.56\times10^{-8} \cdot \bar{E}^3}   \nonumber \\
k_1(\bar{E})  ~/ {\rm m}& = & 6.98\times10^{-7}+1.45\times10^{-2}/(\bar{E}+6.95\times10)^2.
\ea

Note that the {\sc bolsig}+ results for argon is only presented from 5 kV/cm to 180 kV/cm.
Good agreements have been achieved between our flux coefficients and the {\sc bolsig}+ coefficients in electron mobilities, ionization rate. A remarkable difference remains for the transversal diffusion. There is also a good agreement between the two sets for the mean electron energies at low field, but the difference become to noticeable above 50 kV/cm.

\bibliographystyle{elsarticle-num-names}

\end{document}